\documentclass[twocolumn,aps,pre,reprint,floatfix,nofootinbib]{revtex4-1}
\usepackage{graphicx}
\usepackage[caption=false]{subfig}
\usepackage{amsmath}
\usepackage{amssymb}

\usepackage{color}

\usepackage[version=3]{mhchem}

\usepackage{refcount} 
\definecolor{blau}{rgb}{0.1, 0.1, 0.6}
\definecolor{gruen}{rgb}{0.1, 0.6, 0.1}

\DeclareMathOperator{\Tr}{Tr}	

\begin{document}


\title{Superelastic stress-strain behavior in ferrogels of different types of magneto-elastic coupling}

\author{Peet Cremer}
\email{pcremer@thphy.uni-duesseldorf.de}
\affiliation{Institut f{\"u}r Theoretische Physik II: Weiche Materie, 
Heinrich-Heine-Universit{\"a}t D{\"u}sseldorf, D-40225 D{\"u}sseldorf, Germany}
\author{Hartmut L{\"o}wen}
\affiliation{Institut f{\"u}r Theoretische Physik II: Weiche Materie, 
Heinrich-Heine-Universit{\"a}t D{\"u}sseldorf, D-40225 D{\"u}sseldorf, Germany}
\author{Andreas M. Menzel}
\email{menzel@thphy.uni-duesseldorf.de}
\affiliation{Institut f{\"u}r Theoretische Physik II: Weiche Materie, 
Heinrich-Heine-Universit{\"a}t D{\"u}sseldorf, D-40225 D{\"u}sseldorf, Germany}

\date{\today}


\begin{abstract}
Colloidal magnetic particles embedded in an elastic polymer matrix constitute a smart material called ferrogel. 
It responds to an applied external magnetic field by changes in elastic properties, which can be exploited for various applications like dampers, vibration absorbers, or actuators. 
Under appropriate conditions, the stress-strain behavior of a ferrogel can display a fascinating feature: superelasticity, the capability to reversibly deform by a huge amount while barely altering the applied load. 
In a previous work, using numerical simulations, we investigated this behavior assuming that the magnetic moments carried by the embedded particles can freely reorient to minimize their magnetic interaction energy. 
Here, we extend the analysis to ferrogels where restoring torques by the surrounding matrix hinder rotations towards a magnetically favored configuration. 
For example, the particles can be chemically cross-linked into the polymer matrix and the magnetic moments can be fixed to the particle axes. 
We demonstrate that these systems still feature a superelastic regime. 
As before, the nonlinear stress-strain behavior can be reversibly tailored during operation by external magnetic fields. 
Yet, the different coupling of the magnetic moments causes different types of response to external stimuli. 
For instance, an external magnetic field applied parallel to the stretching axis hardly affects the superelastic regime but stiffens the system beyond it. 
Other smart materials featuring superelasticity, e.g.\ metallic shape-memory alloys, have already found widespread applications. 
Our soft polymer systems offer many additional advantages like a typically higher deformability and enhanced biocompatibility combined with high tunability.
\end{abstract}

\maketitle


\section{Introduction}
\label{sec.introduction}

Ferrogels \cite{Zrinyi1995_PolymGelsNetw,Menzel2015_PhysRep,Odenbach2016_ArchApplMech}, also known as soft magnetic materials, magnetic gels, magnetic elastomers, or magnetorheological elastomers, are manufactured by embedding colloidal magnetic particles into an elastic matrix that most often consists of cross-linked polymer. 
This leads to an interplay between magnetic and elastic interactions, allowing to reversibly adjust the material properties via external magnetic fields \cite{Jarkova2003_PhysRevE,Filipcsei2007_AdvPolymSci,Mitsumata2011_PolymChem,Wood2011_PhysRevE,Han2013_IntJSolidsStruct,Mitsumata2013_SoftMatter,Stoll2014_JApplPolymSci,Peroukidis2015_PhysRevE,Peroukidis2015_SoftMatter,Schubert2016_SmartMaterStruct,Wang2016_SmartMaterStruct,Sedlacik2016_ComposStruct}.
On the other hand, dynamically switching the elastic properties allows applications as tunable dampers \cite{Sun2008_PolymTest} or vibration absorbers \cite{Deng2006_SmartMaterStruct}. 
Moreover, shape changes \cite{Filipcsei2007_AdvPolymSci,Stepanov2007_Polymer,Nguyen2010_MacromolChemPhys,Snyder2010_ActaMater,Gong2012_ApplPhysLett} are interesting for the realization of soft actuators \cite{Zhou2005_SmartMaterStruct,Zimmermann2006_JPhysCondensMatter,Boese2012_JIntellMaterSystStruct,Kashima2012_IEEETransMagn,Galipeau2013_ProcRSocA,Allahyarov2014_SmartMaterStruct}.   
Also shape-memory effects have been observed in soft magnetic materials \cite{Nikitin2004_JMagnMagnMater,Stepanov2008_JPhysCondensMatter,Melenev2011_JIntellMaterSystStruct}, opening the way for even more interesting applications. 

Recently we have identified another fascinating feature of soft magnetic materials in a simulation study \cite{Cremer2015_ApplPhysLett}, namely tunable superelasticity. This term was originally introduced in the context of shape-memory alloys \cite{Otsuka2002_MRSBull,Otsuka2005_ProgMaterSci,Liu2013_SciRep}. 
It addresses their special nonlinear stress-strain behavior with a plateau-like regime, where a small additional load leads to a huge additional deformation that is, however, completely reversible.
In shape-memory alloys, the constituents are positioned on regular lattice sites. 
The observed behavior is enabled by a stress-induced transition of the material to a more elongated lattice structure that can accomodate the deformation. 
When the load is released, the shape-memory alloy performs the opposite lattice transition, which renders the whole process reversible.  

In the case of anisotropic soft magnetic gels \cite{Collin2003_MacromolRapidCommun,Bohlius2004_PhysRevE,Filipcsei2007_AdvPolymSci,Guenther2012_SmartMaterStruct,Borin2012_JMagnMagnMater,Han2013_IntJSolidsStruct,Tian2013_MaterResBull}, the superelastic behavior is enabled by stress-induced structural changes.
Such samples can be synthesized by applying a strong external magnetic field during the chemical cross-linking process that forms the elastic matrix. 
Before cross-linking, when the magnetic particles are still mobile, straight chain-like aggregates form along the field direction \cite{Zubarev2000_PhysRevE,Hynninen2005_PhysRevLett,Auernhammer2006_JChemPhys,Smallenburg2012_JPhysCondensMatter}. 
Cross-linking the polymer locks the particle positions into the elastic matrix even after the external field is switched off. 
Our previous numerical study of stretching a magnetic gel containing chain-like aggregates along the direction of the chains revealed the following behavior.
The strong magnetic attractions within the chains first work against the elongation.
However, once the magnetic barriers to detach chained particles are overcome, the material strongly extends.
A part of the stored stress working against the magnetic interactions is released, leading to additional strain without hardly any additional stress necessary. 
This behavior gives rise to to a strongly nonlinear, ``superelastic'' plateau in the stress-strain curve, similar to the phenomenology found for shape-memory alloys.
The strain regime that is covered by this plateau, however, is significantly larger.
Additionally, it is possible to tailor the nonlinear stress-strain behavior by external magnetic fields. 
Combined with the typically higher degree of biocompatibility of soft polymeric materials \cite{ElFeninat2002_AdvEngMater,Liu2007_JMaterChem,Sokolowski2007_BiomedMater,Leng2009_MRSBull,Behl2010_AdvMater}, medical applications \cite{Li2013_AdvFunctMater,Cezar2014_AdvHealthcMater,Cezar2016_ProcNatlAcadSciUSA,Mody2016_JInorgBiochem} might become possible. 

In this previous study \cite{Cremer2015_ApplPhysLett}, we restricted ourselves to the assumption that the magnetic moments of the embedded particles are free to reorient.
First, this is possible when each magnetic moment can reorient within the particle interior, which typically can be observed as the so-called N\'eel mechanism up to particle diameters of to 10--15 nm \cite{Neel1949_AnnGeophys}.
Second, the type of embedding in the elastic matrix can allow the whole particle to rotate, at least quasi-statically, without deforming the matrix, e.g., when in the vicinity of the particles the cross-linking of the polymer matrix is inhibited \cite{Gundermann2014_SmartMaterStruct}.
Finally, yolk-shell colloidal particles feature a magnetic core that can rotate relatively to the nonmagnetic shell surrounding it \cite{Liu2012_JMaterChem,Okada2013_Langmuir}. 

Here, we mainly concentrate on the opposite scenario for spherical, rigid magnetic particles.
That is, the magnetic moments are not free to reorient with respect to the embedding matrix. 
Two ingredients are necessary for this purpose. 
First, the matrix must be anchored to the particle surfaces. 
In reality, this can be achieved when chemically the particles themselves act as cross-linkers of the polymer matrix \cite{Fuhrer2009_Small,Frickel2011_JMaterChem,Ilg2013_SoftMatter,Roeder2014_Macromolecules,Roeder2015_PhysChemChemPhys,Weeber2015_JMagnMagnMater}. 
Second, the magnetic moments must not rotate relatively to the particle frames.
This is the case for magnetically anisotropic monodomain particles that are large enough to block the N\'eel mechanism.
Again we can observe superelastic stress-strain behavior in such systems and again the nonlinearity can be tuned by external magnetic fields.
Yet, the response is altered though, due to the different coupling of the magnetic filler particles to the surrounding matrix. 
An external magnetic field parallel to the chain-like aggregates largely leaves the superelastic behavior intact. 
Still, a sufficiently strong perpendicular field rotates the particles out of the initial alignment configuration and gradually removes the nonlinearity from the stress-strain curve.
However, due to the covalent coupling to the elastic matrix counteracting particle rotations, the necessary field strengths to deactivate superelasticity are much higher when compared to the case of freely reorientable magnetic moments.

In Sec.~\ref{sec.simulation} we begin by introducing our numerical model and our simulation technique for measuring the stress-strain behavior. 
Next, in Sec.~\ref{sec.definition_of_systems}, we define several ferrogel systems with different coupling properties between the particles and the surrounding elastic matrix. 
Afterwards, in Sec.~\ref{sec.results_and_discussion}, we analyze the resulting stress-strain behavior for these different systems and the various mechanisms and effects leading to the emerging superelastic features.
We start with the case of vanishing external magnetic field and then proceed to fields parallel and perpendicular to the chain-like aggregates.
Finally, in Sec.~\ref{sec.conclusions}, we conclude by reviewing our results and discussing possible experimental realizations as well as prospective applications.

\section{Numerical model and simulation procedure}
\label{sec.simulation}

The purpose of our simulations is to determine the nonlinear stress-strain behavior of uniaxial ferrogel systems containing chain-like aggregates. 
To achieve this, we require numerical representations of both the polymer matrix and of the embedded colloidal magnetic particles. 

Let us first discuss our representation of the polymer matrix.
We assume that all molecular details of the cross-linked polymer can be ignored, so that we can treat the matrix as a continuous and isotropic elastic medium. 
We tessellate it into a three-dimensional mesh of sufficiently small tetrahedra.
Spherical magnetic particles are embedded into this mesh by approximating their surfaces as sets of planar triangles, which become faces of the tetrahedral mesh. 
This tessellation was enabled by the mesh generation tool \emph{gmsh} \cite{Geuzaine2009_IntJNumerMethEng}, which is based on Delaunay triangulation \cite{Delaunay1934}.
It allows to set a characteristic length scale parameter controlling the typical length of the tetrahedra edges, for which we used $0.35R$, where $R$ is the radius of the particles. 

\begin{figure}[htb]
	\centering
	\includegraphics[width=1.0\columnwidth]{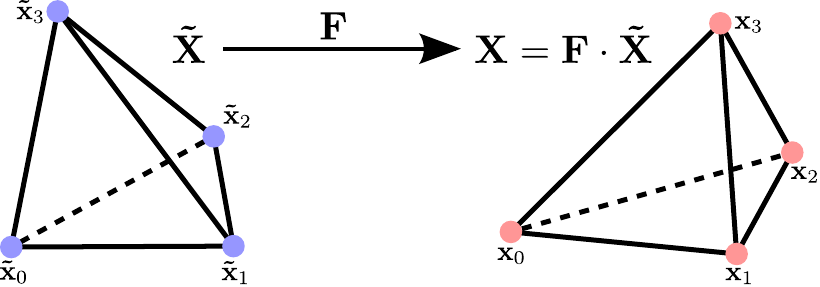}
	\caption{The undeformed state $\mathbf{\tilde{X}}$ of each tetrahedron is determined by the reference node positions $\mathbf{\tilde{x}}_0$, $\mathbf{\tilde{x}}_1$, $\mathbf{\tilde{x}}_2$, $\mathbf{\tilde{x}}_3$ via $\mathbf{\tilde{X}} = ( \mathbf{\tilde{x}}_1 - \mathbf{\tilde{x}}_0,\, \mathbf{\tilde{x}}_2 - \mathbf{\tilde{x}}_0,\, \mathbf{\tilde{x}}_3 - \mathbf{\tilde{x}}_0 )$, while the deformed state $\mathbf{X}$ is characterized by the present node positions $\mathbf{x}_0$, $\mathbf{x}_1$, $\mathbf{x}_2$, $\mathbf{x}_3$ in the form $\mathbf{X} = ( \mathbf{x}_1 - \mathbf{x}_0,\, \mathbf{x}_2 - \mathbf{x}_0,\, \mathbf{x}_3 - \mathbf{x}_0 )$. Both states are connected via the deformation gradient tensor $\mathbf{F}$.}
	\label{fig.deformation_sketch}
\end{figure}

Each tetrahedron of the mesh may deform affinely, which is associated with an elastic deformation energy $U_\textrm{e}$ given by the following nearly-incompressible Neo-Hookean hyperelastic model \cite{Hartmann2003_IntJSolidsStruct}:
\begin{equation}
	\begin{aligned}
		U_\textrm{e} = V_0 &\left[ \left(\frac{\mu}{2} \Tr\left\{\mathbf{F}^t \cdot \mathbf{F}\right\} - 3\right) - \mu\left(\det{\mathbf{F}} - 1\right) \right. \\
			& \left. \quad + \frac{\lambda + \mu}{2} \left(\det{\mathbf{F}} - 1\right)^2 \right] .
	\end{aligned}	
	\label{eq.elasticEnergy_neoHooke}
\end{equation}
Here the elastic properties of the isotropic matrix enter via the Lam\'e coefficients $\mu$ and $\lambda$ \cite{Landau2012_book}. They can also be expressed in terms of the elastic modulus $E$ and the Poisson ratio $\nu$ via $\mu = \frac{E}{2(1 + \nu)}$ and $\lambda = \frac{E \nu}{(1+\nu)(1-2\nu)}$. 
$V_0$ denotes the volume of the tetrahedron in the undeformed state. $\mathbf{F}$ is the deformation gradient tensor prescribing the affine transformation that brings the tetrahedron from its undeformed state to the deformed state. 
The deformed state of the tetrahedron is characterized by the matrix $\mathbf{X} := ( \mathbf{x}_1 - \mathbf{x}_0,\, \mathbf{x}_2 - \mathbf{x}_0,\, \mathbf{x}_3 - \mathbf{x}_0 )$ that contains the current positions $\mathbf{x}_0$, $\mathbf{x}_1$, $\mathbf{x}_2$, $\mathbf{x}_3$ of the four nodes (vertices), see Fig.~\ref{fig.deformation_sketch} for an illustration. 
Similarly, the matrix $\mathbf{\tilde{X}} := ( \mathbf{\tilde{x}}_1 - \mathbf{\tilde{x}}_0,\, \mathbf{\tilde{x}}_2 - \mathbf{\tilde{x}}_0,\, \mathbf{\tilde{x}}_3 - \mathbf{\tilde{x}}_0 )$ determines the undeformed (reference) state of the tetrahedron with node positions $\mathbf{\tilde{x}}_0$, $\mathbf{\tilde{x}}_1$, $\mathbf{\tilde{x}}_2$, $\mathbf{\tilde{x}}_3$. 
Since $\mathbf{F}$ is the affine transformation that connects the deformed state to the reference state, we have $\mathbf{X} = \mathbf{F} \cdot \mathbf{\tilde{X}}$. 
Now the deformation gradient tensor $\mathbf{F}$ can be obtained \cite{Irving2006_GraphModels} by multiplying from the right with $\mathbf{\tilde{X}}^{-1}$ , yielding
\begin{equation}
	\mathbf{F}(\mathbf{X}) = \mathbf{X} \cdot \mathbf{\tilde{X}}^{-1}.
	\label{eq.deformation_gradient_tensor}
\end{equation}
The undeformed reference state never changes, hence the inverse matrix $\mathbf{\tilde{X}}^{-1}$ remains constant and has to be calculated only once. 
This allows to determine the elastic deformation energy $U_e(\mathbf{F(\mathbf{X})})$ in any deformed configuration from the positions of the tetrahedral nodes. 

Calculation of the force $\mathbf{f}_i$ on each node $i$ ($i = 0,1,2,3$) is then straightforward,
\begin{equation}
	\mathbf{f}_i = -\nabla_{\mathbf{x}_i} U_e(\mathbf{F}) = -\frac{\partial U_e(\mathbf{F}) }{\partial \mathbf{F}} \cdot \frac{\partial \mathbf{F}}{\partial \mathbf{x}_i} \, .
	\label{eq.node_force}
\end{equation}
These forces allow us to determine the displacements of the nodes. 
The characterization of the elastic matrix is thus completed. 
In a second step, we turn to the embedded rigid particles. 
Since they are rigid objects, we have to treat nodes attached to particle surfaces in a special way.
The forces on these nodes are transmitted to the corresponding particle, which leads to net forces and torques on the particles.
Rotations and translations of the particles due to these forces and torques are calculated.
They, in turn, determine the displacements of the surface nodes.
We perform a parallel calculation of all node forces in the system by slicing it into different sections, that can be handled separately. 

Next we discuss our representation of the magnetic interactions. 
We assume that all $N$ magnetic particles possess permanent dipolar magnetic moments of equal magnitude $m$. 
This leads us to a total magnetic interaction energy given by
\begin{equation}
	\begin{aligned}    
 		U_m &= \frac{\mu_0}{4\pi} \sum_{i=1}^N \sum_{j = 1}^{i-1} \frac{\mathbf{m}_i \cdot \mathbf{m}_j - 3\left(\mathbf{m}_i \cdot \mathbf{\hat{r}}_{ij} \right)\left(\mathbf{m}_j \cdot \mathbf{\hat{r}}_{ij} \right) }{ r_{ij}^3}	\\ &-\sum_{i=1}^N \mathbf{m}_i \cdot \mathbf{B}.
	\end{aligned}     
	\label{eq.dipole_potential_energy}
\end{equation} 
Here $\mu_0$ is the vacuum permeability, $\mathbf{m}_i$ and $\mathbf{m}_j$ are the magnetic moments of particles $i$ and $j$, respectively, with $|\mathbf{m}_i| = |\mathbf{m}_j| = m$, $\mathbf{r}_{ij} := \mathbf{r}_i - \mathbf{r}_j$ is the separation vector between both particles, $r_{ij}=|\mathbf{r}_{ij}|$ is its magnitude, $\mathbf{\hat{r}}_{ij}=\mathbf{r}_{ij}/r_{ij}$,  and $\mathbf{B}$ is an externally applied magnetic field. 

The magnetic dipolar interaction can be strongly attractive at short distances, when the magnetic moments of interacting particles are in a head-to-tail configuration. 
In order to prevent an unphysical interpenetration of the particles due to such an attraction, we additionally introduce a steric repulsion between the particles that counteracts the attraction at short distances.
The WCA potential \cite{Weeks1971_JChemPhys} 
\begin{equation}
	U_{\text{wca}} = \begin{cases}
		4\epsilon\left[ \left(\frac{\sigma}{r}\right)^{12}-\left(\frac{\sigma}{r}\right)^{6}+\frac{1}{4} \right], & \text{if } r \leq 2^{1/6}\sigma, \\
		0, & \text{if } r> 2^{1/6}\sigma,
	\end{cases}
	\label{eq.wca}
\end{equation}
is hard and finite-ranged and commonly used to represent steric repulsions. 
Its strong scaling with the particle distance compared to the dipolar interactions ($r^{-12}$ vs. $r^{-3}$) makes it the dominating contribution for short distances. 
By setting $\epsilon = \frac{\mu_0}{4\pi} \frac{m^2}{32 R^3}$ and $\sigma=2R$, the dipolar force between two particles, with their magnetic moments aligned in the most attractive head-to-tail configuration, is exactly balanced by the repulsive WCA interaction when they are at contact. 

All these ingredients together express the total energy of the system. 
It is a function of the node positions of the tetrahedral mesh, the particle positions, the particle orientations, and the orientations of the magnetic moments of the particles. 
We equilibrate our systems by performing an energy minimization with respect to these degrees of freedom. 
As a numerical scheme, we employ the \emph{FIRE} algorithm \cite{Bitzek2006_PhysRevLett}, using the forces and torques resulting from Eqs.~\eqref{eq.elasticEnergy_neoHooke}--\eqref{eq.wca} to drive the system towards its energetic minimum. 
\emph{FIRE} is a molecular dynamics scheme that uses adaptive time steps and modifies the velocities resulting from the forces and torques to achieve a quicker relaxation. 
There are several parameters controlling these modifications of velocities and time step, for which we use the values suggested in Ref.~\cite{Bitzek2006_PhysRevLett}. 
Numerical stability is ensured by an upper bound $\Delta t_\text{max}$ for the time step, which we have set to $\Delta t_\text{max} = 0.01$. 
From our experience, this rather simple minimization scheme is quite competitive with more sophisticated schemes like nonlinear conjugate gradient \cite{Hager2006_PacJOptim} that we employed in our earlier work in Ref.~\cite{Cremer2015_ApplPhysLett}. 
In extreme situations of deformation, unphysical behavior may result, such as the inversion of individual tetrahedra or their penetration into the spherical particles.

The physical input parameters for our simulations are the elastic modulus $E$ and Poisson ratio $\nu$ of the matrix, the magnitude $m$ of the magnetic moments and eventually the external magnetic field $\mathbf{B}$. 
We measure forces $F$ in units of $F_0 = ER^2$, magnetic moments in units of $m_0 = R^3 \sqrt{\frac{4\pi}{\mu_0} E}$, and magnetic field strength in units of $B_0 = \sqrt{\frac{\mu_0}{4\pi}E}$. 
Throughout his work, we fix the material parameters by choosing $\nu = 0.495$ and $m = 10\,m_0$. 

Besides the material properties, the behavior of a sample depends on its shape \cite{Diguet2010_JMagnMagnMater,Allahyarov2014_SmartMaterStruct} and on the internal distribution of particles \cite{Zubarev2013_PhysicaA,Pessot2014_JChemPhys,Gundermann2016_unpublished}.
Our characteristic numerical probes are small three-dimensional systems of magnetic particles embedded into an initially rectangular box of elastic material.
The box dimensions are $22.5R \times 10.4R \times 10.4 R$, containing $96$ identical spherical particles.
These particles are arranged into $12$ chain-like aggregates of $8$ particles each. 
All chains are aligned parallel to the long edge of the box (the $x$-direction). 
Neighboring particles in the same chain are initially separated by a finite gap of elastic material of thickness $R/2$. 
The positions of the chains are chosen at random, with the constraint that they shall not overlap and have a minimum distance of $R/4$ to the box boundaries. 
This results in a configuration of chains shifted with respect to each other along their axes by a maximum amount of $2.5R$. 
Since this maximum shift equals the particle diameter plus the gap thickness, there is no statistical preference of any particular particle-gap configuration between two chains. 
Our results are based on $20$ different systems created in this manner, each with a unique particle configuration. 
About 250000 tetrahedra result in each case from the mesh generation.  

To measure the uniaxial stress-strain behavior of such a numerical system, we quasi-statically stretch it along the chain direction, using the following protocol. 
We define two numerical clamps, on the two faces where chains start and end. 
In our geometry, these faces are normal to the $x$-direction. 
All particles on the chain ends are subject to the corresponding numerical clamp.
Particles within the clamps may rotate. 
They may also translate in the $y$- and $z$-direction, however, with the constraint that the center-of-mass displacement of all particles in a clamp is zero.
This keeps the centers of the clamps fixed in the $yz$-plane and prevents an overall rotation of the long axis of the system.
Finally, we prevent global rotations of the whole system around its long axis at all times. 
For this purpose, at each timestep, we determine the global rotational modes from which the rotation is eliminated. 
Overall, this definition of the clamps differs from our approach in Ref.~\cite{Cremer2015_ApplPhysLett}. 
There, the clamps consisted of the complete outer $10\%$ of the system at both ends, that is, besides the particles also all matrix mesh nodes in these volumes were included. 

After switching on the magnetic moments, we perform an initial equilibration process. 
During this period, the clamps are allowed to relatively translate along the $x$-axis. 
However, the relative distance between the particles in a clamp is kept constant along the $x$-direction. 
Due to this initial equilibration, we can observe an initial matrix deformation and define the resulting state as unstretched. 
This sets the equilibrium distance $L_0$ between the clamps as the $x$-separation between the innermost clamped particles. 
To apply a uniaxial strain, we increase the distance between both clamps in small steps, displacing all clamped particles uniformly. 
So we can define the uniaxial strain as $\epsilon_{xx} = \Delta L / L_0$, where $\Delta L$ is the momentary increase in the distance between both clamps. 
After each step, we equilibrate the sample again under the constraint of keeping the $x$-positions of the clamped particles fixed.  
Subsequently, we can extract the force $F$ that has to be applied to the clamps to maintain the system in the prescribed strained state.  
We continue this stress-strain measurement up to a maximum strain of $\Delta L / L_0 = 150\%$ and then gradually unload the system again. 
To check the reversibility of the deformation, we perform several loading and unloading cycles.

\section{Definition of the numerical systems}
\label{sec.definition_of_systems}

\begin{figure}[b!]
	\centering
	\subfloat{\label{fig.snapshot_free_equi}} 
	\subfloat{\label{fig.snapshot_covrightright_equi}} 
	\includegraphics[width = 1.0\columnwidth]{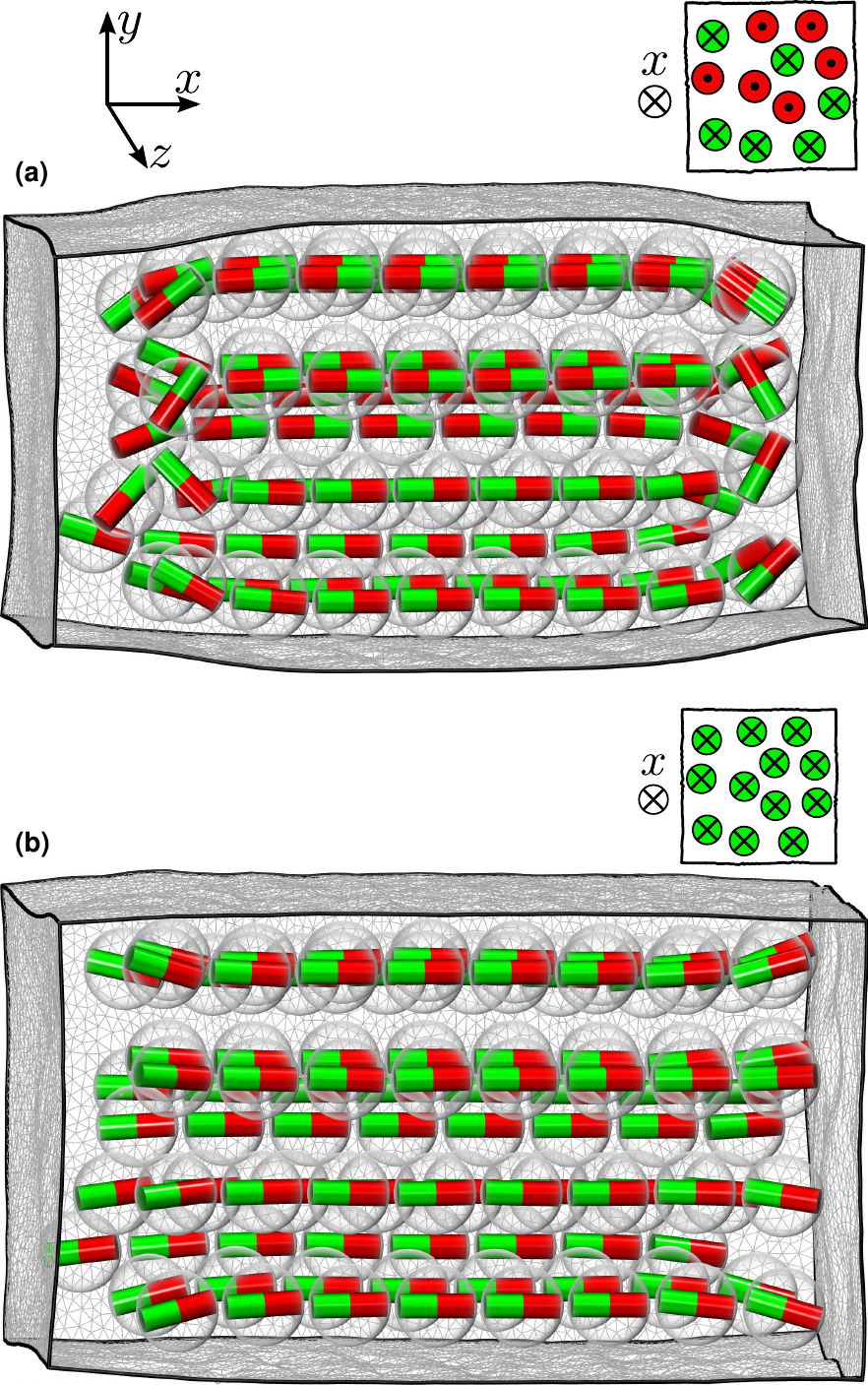}
	\caption{Snapshots of characteristic samples containing chain-like aggregates in the equilibrated unstretched state. The two displayed systems are generated from the same initial placement of the rigid embedded particles. Yet, the way of subsequent anchoring of the magnetic moments, here indicated by small bar magnets, is different, leading to the two different equilibrated states. The matrix was tessellated into a mesh of tetrahedra, those faces of which that constitute the overall system boundaries are depicted explicitly. \protect\subref{fig.snapshot_free_equi} \emph{Free} system, where the magnetic moments can rotate freely with respect to the carrying particles. This leads to opposite alignment of the magnetic moments in different chains, as indicated in the top right for the cross-sectional center plane perpendicular to the chain axes. \protect\subref{fig.snapshot_covrightright_equi} Snapshot for the \emph{cov$\,\rightrightarrows$} system, where the magnetic moments are fixed to the particle axes, likewise including a cross-sectional cut. The snapshot for the \emph{cov$\,\leftrightarrows$} system is by definition again the one shown in \protect\subref{fig.snapshot_free_equi}, because in this system the magnetic moments are fixed to the particle axes only after the initial equilibration in the \emph{free} system.}
	\label{fig.snapshots_equi}
\end{figure}

Within our numerical samples defined above, we distinguish between two scenarios of how the magnetic moments are coupled to the surrounding matrix via their carrying particles. 
Systems showing the first one, which we term \emph{free} systems, feature magnetic moments that can freely rotate relatively to the particle frames and surrounding matrix, see also Ref.~\cite{Cremer2015_ApplPhysLett}. 
In this case, a reorientation of a magnetic moment does not directly induce a deformation of the matrix surrounding the carrying particle.  
Computationally, we treat this system by keeping the orientations of the magnetic moments and the orientations of the carrying particles as separate degrees of freedom.
During the initial equilibration, within the particles constituting one chain, the magnetic moments tend to align parallel to the chain axis.  
The magnetic moments within neighboring chains have the tendency to align in opposite directions to minimize the overall magnetic interaction energy. 
Figure~\ref{fig.snapshot_free_equi} illustrates this situation by showing a snapshot of an equilibrated \emph{free} system before stretching.
A cut along the cross-sectional center plane perpendicular to the chain axes in Fig.~\ref{fig.snapshot_free_equi} stresses the different alignment of the magnetic moments in different chains.

In the opposite scenario, we assume that the magnetic moments are fixed to the axes of the carrying particles, while the particles are covalently embedded into the elastic matrix. 
A torque on a magnetic moment is then equivalent to a torque on the carrying particle, which in turn leads to a deformation of the surrounding matrix.
We mark these systems by the term \emph{cov} and represent them computationally by rigidly coupling the magnetic moment orientations to the particles.

Consequently, the initial orientations of the magnetic moments have a determining influence on the structure of the \emph{cov} samples and, thus, on their stress-strain behavior. 
We distinguish  between two sub-scenarios and term the corresponding systems \emph{cov$\,\rightrightarrows$} and \emph{cov$\,\leftrightarrows$}. 
In the \emph{cov$\,\rightrightarrows$} systems, we define all magnetic moments in the sample to initially point into the same direction parallel to the chains. 
During the initial equilibration, the orientations of the magnetic moments barely change as particle rotations are energetically expensive. 
The magnetic moments within all chains are still aligned in the same direction, see Fig.~\ref{fig.snapshot_covrightright_equi} for a snapshot. 
This is different in the \emph{cov$\,\leftrightarrows$} systems: here we take the equilibrated state from the \emph{free} systems, but then fix the magnetic moments to the particle axes before stretching the sample. 
As a result, the magnetic moments are rigidly anchored to the carrying particles and are arranged into the chains with alternating alignment, see again Fig.~\ref{fig.snapshot_free_equi} for a snapshot. 
The \emph{cov$\,\leftrightarrows$} system constitutes an in-between case of the \emph{free} and \emph{cov$\,\rightrightarrows$} systems. 
We can, therefore, use it to test separately the effect of the two main modifications from the \emph{free} to the \emph{cov$\,\rightrightarrows$} system: anchoring the magnetic moments to the particle frames (\emph{free} to \emph{cov$\,\leftrightarrows$}) and having all magnetic moments point into the same direction (\emph{cov$\,\leftrightarrows$} to \emph{cov$\,\rightrightarrows$}). 

If we would apply an external magnetic field before the magnetic moments are anchored, we would destroy the alternating chain morphology that we want to study. 
Thus, when studying the influence of an external magnetic field on these alternating chain systems, we apply it after the magnetic moments have been anchored. 
Subsequently, we reequilibrate the systems under these new conditions before performing the stress-strain measurement.

\section{Results and discussion}
\label{sec.results_and_discussion}

In the following, we will present and discuss our results for the three systems \emph{free}, \emph{cov$\,\rightrightarrows$}, and \emph{cov$\,\leftrightarrows$} as defined above. 
We begin with vanishing external magnetic field and then proceed to the situation of magnetic fields applied parallel and perpendicular to the stretching direction. 
For each system and each magnetic field, we show snapshots as well as the uniaxial stress-strain curves and discuss the various mechanisms that lead to our results. 

Important insight can be gained by statistically analyzing the orientations of the magnetic moments in the systems. 
We evaluate them  by considering the nematic order parameter $S_m$, which is defined as the largest eigenvalue of the nematic order parameter tensor \cite{deGennes1995_book} 
\begin{equation}
	\mathbf{Q}_m = \frac{1}{N} \sum_{i = 1}^N \left( \frac{3}{2} \mathbf{\hat{m}}_i \otimes \mathbf{\hat{m}}_i - \frac{1}{2} \mathbf{\hat{I}} \right) \, .
	\label{eq.nematic_tensor_order_parameter}
\end{equation}
Here the $\mathbf{\hat{m}}_i$ are the magnetic moment orientations of the $N$ particles in the system, $\otimes$ marks the dyadic product, and $\mathbf{\hat{I}}$ is the unity matrix.
$S_m$ measures the degree of alignment of the input orientations without distinguishing between an orientation $\mathbf{\hat{m}}_i$ and its opposing orientation $-\mathbf{\hat{m}}_i$. 
Perfect alignment leads to $S_m = 1$, while in the absence of global orientational order $S_m = 0$. 

In addition to the magnetic order in the systems, also the structural order contains useful information.
It can be quantified in a very similar way by defining another nematic order parameter $S_r$ for the orientations $\mathbf{\hat{r}}_{i}$ of the separation vectors from each particle $i$ to its nearest-neighbor. 

As will be revealed later in more detail, in the \emph{free} system a ``flipping mechanism'' \cite{Cremer2015_ApplPhysLett} plays an important role. 
``Flips'' refer to events during which some magnetic moments suddenly change their direction with respect to the stretching axes from parallel towards perpendicular.
They are induced by the stress-induced structural change of the system. 
To appropriately characterize this flipping mechanism, we define special modified nematic order parameters $\tilde{S}_m$ and $\tilde{S}_r$ as described here for $\tilde{S}_m$. 
First, to get rid of the distinction between different perpendicular directions, we determine the projections $\hat{m}_i^\parallel$ of the magnetic moment orientations $\mathbf{\hat{m}}_i$ onto the stretching axis as well as the projections $\hat{m}_i^\perp$ into the plane perpendicular to the stretching axis. 
Then we define a two-dimensional nematic order parameter tensor as 
\begin{equation}
	\mathbf{\tilde{Q}}_m = \frac{1}{N} \sum_{i = 1}^N \begin{pmatrix}
		2 \big(\hat{m}_i^\parallel\big)^2 - 1			&	2 \hat{m}_i^\parallel \hat{m}_i^\perp	\\[.2cm]
		2 \hat{m}_i^\parallel \hat{m}_i^\perp	&	2 \big(\hat{m}_i^\perp\big)^2 - 1				\\
	\end{pmatrix}
	\label{eq.nematic_tensor_order_parameter_projected}
\end{equation}
and obtain $\tilde{S}_m$ as the largest eigenvalue of this tensor.  
The calculation of $\tilde{S}_r$ is analogous.

\subsection{Vanishing external magnetic field ($\mathbf{B} = 0$)}
\label{subsec.vanishing_field}

We now start by quasistatically stretching the three systems along the chain axes in the absence of an external magnetic field. 
The elongation is stepwise increased to a maximum and then, in the inverse way, reduced back to zero. 
The necessary forces on the clamps are recorded. 

\begin{figure}[t]
	\centering
	\includegraphics[width = 1.0\columnwidth]{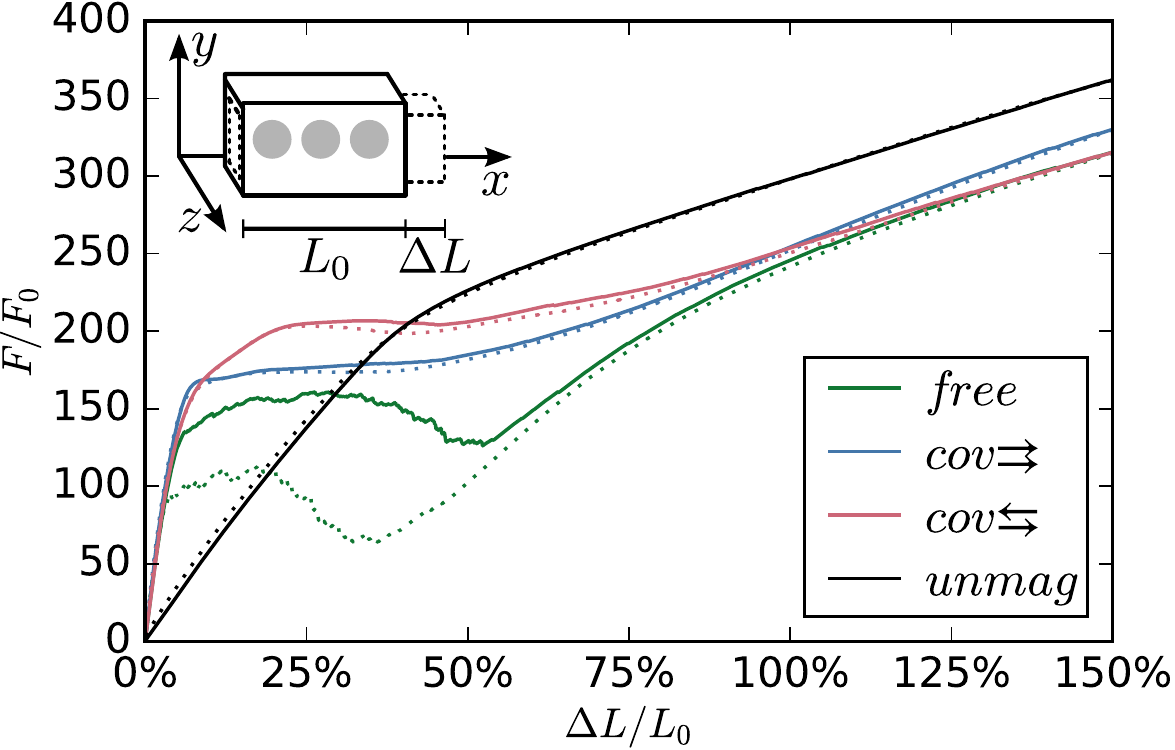}
	\caption{Uniaxial stress-strain curves for the \emph{free}, \emph{cov$\,\rightrightarrows$}, and \emph{cov$\,\leftrightarrows$} systems as well as for a corresponding system containing unmagnetized particles (\emph{unmag}) when stretching along the axes of the chain-like aggregates. The magnetized samples show a superelastic plateau between $\Delta L / L_0 \approx 10\%$ and $\Delta L / L_0 \approx 50\%$. In this regime, they can be deformed by a significant amount by only barely increasing the load. In contrast to that, the curve for the unmagnetized case lacks this appealing feature. The solid lines signify loading and the dotted lines unloading processes, as in all subsequent figures. In the \emph{free} system, our curves show pronounced hysteresis.}
	\label{fig.nofieldcomp_stressstrain}
\end{figure}

Figure~\ref{fig.nofieldcomp_stressstrain} shows the strongly nonlinear stress-strain behaviors resulting for the three systems.
In the beginning, all systems show an almost identically steep increase of the stress with the imposed strain. 
Then, from a strain of about $\Delta L / L_0 \approx 10\%$ up to  $\Delta L / L_0 \approx 50\%$, a pronounced superelastic plateau follows. 
In this regime, a small increase in the applied load leads to a huge deformation that is completely reversible.
The shape of the superelastic plateau differs among the systems. 
In the \emph{cov$\,\rightrightarrows$} and \emph{cov$\,\leftrightarrows$} systems the plateau is almost completely flat.
However, in our strain-controlled measurements we find a regime of negative slope \cite{Chernenko2003_JApplPhys} for the \emph{free} system. 
Moreover and in contrast to the other systems, we here observe considerable hysteresis for the \emph{free} system in the strain interval containing the superelastic plateau.      
In all cases, subsequent to the plateau, the slope partially recovers, becomes relatively constant, and does not differ much among the different systems. 

The main mechanism responsible for the nonlinearities in all systems is a stress-induced detachment mechanism \cite{Cremer2015_ApplPhysLett}. 
We briefly illustrate how it can lead to the change from the steep slope at the origin of the stress-strain curve to the subsequent superelastic plateau.
Consider again the unstretched states depicted in Fig.~\ref{fig.snapshots_equi}. 
In these states, the chains are contracted because of the mutual attraction between the magnetic moments of neighboring particles.  
Thus, the elastic material in the gaps between particles is pre-compressed  and the particles are close to each other.
In this situation, the dipolar attraction is strong, since its interaction energy, see Eq.~\eqref{eq.dipole_potential_energy}, scales with the inverse cube of the distance.
To stretch the system, work has to be performed against this strong attraction between the particles, which accounts for the steep initial increase in the stress-strain curve.
However, when a section of a chain is detached a little from the remainder, the attraction between both parts weakens considerably. 
Therefore, once overcoming the magnetic barrier, the displaced chain section can be detached from the remainder of the chain. 
Such a detachment event releases the energy stored in the gap between the detached particles and allows a sudden elongation of the system. 

\begin{figure*}[htb]
	\centering
	\centering
	\subfloat{\label{fig.snapshot_free_35}} 
	\subfloat{\label{fig.snapshot_covrightright_100}} 
	\subfloat{\label{fig.snapshot_covleftright_100}} 
	\subfloat{\label{fig.snapshot_unmag_100}} 
	\includegraphics[width = 1.0\textwidth]{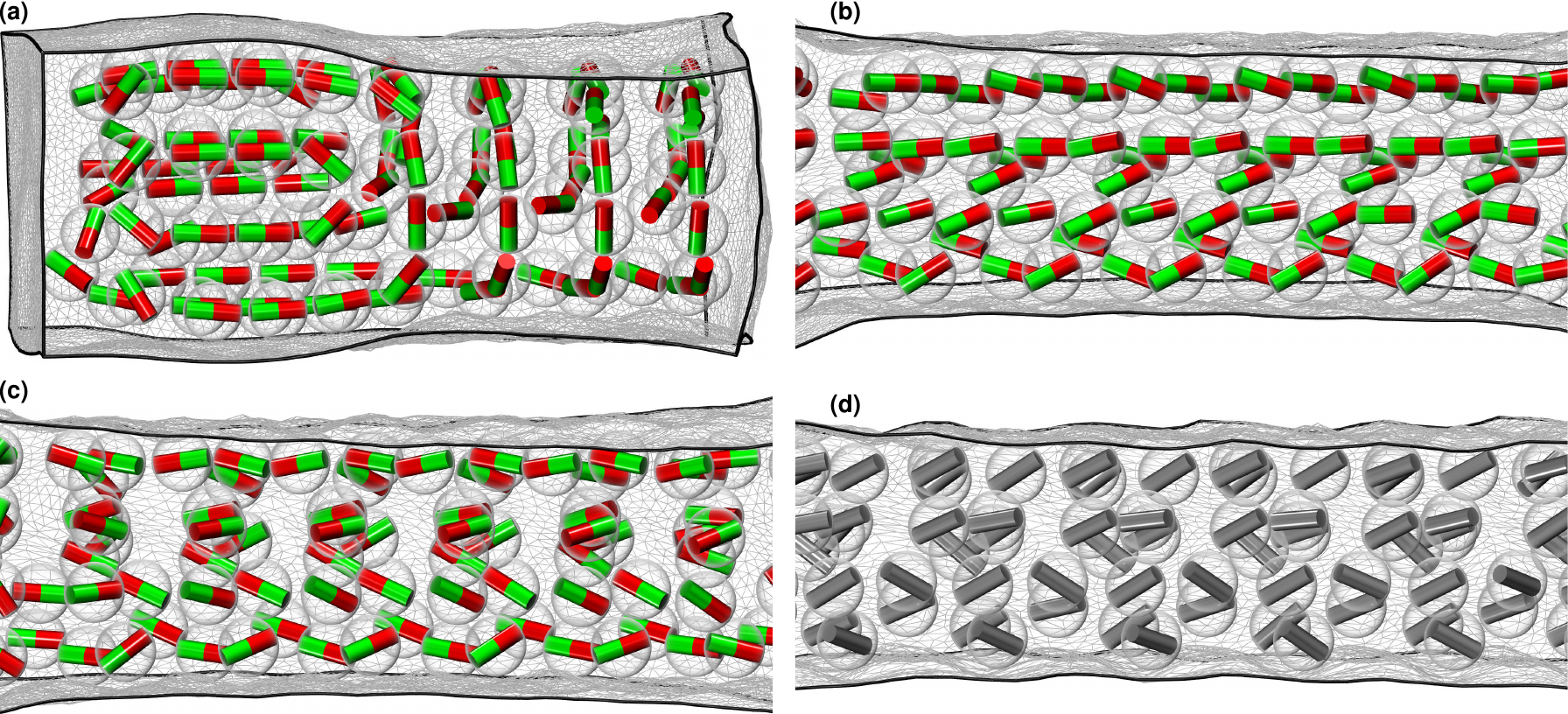}
	\caption{\protect\subref{fig.snapshot_free_35} Snapshot of a \emph{free} sample stretched by $35\%$. The freely rotating magnetic moments in this system can minimize their magnetic interaction by aligning along the direction of shortest interparticle distance. When the sample is stretched, the perpendicular direction becomes more and more favored, because the interparticle distance within the chains is increased, while near-incompressibility of the sample forces neighboring chains to approach each other. In the depicted situation, about half of the particles are detached from the chains, their magnetic moments having performed a flip from a direction parallel to the stretching axis towards perpendicular.   
In the \protect\subref{fig.snapshot_covrightright_100} \emph{cov$\,\rightrightarrows$} and \protect\subref{fig.snapshot_covleftright_100} \emph{cov$\,\leftrightarrows$} systems, rotations of the magnetic moments necessitate rotations of the carrying particles, causing restoring torques by the surrounding matrix.
Still, significant particle rotations can be observed in these samples stretched by $100\%$ with respect to the unstretched states in Fig.~\ref{fig.snapshots_equi}, caused, however, primarily by inhomogeneous deformations of the surrounding matrix due to the particle embedding. 
\protect\subref{fig.snapshot_unmag_100} Snapshot of an unmagnetized (\emph{unmag}) system starting from the same configuration. The bars indicate the initially horizontal particle axes to illustrate the particle rotations. They show a similar pattern as the systems in \protect\subref{fig.snapshot_covrightright_100},\protect\subref{fig.snapshot_covleftright_100} although magnetic interactions are absent.}
	\label{fig.snapshots_nofield}
\end{figure*}

Figure~\ref{fig.snapshot_free_35} shows a snapshot of a \emph{free} sample stretched by $35\%$, illustrating this process. 
In the depicted situation, some particles are detached from the chains with increased particle separation, while smaller segments are still intact. 
Each individual detachment event corresponds to a small localized drop in the stress-strain curve. 
In a very small and ordered system, this would lead to a spiky appearance of the stress-strain relation as we have demonstrated for a single chain in Ref.~\cite{Cremer2015_ApplPhysLett}. 
However, averaging over the many detachment events that occur in a larger, inhomogeneous system with many parallel chains yields a smooth superelastic plateau as in Fig.~\ref{fig.nofieldcomp_stressstrain}.
Upon unloading the system, the individual particles can simply reattach, reform the chains, and restore the energy in their separating gaps, so that the detachment mechanism is reversible. 

The second mechanism contributing to the observed superelasticity is the flipping mechanism. 
It only plays a significant role in the \emph{free} system. 
In the unstretched sample, the magnetic moments align along the chain axes in a head-to-tail configuration to minimize their magnetic interaction energy, see Fig.~\ref{fig.snapshots_equi}. 
This situation changes when the sample is sufficiently stretched in the direction parallel to the chains. 
The distances between particles in the same chain eventually increase, see Fig.~\ref{fig.snapshot_free_35} and Ref.~\cite{Cremer2015_ApplPhysLett}. 
Meanwhile, volume preservation in our nearly incompressible systems enforces a contraction in the perpendicular direction, driving different chains closer to each other. 
Eventually, the interparticle distances in the parallel and perpendicular directions become approximately equal for subsets of particles. 
For the involved magnetic moments this means a sudden change in their preferred orientation from parallel to the stretching axis towards perpendicular. 
In the \emph{free} system, the moments can easily seize this opportunity to minimize their magnetic interaction energy by sudden reorientation. 
This constitutes a flip event.

Flips are associated with drops in the stress-strain curve for the following reason. 
As long as the magnetic moments participating in a flip event are still aligned parallel to the stretching direction, their magnetic interaction energy increases with the strain.
However, once the flip has occurred and they have aligned towards perpendicular, their magnetic interaction energy decreases with the stretching.   
Therefore, during a flip event, the slope of the magnetic interaction energy suddenly changes for the participating magnetic moments.  
Since the stress is the derivative of the energy with respect to the strain, this causes a drop in the stress-strain curve. 
Or, discussing the situation directly in the force picture: as long as the magnetic moments align along the stretching axis, they counteract the elongation, which requires a higher stretching force; once they flip, they repel each other along the stretching axis, which supports the elongation. 
In an inhomogeneous sample, flips are local events and can occur over a wide range of global strain magnitudes. 
As a result, the individual drops are smoothened out in the stress-strain curves resulting from our characteristic systems. 

Consider again the snapshot in Fig.~\ref{fig.snapshot_free_35}. 
Compared to the particles in the still intact chain parts, the detached particles have a larger interparticle distance in the stretching direction and their magnetic moments indeed prefer an orientation towards perpendicular to that direction.  
When a detachment event occurs, the corresponding sample section elongates, which can in turn trigger flip events. 
Conversely, a reorientation of magnetic moments towards a perpendicular direction can induce detachment.
So in our characteristic \emph{free} systems, the detachment and flipping mechanisms are intertwined.  
Yet, considering suitable idealized model situations, both mechanisms can be studied in isolation, see Ref.~\cite{Cremer2015_ApplPhysLett}. 
The interplay between both mechanisms supports the hysteresis observed in our stress-strain curves for the \emph{free} system, see Fig.~\ref{fig.nofieldcomp_stressstrain}.
The magnetic attractions pull the particles together along the orientation of the magnetic moments, which in turn self-strengthens the magnetic interaction. 
In this way, an energetic barrier is created that needs to be overcome every time the magnetic moments are pulled apart and flip, either during initial stretching, or in the flipped state during unloading.

\begin{figure}[htb]
	\centering
	\includegraphics[width=1.0\columnwidth]{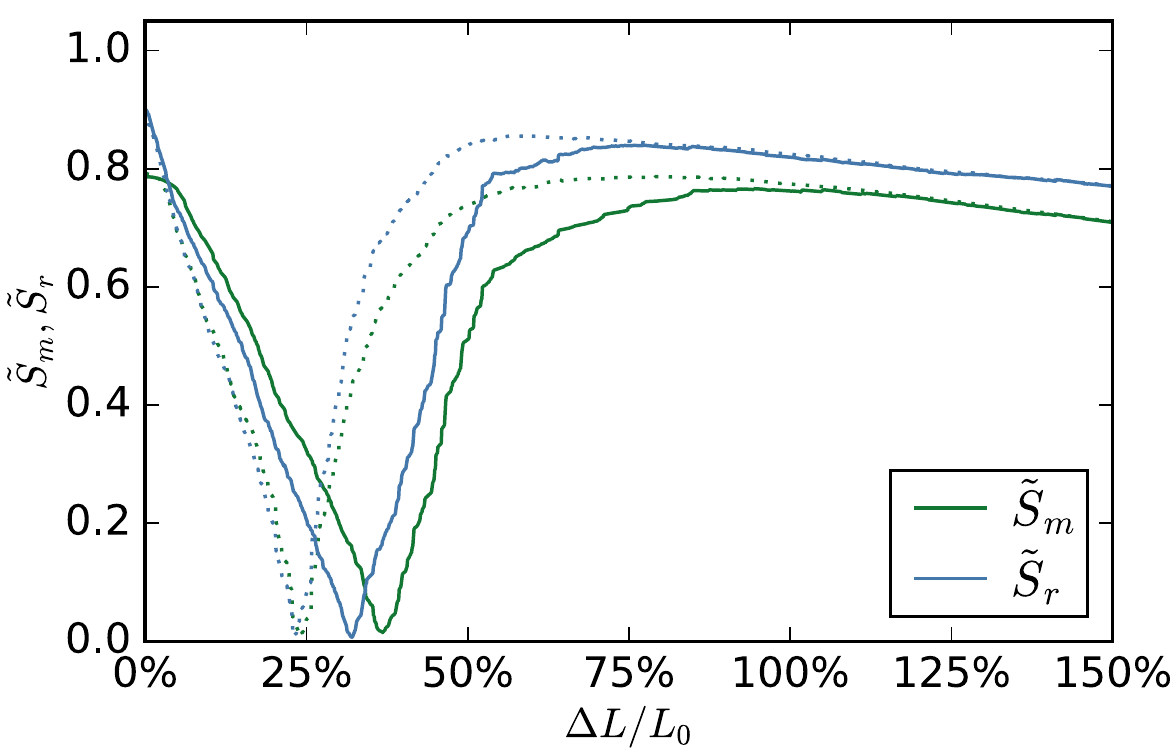}
	\caption{Degrees of magnetic order $\tilde{S}_m$ and structural order $\tilde{S}_r$ for the \emph{free} system, following the definition in Eq.~\eqref{eq.nematic_tensor_order_parameter_projected}. For vanishing strain, alignment along the initial anisotropy axis is preferred both magnetically and structurally. When the strain is increased, detachment and flip events occur and the system enters a mixed state where the parallel direction becomes less dominant in favor of directions perpendicular to the stretching axis. The minimum is reached at a strain of $\Delta L / L_0 \approx 35\%$, corresponding to the situation depicted in Fig.~\ref{fig.snapshot_free_35}. Subsequently, the degrees of order increase again until all possible detachments and flips have occurred. The hysteretic behavior observed for the stress-strain curves in Fig.~\ref{fig.nofieldcomp_stressstrain} shows up as well in the order parameters.}
	\label{fig.free-nofield_nematicOrder-projected}
\end{figure}

We can further quantify the flipping mechanism by statistically analyzing the orientations of the magnetic moments. 
Let us evaluate the nematic order parameters $\tilde{S}_m$ and $\tilde{S}_r$ defined in Eq.~\eqref{eq.nematic_tensor_order_parameter_projected}  as a function of the imposed strains.  
$\tilde{S}_m$ measures the degree of alignment of the magnetic moments and $\tilde{S}_r$ does the same for the separation vectors between nearest-neighboring particles.
The result is plotted in Fig.~\ref{fig.free-nofield_nematicOrder-projected}.
For low strains, magnetic moments are aligned parallel to the stretching axis, because this is the direction of smallest interparticle distance.  
Consequently the system is in a state of high magnetic and structural order, reflected by the high levels of $\tilde{S}_m$ and $\tilde{S}_r$.   
Upon increasing the strain, the interparticle distances in the stretching direction increase, particles are detached and magnetic moments flip, taking the system into a mixed state. 
$\tilde{S}_m$ and $\tilde{S}_r$ simultaneously decrease and reach a minimum at $\Delta L / L_0 \approx 35\%$, where they almost vanish. 
This state is depicted in the snapshot in Fig.~\ref{fig.snapshot_free_35}, where about half of the particles are detached from the chains with their magnetic moments flipped towards a perpendicular direction. 
From there on, both $\tilde{S}_m$ and $\tilde{S}_r$ increase again until finally all particles are detached and all magnetic moments have flipped. 
The strain regime where the order parameters change significantly coincides with the position of the superelastic plateau in the stress-strain curve in Fig.~\ref{fig.nofieldcomp_stressstrain}. 
Finally, at the highest strains, both order parameters again decrease slightly when the lateral contraction of the system squeezes the particles together.
This causes them to evade each other when they come too close and makes them shift relatively to each other along the stretching axis, which disturbs the perpendicular alignment.
Also for the order parameters, we here observe again the hysteresis discussed already before in the context of the stress-strain curve.

Let us now come back to the \emph{cov$\,\rightrightarrows$} and \emph{cov$\,\leftrightarrows$} systems where the magnetic moments cannot rotate relatively to the particle frames.
Then magnetic reorientations cost a significant amount of elastic energy, as this requires a corotation of the elastic matrix directly anchored to the particle surfaces. 
Figures~\ref{fig.snapshot_covrightright_100},\subref*{fig.snapshot_covleftright_100} show snapshots of corresponding samples at a strain of $100\%$. 
There we can nonetheless observe particle rotations. 
These particle rotations, however, do not apparently lead to a configuration that minimizes the magnetic interaction energy. 
In fact, the primary reason for these rotations is not the magnetic interaction between particles, but inhomogeneities in the stiffness across the system. 
We recall that the particles in our systems are rigid inclusions of finite extension. 
Consequently, the particles are local sources of elevated rigidity within the soft elastic matrix. 
Already in an unmagnetized system, such rigid inclusions lead to an overall stiffer elastic behavior of the whole system \cite{Einstein1906_AnnPhysBerlin,Einstein1911_correction_AnnPhysBerlin,LopezPamies2013_JMechPhysSolids}. 
In our case, an increase of a factor of $\sim 7$ in the elastic modulus was observed \cite{Cremer2015_ApplPhysLett}. 
Placing the particles into the randomly shifted chains when designing our systems adds a certain randomness to the distribution of our localized rigidities. 
When stretching the systems, the inhomogeneous distribution of rigidity can lead to local shear strains that rotate the embedded rigid particles. 
Of course, this does not require  the particles to be magnetized and occurs in unmagnetized systems ($m = 0$) just as well.
In Fig.~\ref{fig.snapshot_unmag_100} we show a snapshot of an unmagnetized system stretched by $100\%$ for demonstration. 
There we indicate the initially horizontal particle axes by bars to visualize the particle rotations. 
The resulting patterns of particle rotation are qualitatively similar to the ones in the \emph{cov$\,\rightrightarrows$} and \emph{cov$\,\leftrightarrows$} systems. 

\begin{figure}[t!]
	\centering
	\subfloat{\label{fig.nofieldcomp_nematicOrder}} 
	\subfloat{\label{fig.nofieldcomp_nematicOrder-nN}} 
	\includegraphics[width = 1.0\columnwidth]{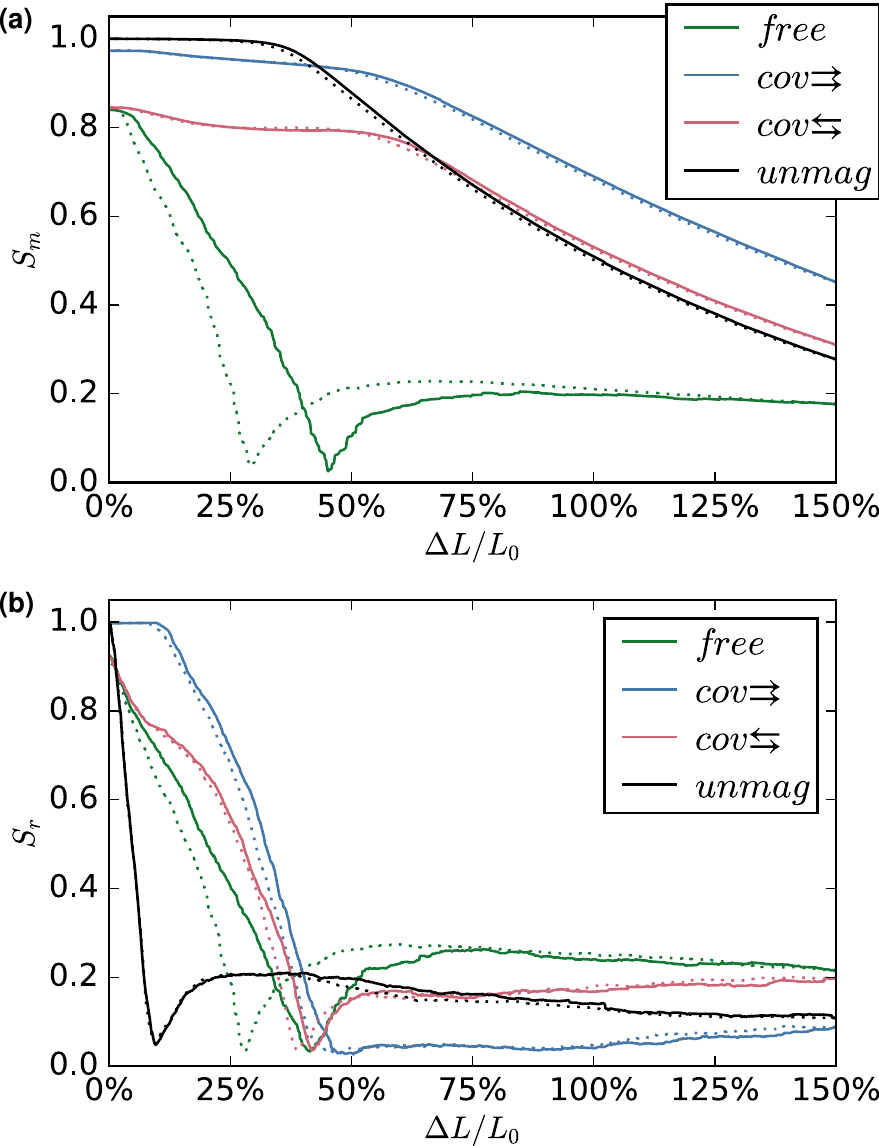}
	\caption{ \protect\subref{fig.nofieldcomp_nematicOrder} Nematic order parameter $S_m$ according to Eq.~\eqref{eq.nematic_tensor_order_parameter} for the magnetic moment orientations of the \emph{free}, \emph{cov$\,\rightrightarrows$}, \emph{cov$\,\leftrightarrows$} systems, as well as for an unmagnetized (\emph{unmag}) system as function of the imposed strain $\Delta L / L_0$. In the latter three systems, there is a regime of high magnetic order at low strains. At a strain of  $\Delta L / L_0 \approx 35\%$ in the \emph{unmag} system and $\Delta L / L_0 \approx 50\%$ in the \emph{cov$\,\rightrightarrows$} and \emph{cov$\,\leftrightarrows$} systems, there is a crossover to a regime of declining order, as inhomogeneous stresses begin to rotate the particles. In the \emph{free} system, again a  minimum indicates the occurrence of flip events. The recovery of $S_m$ beyond the minimum shows, that there is one globally preferred perpendicular direction emerging subsequent to flipping. \protect\subref{fig.nofieldcomp_nematicOrder-nN} Nematic order parameter $S_r$ for the nearest-neighbor separation vectors in the same systems. All curves have a minimum at the point where the preferred direction switches from parallel to the stretching axis towards perpendicular. In the magnetized systems, this minimum is postponed to higher strains. In these systems, the detachment barrier and magnetic interactions along the chains stabilize the chain structure, which is then preserved up to higher strains.}
	\label{fig.nofieldcomp_nematicOrders}
\end{figure}

Again we use statistical analysis to further quantify the particle rotations. 
Due to the different mechanism when compared to the flipping process, we are here only interested in the degree of alignment along the initial anisotropy axis. 
Therefore, we use the nematic order parameter $S_m$ defined in Eq.~\eqref{eq.nematic_tensor_order_parameter} for quantification. 
The results are plotted in Fig.~\ref{fig.nofieldcomp_nematicOrder} as a function of the imposed strain for the \emph{free}, \emph{cov$\,\rightrightarrows$}, and \emph{cov$\,\leftrightarrows$} systems, as well as for the unmagnetized (\emph{unmag}) case. 
Let us first consider the \emph{unmag} system. 
Up to a strain of $\Delta L / L_0 \approx 35\%$, particle rotations barely seem to occur, as $S_m$ stays close to $1$. 
Then, there is a crossover to a regime of approximately linear decay of $S_m$. The particles rotate more and more away from the initial axes of alignment as a consequence of the inhomogeneous stiffness. 
The behavior in the \emph{cov$\,\rightrightarrows$} and \emph{cov$\,\leftrightarrows$} systems is very similar, the crossover to the regime of declining order merely occurs at a higher strain of $\Delta L / L_0 \approx 50\%$, which also roughly marks the end of the superelastic plateau in Fig.~\ref{fig.nofieldcomp_stressstrain}. 
In these systems, the dipolar magnetic interactions along the initial, still intact chains counteract particle rotations and stabilize the alignment up to higher strains. 
When the detachment of the particles from the chains has been completed at the end of the superelastic plateau, this stabilizing magnetic interaction disappears, rendering the particles susceptible to shear stresses originating from the system inhomogeneity. 
The curve for the \emph{cov$\,\leftrightarrows$} system is always below the one for \emph{cov$\,\rightrightarrows$}, because already the initial unstretched state is less ordered, see again Fig.~\ref{fig.snapshots_equi}. 

The behavior of $S_m$ for the \emph{free} system is obviously completely different and should rather be compared with $\tilde{S}_m$ in Fig.~\ref{fig.free-nofield_nematicOrder-projected}. 
$S_m$ shows a rapid initial decay up to a minimum and afterwards recovers to reach a relatively low but constant level. 
This is despite the fact that $S_m$, unlike $\tilde{S}_m$, distinguishes between different directions perpendicular to the stretching direction.   
Therefore, beyond the superelastic plateau, one particular axis perpendicular to the stretching axis must emerge along which the magnetic moments preferably align.  
Such a direction forms as nearby flipped magnetic moments tend to align by magnetic dipolar interaction. 
In turn, this favors further contraction along such an emerging axis of alignment, providing a self-supporting mechanism. 
Inherent structural inhomogeneities will affect this mechanism. 

The same analysis as for $S_m$ can be conducted for the nematic order parameter $S_r$ of the separation vectors between nearest-neighbors. 
It is plotted for all systems in Fig.~\ref{fig.nofieldcomp_nematicOrder-nN}. 
$S_r$ starts at a high value for all systems, because in the unstretched state the nearest-neighbor of each particle is always along the chain. 
The more the sample is stretched, the more the distances along the chains increase, while the distances between separate chains decrease due to volume preservation.
Thus, it becomes increasingly likely that the nearest-neighbor for a particle is a member of a different chain. 
In the \emph{unmag} system, there is no stabilizing attractive interaction keeping the chains together.
So the minimum, where nearest-neighbor directions predominantly switch, is reached relatively soon. 
In the other systems, however, the magnetic attraction makes the chains subject to the detachment mechanism. 
Segments detach from the chains, while the remainder of the chains remains intact. 
As a result, partial structural order is preserved up to much higher strains. 
Again, the strain regime where $S_r$ changes a lot due to the changes in structural order coincides with the strain interval of the superelastic plateau in the stress-strain curves in Fig.~\ref{fig.nofieldcomp_stressstrain}.

\subsection{External magnetic field along the stretching axis ($\mathbf{B} = B_x \mathbf{\hat{x}}$)}
\label{subsec.parallel_field}
\begin{figure*}[bth]
	\centering
	\subfloat{\label{fig.free_Bx_stressstrain}} 
	\subfloat{\label{fig.snapshot_free-Bx1_100}} 
	\subfloat{\label{fig.free_Bx_nematicOrder}} 
	\subfloat{\label{fig.free_Bx_nematicOrder-nN}} 
	\includegraphics[width = 1.0\textwidth]{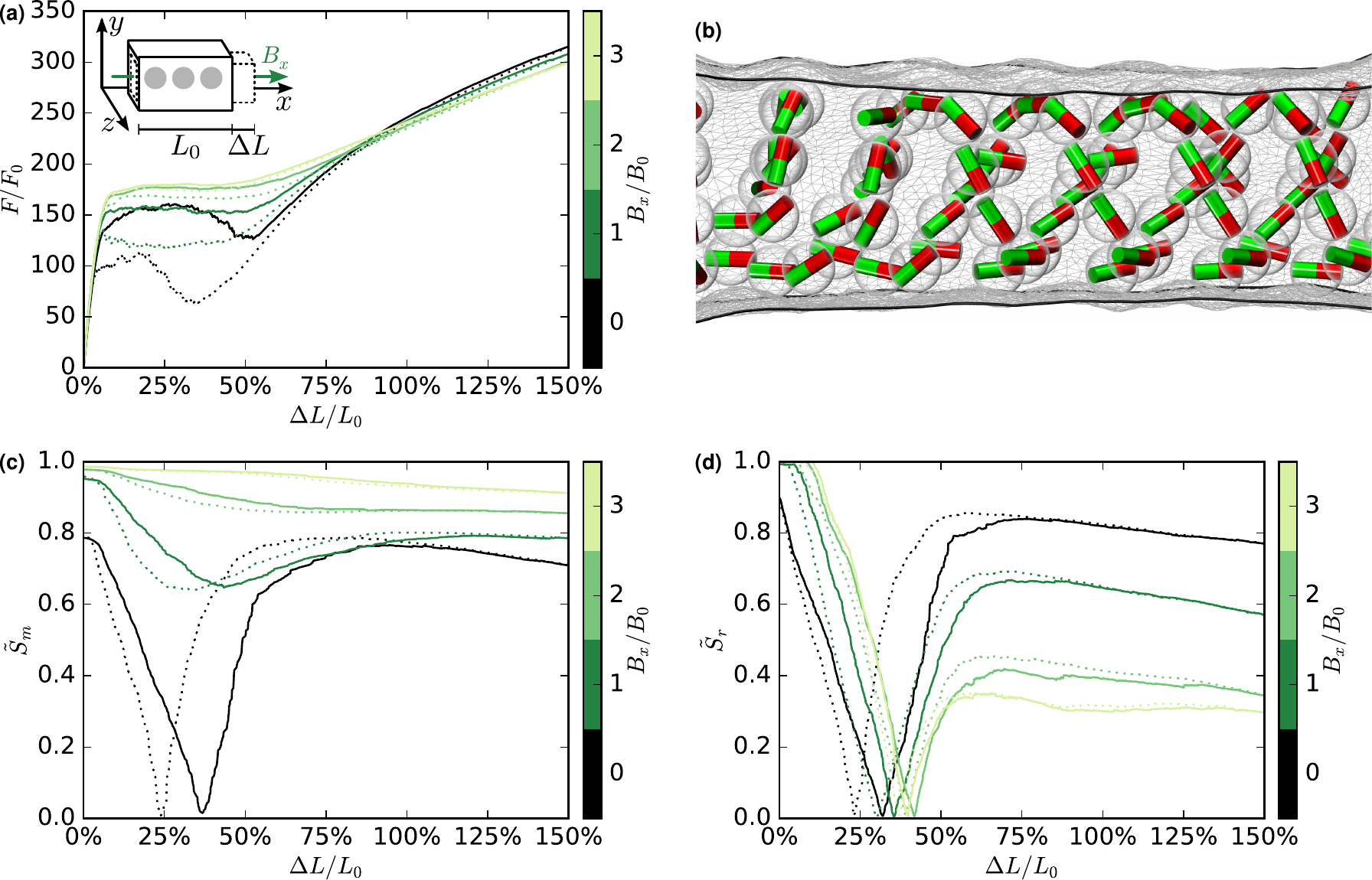}
	\caption{Results for the \emph{free} system under the influence of an external magnetic field of varying strength, applied parallel to the stretching axis. \protect\subref{fig.free_Bx_stressstrain} Uniaxial stress-strain behavior$^1$. The external field gradually deactivates the flipping mechanism. As a result the superelastic plateau is flattened, the dip at $\Delta L / L_0 \approx 50\%$ and the hysteresis are removed until the behavior resembles the one for the \emph{cov$\,\rightrightarrows$} system in Fig.~\ref{fig.nofieldcomp_stressstrain} for vanishing external magnetic field. \protect\subref{fig.snapshot_free-Bx1_100} Snapshot showing a \emph{free} system subject to an external field of $B_x = 1 B_0$ at a strain of $\Delta L / L_0 = 100\%$. Even in this highly strained state, the magnetic moments assume oblique angles instead of performing full flips towards a perpendicular direction. 
\protect\subref{fig.free_Bx_nematicOrder} Degree of magnetic order $\tilde{S}_m$ and \protect\subref{fig.free_Bx_nematicOrder-nN} degree of structural positional order $\tilde{S}_r$ as defined by Eq.~\eqref{eq.nematic_tensor_order_parameter_projected}, indicating the deactivation of the flipping mechanism with increasing $B_x$. The minimum in $\tilde{S}_m$ is gradually removed by the parallel external magnetic field. Meanwhile, the minimum in $\tilde{S}_r$ is shifted slightly.}
	\label{fig.free_Bx}
\end{figure*}

Applying an external magnetic field parallel to the chain and stretching axis (the $x$-direction) when recording the stress-strain behavior changes the situation fundamentally in all three systems \emph{free}, \emph{cov$\,\rightrightarrows$}, and \emph{cov$\,\leftrightarrows$}.
In the \emph{free} system, turning on the field after the initial equilibration causes all magnetic moments to point into the same direction along the field as opposed to the situation in Fig.~\ref{fig.snapshot_free_equi}. 
There, the magnetic moments carried by particles in different chains can show opposite magnetic alignment. 
In the \emph{free} system as well as in the \emph{cov$\,\rightrightarrows$} system, the field also introduces an additional energetic penalty for the rotation of magnetic moments away from the chain axes. 
The detachment mechanism is not impeded by this, as it relies on the strong magnetic attraction between neighboring particles within the same chain and the storage of elastic energy within the compressed gap material. 
The magnetic moments are not rotated away from the alignment along the chain axes during this process. 
In contrast to that, the flipping mechanism is based on reorientations away from the direction of the applied magnetic field and is, therefore, affected by the aligning magnetic field. 
In the \emph{cov$\,\leftrightarrows$} system featuring anchored magnetic moments of opposite alignment, the external magnetic field has a particularly interesting effect. 
Roughly half of the magnetic moments are aligned with the field. 
The remaining moments are misaligned and the corresponding particles would need to rotate by about 180 degrees to minimize the interaction energy with the external magnetic field.

%
\addtolength{\skip\footins}{-4pt}
\footnotetext[1]{In Ref.~\cite{Cremer2015_ApplPhysLett}, the magnetic field strengths in the figures containing stress-strain curves were not scaled correctly. Instead of $10 B_0$, $20 B_0$, $30 B_0$ it should read $1 B_0$, $2 B_0$, $3 B_0$, respectively.}
Figure~\ref{fig.free_Bx} revisits our results for the \emph{free} systems for various applied magnetic field strengths. 
The stress-strain curves in Fig.~\ref{fig.free_Bx_stressstrain} illustrate the tunability of the material\footnotemark[1].
Already a small external magnetic field of $B_x = 1 B_0$ removes the dip at $\Delta L / L_0 \approx 50\%$, flattens the superelastic plateau, and also reduces the hysteresis considerably. 
As noted in Ref.~\cite{Cremer2015_ApplPhysLett}, the dip was mainly generated by flipping of magnetic moments.
When a stronger field is applied, the shape of the superelastic plateau becomes almost identical to the one for the \emph{cov$\,\rightrightarrows$} system in the case of vanishing external magnetic field, see Fig.~\ref{fig.nofieldcomp_stressstrain}. 

The snapshot in Fig.~\ref{fig.snapshot_free-Bx1_100} shows a \emph{free} system for $B = 1 B_0$ at a strain of $\Delta L / L_0 = 100\%$. 
It reveals, that the magnetic moments do not perform complete flips anymore and instead show oblique orientation angles. 
In summary, the flipping transition and the connected bumps in the superelastic plateau together with the hysteresis can be deactivated by the field. 

The plot in Fig.~\ref{fig.free_Bx_nematicOrder} of the nematic order parameter $\tilde{S}_m$ quantifying the magnetic order in the system provides further evidence that the field impedes the flipping mechanism. 
An external magnetic field of $B_x = 1 B_0$ is sufficiently strong to smoothen the sharp local minimum in $\tilde{S}_m$ corresponding to the transition from a state of parallel towards perpendicular magnetic alignment with respect to the stretching axis. 
Stronger fields enforce a parallel alignment, remove the local minimum in $\tilde{S}_m$ and thus deactivate the flipping mechanism.
Only the detachment mechanism remains active.
Meanwhile, the structural positional order in the sample does not seem to be influenced significantly by the external magnetic field, as the plots of the nematic order parameter $\tilde{S}_r$ for the separation vectors between nearest-neighbors in Fig.~\ref{fig.free_Bx_nematicOrder-nN} suggest. 
The minimum where the most likely nearest-neighbor direction switches from parallel towards perpendicular is shifted slightly. 
Beyond the minimum, $\tilde{S}_r$ decreases with increasing $B_x$. 
This results from an arising competition between two effects. 
On the one hand, due to overall volume preservation, the particles are driven together along the direction perpendicular to the stretching axis as before.
On the other hand, flips are hindered by the external magnetic field, or even suppressed completely. 
Therefore, the magnetic moments cannot support the perpendicular approach anymore as efficiently, or even counteract it due to the magnetic repulsion when the magnetic moments are forced into the direction of the external magnetic field. 
This also largely removes the hysteresis from our curves.   

\begin{figure*}[htb]
	\centering
	\subfloat{\label{fig.covrightright_Bx_stresstrain}} 
	\subfloat{\label{fig.snapshot_covrightright-Bx10_100}} 
	\subfloat{\label{fig.covrightright_Bx_nematicOrder}} 
	\subfloat{\label{fig.covrightright_Bx_nematicOrder-nN}} 
	\includegraphics[width = 1.0\textwidth]{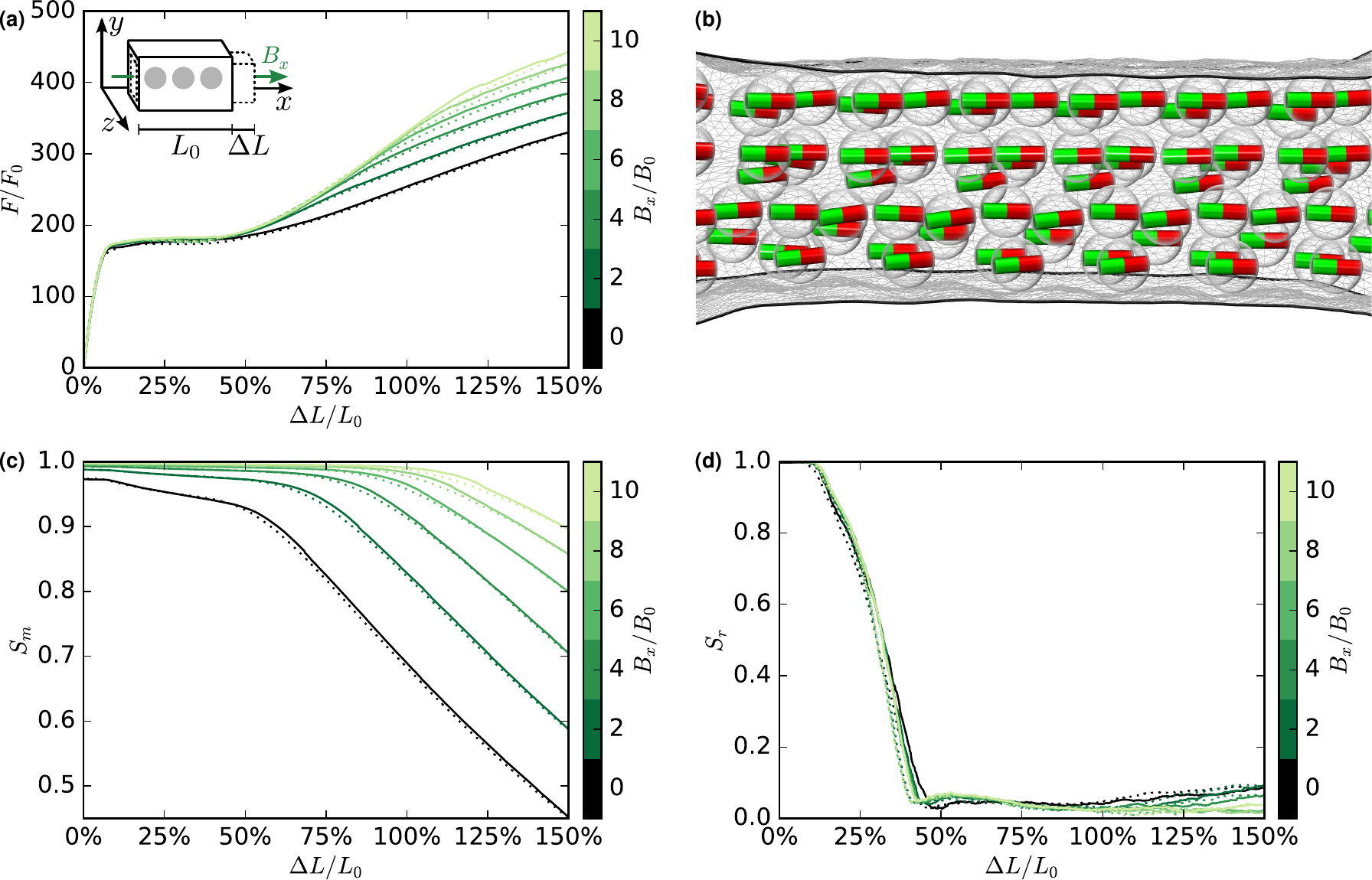}
	\caption{Same as Fig.~\ref{fig.free_Bx}, but for the \emph{cov$\,\rightrightarrows$} system. \protect\subref{fig.covrightright_Bx_stresstrain} The stress-strain curves for different external magnetic field strengths are almost identical up to the end of the superelastic plateau. Beyond this point, higher field strengths increase the stiffness until at very high strains the slopes become similar again. \protect\subref{fig.snapshot_covrightright-Bx10_100} Snapshot of a system at a strain of $\Delta L / L_0 = 100\%$ illustrating that a field of $B_x = 10 B_0$ can effectively prevent the particle rotations favored by local shears due to the elastic inhomogeneities. Here, the internal shear stresses of the system cannot relax via particle rotations and the parallel magnetic moments repel each other in the direction perpendicular to the stretching axis, both effects stiffen the system against further elongation. \protect\subref{fig.covrightright_Bx_nematicOrder} Nematic order parameter $S_m$ for the magnetic moment orientations. The external magnetic field can postpone the crossover to the regime of decreasing orientational order, allowing for particle rotations and magnetic moment reorientations only at very high strains.  \protect\subref{fig.covrightright_Bx_nematicOrder-nN} Here, the nematic order parameter $S_r$ for the nearest-neighbor separation vectors is barely sensitive to a change in the external magnetic field strength.} 
	\label{fig.covrightright_Bx}
\end{figure*}

Let us discuss the \emph{cov$\,\rightrightarrows$} system next. 
The results are summarized in Fig.~\ref{fig.covrightright_Bx}. 
Figure~\ref{fig.covrightright_Bx_stresstrain} shows the corresponding stress-strain behavior.  
Up to the end of the superelastic plateau, the curves for different external magnetic field strengths hardly differ. 
This is not surprising, since we have established before that the flipping mechanism plays no role for these systems and that the detachment mechanism is not impeded by an external magnetic field parallel to the chains. 
However, beyond the plateau, where we have a regime of relatively constant increase of the stress with the imposed strain, we can observe a stiffening of the system when a higher field strength is applied.
Only at very high strain, the slopes for all different field strengths become similar again.
The explanation for this stiffening influence of the external magnetic field is the suppression of magnetic moment reorientations and, thus, in this \emph{cov$\,\rightrightarrows$} system, of particle rotations. 
We have seen, however, in Fig.~\ref{fig.snapshot_covrightright_100} that such particle rotations would arise in the absence of a magnetic field to minimize the elastic energy. 
Suppressing them increases the necessary mechanical energy input into the system.  
The snapshot in Fig.~\ref{fig.snapshot_covrightright-Bx10_100} shows a sample with an applied field of $B_x = 10 B_0$ at a strain of $\Delta L / L_0 = 100\%$ for comparison with the analogous situation in Fig.~\ref{fig.snapshot_covrightright_100} for $B_x = 0$.

For a more quantitative analysis of the rotation effects, we evaluate the nematic order parameter $S_m$ of the orientations of the magnetic moments as a function of the imposed strain, see Fig.~\ref{fig.covrightright_Bx_nematicOrder}.
We can distinguish between two major regimes. 
In the first one, the overall strain is still too low to induce significant local shear deformations due to the inhomogeneities, thus, the particles rotate only slightly and $S_m$ remains on a high and relatively constant level. However, in the second regime, we can observe an approximately linear decay in $S_m$ as the particles begin to significantly rotate. 
In the absence of an external magnetic field, the crossover between both regimes occurs at the end of the superelastic plateau.
There, the particles are detached from the chains.
This reduces the aligning magnetic interactions and the particles become susceptible to rotations due to the elastic inhomogeneities in the system. 
Interestingly, increasing the strength of the external magnetic field can postpone the crossover far beyond this point by supporting the magnetic moment orientations along the field direction. 
This stiffens the system in two ways. 
First, the inhomogeneity shear stresses are prevented from relaxing via the favored channel: the rotation of particles. 
Second, the magnetic moments in the system keep repelling each other perpendicular to the stretching axis, which works against their perpendicular approach. 
The stronger the external magnetic field strength, the longer the embedded particles can resist a rotation, maintaining the stiffening effect.
For all considered magnetic field strengths, the particles eventually begin to rotate, as indicated by the crossover in $S_m$. 
Therefore, the slopes of the stress-strain curves become similar again at the maximum strain.    

Finally, we show for completeness in Fig.~\ref{fig.covrightright_Bx_nematicOrder-nN} the nematic order parameter $S_r$ for the nearest-neighbor separation vectors  as a function of the imposed strain. 
Here, the curves for different magnetic field strengths are largely similar. 

\begin{figure*}[htb]
	\centering
	\subfloat{\label{fig.covleftright_Bx_stresstrain}} 
	\subfloat{\label{fig.snapshot_covleftright-Bx6_30}} 
	\subfloat{\label{fig.covleftright_Bx_nematicOrder}} 
	\subfloat{\label{fig.covleftright_Bx_nematicOrder-nN}} 
	\includegraphics[width = 1.0\textwidth]{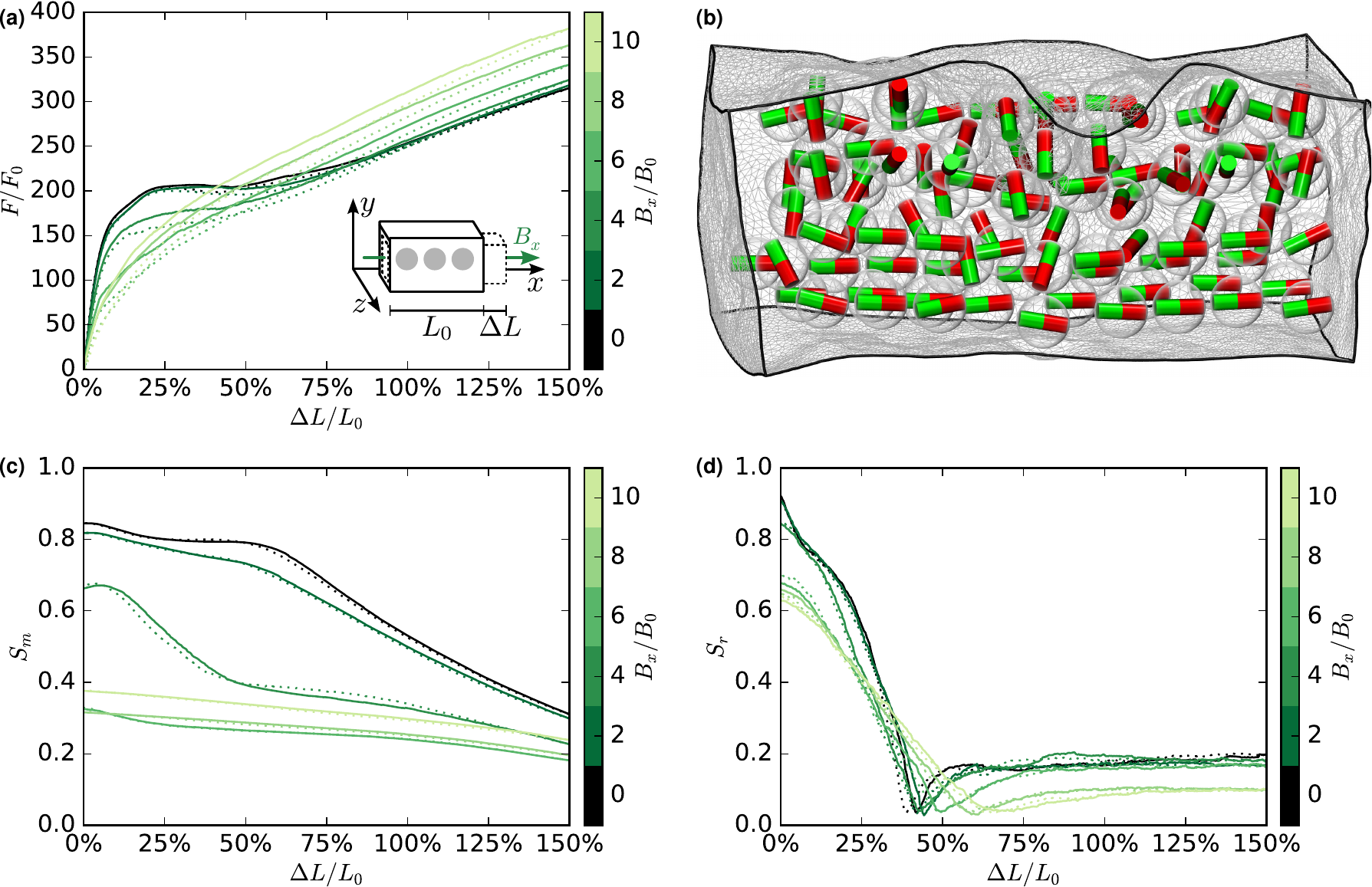}
	\caption{Same as Fig.~\ref{fig.free_Bx} but for the \emph{cov$\,\leftrightarrows$} system. \protect\subref{fig.covleftright_Bx_stresstrain} Uniaxial stress-strain behavior. Applying an external magnetic field parallel to the stretching axis gradually removes the pronounced nonlinearity. \protect\subref{fig.snapshot_covleftright-Bx6_30} Snapshot of a \emph{cov$\,\leftrightarrows$} system under the influence of an external magnetic field of $B_x = 6 B_0$ at a strain of $\Delta L / L_0 = 30\%$. The particles carrying the misaligned magnetic moments are strongly rotated towards the external magnetic field and distort their environment in the process, which also affects the chains containing the particles of aligned magnetic moments. As a result, the detachment mechanism is mostly deactivated.  \protect\subref{fig.covleftright_Bx_nematicOrder} Nematic order parameter $S_m$ for the magnetic moment orientations. Increasing the strength of the external magnetic field first lowers the overall $S_m$ due to the rotations of particles carrying misaligned magnetic moments and due to the resulting distortions of the rest of the system. At high field strengths, $S_m$ increases slightly with $B_x$, as the orientations of the aligned magnetic moments are stabilized. \protect\subref{fig.covleftright_Bx_nematicOrder-nN} The structural order in the system measured by $S_r$ is not influenced strongly as long as $B_x \lesssim 4 B_0$. Beyond that field strength, however, it significantly decreases because of the induced rotations of the particles carrying misaligned magnetic moments.}
	\label{fig.covleftright_Bx}
\end{figure*}

Now we come to the \emph{cov$\,\leftrightarrows$} system and present the results in Fig.~\ref{fig.covleftright_Bx}. 
Before the external magnetic field is applied, these systems are in a state like the one depicted in Fig.~\ref{fig.snapshot_free_equi}. 
Roughly half of the magnetic moments are aligned along to the magnetic field direction, while the other half is oppositely aligned and tends to reorient to minimize the magnetic interaction energy with the external field.  
This has implications on the stress-strain behavior, as illustrated in Fig.~\ref{fig.covleftright_Bx_stresstrain}. 
For small field strengths ($B_x = 2B_0$), the behavior barely changes compared to the case of vanishing external magnetic field. 
Then for intermediate fields of $B_x = 4B_0$, the steep increase at low strains as well as the superelastic plateau become less pronounced. 
Starting from a field of  $B_x = 6B_0$, the superelastic features vanish altogether. 
An explanation is given in the following.
As long as the external field strength is low enough ($B_x = 2B_0$), the energy cost of misalignment is not particularly large for the magnetic moments in the metastable configuration antiparallel to the field. 
However, when increasing the external field, due to imperfections in the initial antiparallel alignment, at some point the magnetic particles can be rotated by a significant amount.
Then, the torques due to the external field get amplified, causing the particles to rotate even further. 
At this stage, the reorientations of the misaligned moments together with their carrying particles begin to distort the sample substantially. 
Obviously, for the corresponding chains, the detachment mechanism will seize to function at this point, but also the chains containing aligned magnetic moments in the neighborhood will be disturbed. 
This chaotic situation is depicted in the snapshot in Fig.~\ref{fig.snapshot_covleftright-Bx6_30} for an external magnetic field of $B_x = 6B_0$ and a strain of $30\%$.
One can still identify the particles that have been aligned along the field direction, but the corresponding chains are distorted. 
As a result, the detachment mechanism is disabled and the superelastic plateau vanishes. 

The plots of the nematic order parameters $S_m$ and $S_r$ in Figs.~\ref{fig.covleftright_Bx_nematicOrder},\subref*{fig.covleftright_Bx_nematicOrder-nN} support this picture. 
For small magnetic field strength of $B_x = 2 B_0$, $S_m$ is still very similar to the case of vanishing magnetic field. 
Further increasing the field strength up to $B_x = 6 B_0$ promotes magnetic disorder in the system, leading to an overall low level of $S_m$. 
From there on, the level of $S_m$ slightly increases with the magnetic field strength as the orientations of the aligned magnetic moments are stabilized by the field.
The structural order measured by $S_r$ does not change too much as long as $B_x \lesssim 4 B_0$. 
Starting from $B_x \gtrsim 6B_0$, however, the misaligned magnetic moments are rotated significantly and distort the system.
The increased magnetic order indicated by a higher level of $S_m$ apparently cannot prevent the structure from becoming more disturbed, so that $S_r$ is still lowered further.  

In conclusion, the effect of an external magnetic field applied parallel to the stretching axis varies substantially among the different systems.
In the \emph{free} system, the main effect is the deactivation of the flipping mechanism, which makes the stress-strain behavior almost identical to the one of the \emph{cov$\,\rightrightarrows$} system in the absence of an external magnetic field.
Within the \emph{cov$\,\rightrightarrows$} system the superelasticity is barely affected. 
However, the external magnetic field stabilizes the particle orientations at strains beyond the superelastic plateau and thereby stiffens the stress-strain behavior.
Finally, in the \emph{cov$\,\leftrightarrows$} system the field promotes a strongly disturbed structure by rotating particles carrying magnetic moments misaligned with the field. 
As a consequence, the detachment mechanism is disabled and the superelastic plateau vanishes from the stress-strain curves. 

\subsection{External magnetic field perpendicular to the stretching axis ($\mathbf{B} = B_y \mathbf{\hat{y}}$)}
\label{subsec.perpendicular_field}
\begin{figure*}[t!]
	\centering
	\subfloat{\label{fig.free_By_stressstrain}} 
	\subfloat{\label{fig.snapshot_free-By2_equi}} 
	\subfloat{\label{fig.free_By_nematicOrder}} 
	\subfloat{\label{fig.free_By_nematicOrder-nN}} 
	\includegraphics[width = 1.0\textwidth]{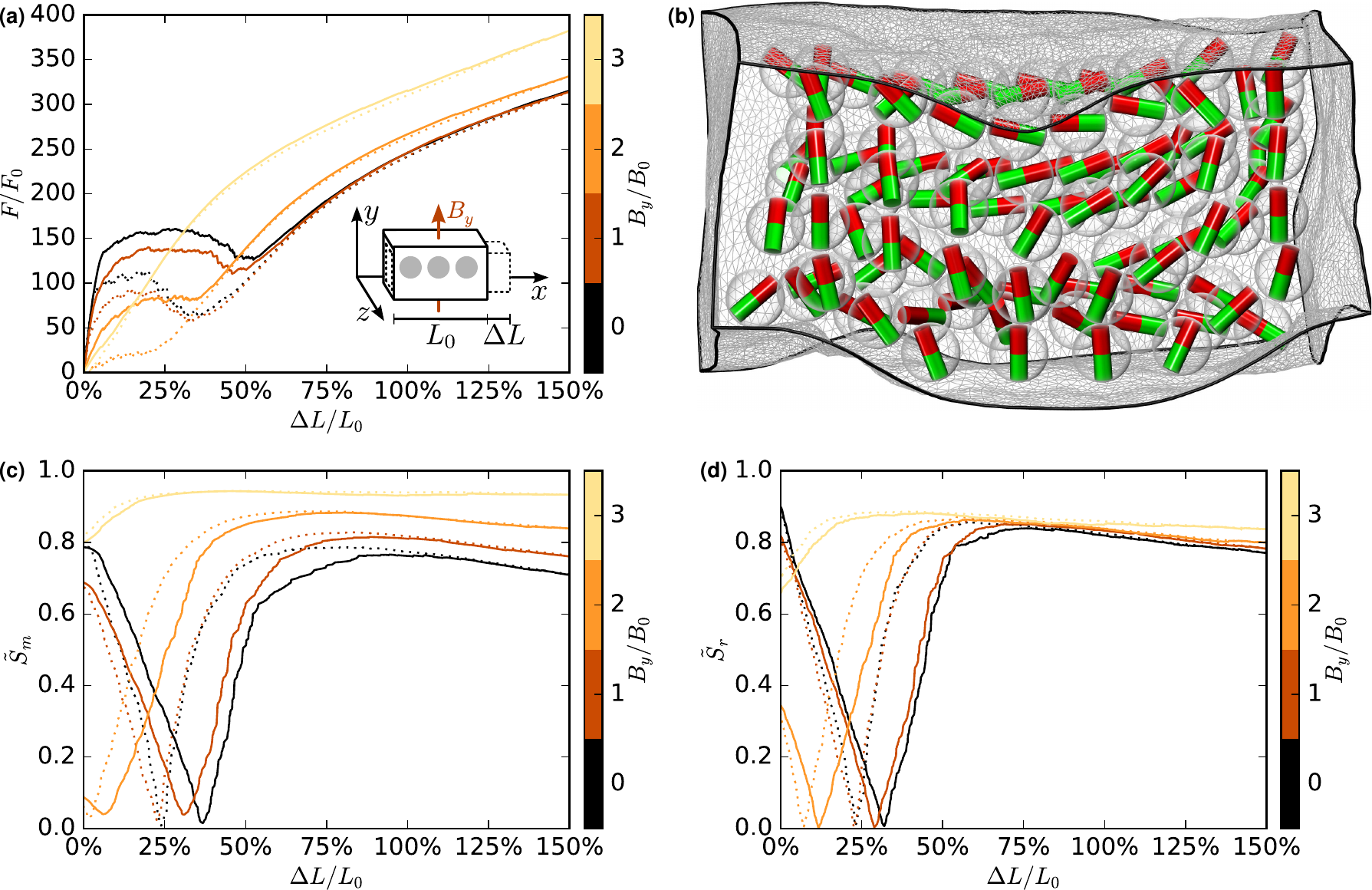}
	\caption{ Results for the \emph{free} system under the influence of an external magnetic field of varying strength perpendicular to the stretching axis. \protect\subref{fig.free_By_stressstrain} The superelastic stress-strain behavior can be readily tuned$^1$. Increasing the field gradually removes the superelasticity and lowers the slope of the initial steep increase. A field of $B_y = 3 B_0$ is already strong enough to remove all superelastic nonlinearities. \protect\subref{fig.snapshot_free-By2_equi} Snapshot of an unstretched sample with an applied external magnetic field of $B_y = 2 B_0$. A significant portion of the particles is already detached, their carried magnetic moments already flipped. As a consequence, the detachment and flipping mechanism have less impact on the stress-strain behavior, and superelastic as well as hysteretic features are reduced. \protect\subref{fig.free_By_nematicOrder} Degree of magnetic order $\tilde{S}_m$ and \protect\subref{fig.free_By_nematicOrder-nN} degree of structural order $\tilde{S}_r$ using the definition in Eq.~\eqref{eq.nematic_tensor_order_parameter_projected}. Both order parameters are again strongly correlated. Increasing the magnetic field strength shifts the local minimum marking the regime of mixed orientations to lower strains. That is, the threshold strains for detachment and flip events are lowered, with many events having occurred already in the unstretched state. This limits the amount of events that can still take place when the sample is stretched. At $B_y = 3 B_0$, the pronounced minima of $\tilde{S}_m$ and $\tilde{S}_r$ have vanished as all magnetic moments are already reoriented in the unstretched state. Therefore, there are no remaining flip or detachment events already in the unstretched state and, as a consequence, superelasticity is switched off.}
	\label{fig.free_By}
\end{figure*}

An external magnetic field applied perpendicular to the stretching axis (here the $y$-axis) attempts to rotate the magnetic moments away from their attractive head-to-tail configuration within the chains. 
This influence is strongest in the \emph{free} system, where the magnetic moments are free to reorient to minimize their magnetic energy.
In the \emph{cov$\,\rightrightarrows$} and \emph{cov$\,\leftrightarrows$} systems, however, rotations of the magnetic moments are counteracted by restoring torques on the embedded particles due to the induced deformation of the surrounding matrix.

Let us again discuss the \emph{free} system first.  
We present the results in the same fashion as before for the parallel field.
Figure~\ref{fig.free_By_stressstrain} shows the resulting stress-strain behavior\footnotemark[1].
The perpendicular field has two effects. 
First, it influences the superelasticity, causing the plateau to be confined to a smaller strain interval. 
Second, it lowers the initial slope of the stress-strain curve. 
At a high enough magnetic field strength, the superelastic nonlinearities are switched off completely together with the hysteresis, and the stress-strain curve becomes ordinary. 

To understand this behavior, it is first noted that the perpendicular magnetic field shifts the flipping mechanism to smaller strains. 
This is intuitive, as the external magnetic field energetically supports flips to a direction perpendicular to the stretching axis. 
Analysis of the nematic order parameters $\tilde{S}_m$ and $\tilde{S}_r$ in Figs.~\ref{fig.free_By_nematicOrder},\subref*{fig.free_By_nematicOrder-nN}, respectively, confirms this expectation.
The regime of mixed orientations centered around the minimum in $\tilde{S}_m$ is shifted to lower strains by the field. 
In this regime, some of the magnetic moments are still aligned along the chains, while others have already flipped. 
Meanwhile, $\tilde{S}_r$ remains strongly correlated with $\tilde{S}_m$.  
This indicates that the external magnetic field does not only influence the flipping mechanism, but also the detachment mechanism. 
As noted before, flip events trigger detachment events and vice versa.
Reoriented magnetic moments do not feel a strong attraction along the stretching axis that could keep the carrying particles attached to the chains.  
So the threshold strains for both mechanisms are lowered at the same time. 

This shift of threshold strains can cause the system to enter a mixed state already without any external strain imposed.
The snapshot in Fig.~\ref{fig.snapshot_free-By2_equi} shows a situation of $B_y = 2B_0$. 
Although the system is unstretched in the depicted case, a significant amount of particles has already detached from the chains. 
Their magnetic moments are aligned along the field direction, perpendicular to the chain axis. 
So the fraction of particles that can still perform detachment or flip events is lowered.  
As a result, the features corresponding to both mechanisms are less pronounced in the stress-strain curves. 
Also the initial slope is lower, because the overall magnetic attraction along the stretching direction cannot counteract the elongation as strongly. 
Consequently, the superelastic plateau spans a smaller strain interval. 

\begin{figure*}[t!]
	\centering
	\subfloat{\label{fig.covrightright_By_stresstrain}} 
	\subfloat{\label{fig.snapshot_covrightright-By10_equi}} 
	\subfloat{\label{fig.covrightright_By_nematicOrder}} 
	\subfloat{\label{fig.covrightright_By_nematicOrder-nN}} 
	\includegraphics[width = 1.0\textwidth]{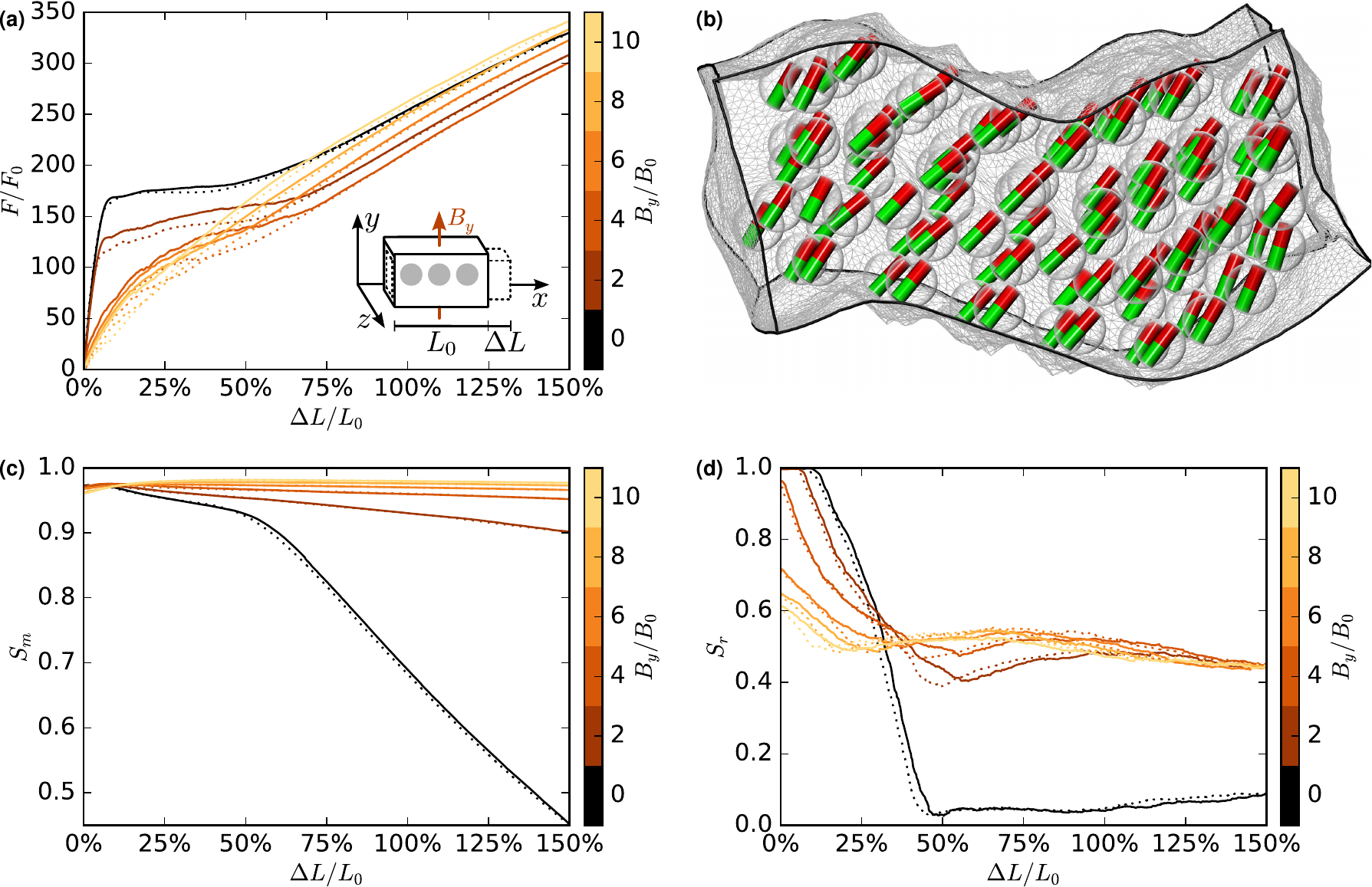}
	\caption{Same as Fig.~\ref{fig.free_By}, but for the \emph{cov$\,\rightrightarrows$} system. \protect\subref{fig.covrightright_By_stresstrain} The superelasticity in the stress-strain behavior can again be deactivated by a perpendicular external magnetic field, but only at significantly higher field strengths. \protect\subref{fig.snapshot_covrightright-By10_equi} Snapshot showing the unstretched state of a system under the influence of a field of $B_y = 10 B_0$. The system enters a new state of global polar magnetic order, with magnetic moments aligned oblique to the external magnetic field. Energetically expensive rotations of individual particles are avoided, instead whole chain segments rotate as one unit. \protect\subref{fig.covrightright_By_nematicOrder} Plot of the nematic order parameter $S_m$ for the magnetic moment orientations demonstrating that already a moderate magnetic field strength can maintain a state of global polar magnetic order up to the maximum elongation. \protect\subref{fig.covrightright_By_nematicOrder-nN} Nematic order parameter $S_r$ for the nearest-neighbor separation vectors. When the external magnetic field is weak, $S_r$ is high at low strains and then drops to a low and relatively constant level. A strong field removes this large drop so that a relatively constant intermediate level of structural order remains at all strains. This indicates again the tendency of whole chain segments to rotate as one unit, creating a principal axis of structural order oblique to the external magnetic field direction and the initial chain axes.}
	\label{fig.covrightright_By}
\end{figure*}

We now proceed to the results for the \emph{cov$\,\rightrightarrows$} system shown in Fig.~\ref{fig.covrightright_By}.
In the case of vanishing external magnetic field, this system features global magnetic order in the $x$-direction, see again Fig.~\ref{fig.snapshot_covrightright_equi}.
Applying an external magnetic field perpendicular to the stretching axis leads to a new state of rotated global polar magnetic order. 
Figure~\ref{fig.snapshot_covrightright-By10_equi} shows a snapshot of an unstretched system subject to a strong external magnetic field of $B_y = 10 B_0$.
The magnetic moments, together with the carrying particles, are rotated towards a configuration of collective polar alignment oblique to the external magnetic field.  
This occurs against the strong magnetic attractions within each chain and the necessary elastic deformation of the matrix between the particles. 
The rotations of individual particles are energetically expensive. 
In fact, the system partially avoids these expensive rotations by allowing chain segments to rotate as a whole towards the field.  
Undulations and buckling of the chains \cite{Huang2016_SoftMatter} then lead to a compromise between the minimization of the elastic and magnetic parts of the total energy. 

Either way, the magnetic dipolar attraction between neighboring particles along the stretching direction is weakened, which impedes the detachment mechanism. 
So the influence of the perpendicular external magnetic field on the stress-strain behavior is again a gradual removal of the superelastic plateau, as illustrated in Fig.~\ref{fig.covrightright_By_stresstrain}. 
A stiffening of the stress-strain behavior beyond the superelastic plateau, as in the case of a parallel external magnetic field, however, cannot be observed. 
Contrary to the parallel magnetic field, the perpendicular magnetic field breaks the uniaxial symmetry of the system and offers a distinctive direction for the particles to rotate towards. 
As can be deduced from the nematic order parameter $S_m$ of the magnetic moments plotted in Fig.~\ref{fig.covrightright_By_nematicOrder}, the perpendicular external field aligns the particles very effectively even up to the highest considered strains. 
Differences in the rotations of the particles due to elastic inhomogeneities can, thus, be prevented. 
A field of $B_y = 2 B_0$, is already quite successful in this respect, using stronger fields does not significantly increase the effect much further.
The mutual repulsion between the parallel magnetic moments does not counteract an elongation of the system any more. 
Thus, there is no significant stiffening of the stress-strain behavior when changing the external magnetic field strength. 

We also plot the nematic order parameter $S_r$ of the nearest-neighbor separation vectors in Fig.~\ref{fig.covrightright_By_nematicOrder-nN}. 
For $B_y = 0$, $S_r$ is at a high level for low strains, where it is most likely that the nearest-neighbor of a particle is located along the stretching axis within the same chain. 
Then $S_r$ quickly drops as the chains are stretched out and subsequently remains at a low level.
When a perpendicular magnetic field is applied, such a drop of $S_r$ never occurs. 
It remains likely that the nearest-neighbor of a particle is within the same chain for the whole considered range of strains. 
This reflects again the tendency of whole chain segments to rotate as one unit towards the field, staying structurally intact and creating the partial structural order reflected by $S_r$. 

\begin{figure*}[t!]
	\centering
	\subfloat{\label{fig.covleftright_By_stresstrain}} 
	\subfloat{\label{fig.snapshot_covleftright-By10_equi}} 
	\subfloat{\label{fig.covleftright_By_nematicOrder}} 
	\subfloat{\label{fig.covleftright_By_nematicOrder-nN}} 
	\includegraphics[width = 1.0\textwidth]{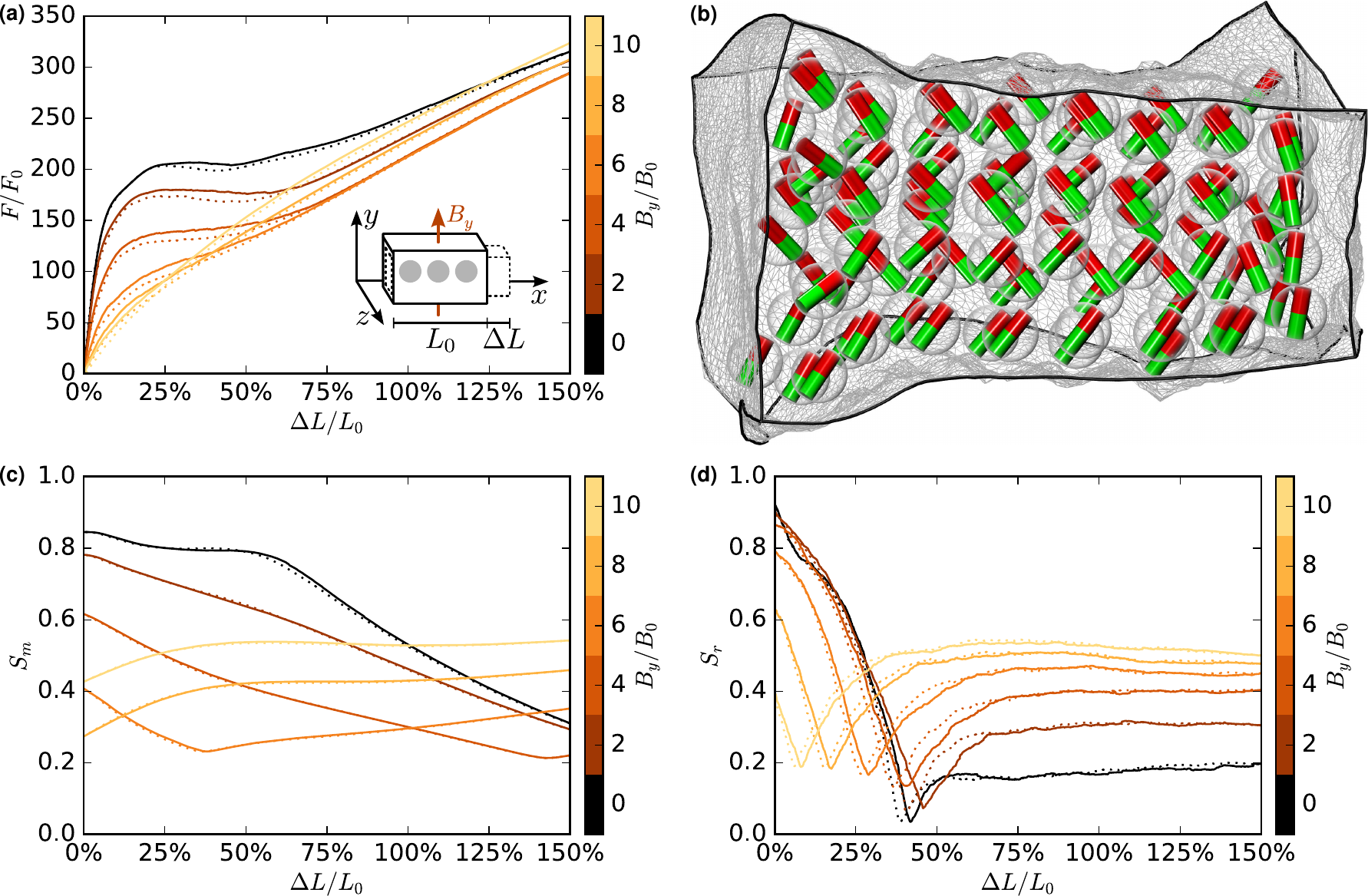}
	\caption{Same as Fig.~\ref{fig.free_By}, but for the \emph{cov$\,\leftrightarrows$} system. \protect\subref{fig.covleftright_By_stresstrain} The stress-strain behavior responds to the external magnetic field in a very similar way as for the \emph{cov$\,\leftrightarrows$} system. Increasing the field strength gradually removes the superelastic nonlinearity. \protect\subref{fig.snapshot_covleftright-By10_equi} Snapshot of an unstretched system with an applied external magnetic field of $B_y = 10 B_0$. There are two competing polarities for the magnetic moments, sharing a common $y$-component but with opposite $x$-components. \protect\subref{fig.covleftright_By_nematicOrder} Quantification of the magnetic order in the system via the nematic order parameter $S_m$ for the magnetic moment orientations. When the magnetic field strength and the strain are low, the two opposing polarities that are not aligned along a common axis compete, and $S_m$ is a decreasing function of the strain. The higher the magnetic field strength and the higher the strain, the more the magnetic moments are rotated. Eventually, the magnetic field direction is preferred over the stretching axis by both polarities and $S_m$ becomes an increasing function of the strain. For $B_y \gtrsim 8 B_0$ this is already the case in the unstretched state, which is consistent with the observation that the corresponding stress-strain curves do not show superelasticity anymore. \protect\subref{fig.covleftright_By_nematicOrder-nN} Nematic order parameter $S_r$ for the nearest-neighbor separation vectors, quantifying the structural order. The minimum in $S_r$ shifts to lower strains when increasing the field strength and the overall value beyond the minimum is increased. This is simply a consequence of the particle rotations that lead to less magnetic attraction between particles along the stretching axis and to more attraction along the magnetic field direction.}
	\label{fig.covleftright_By}
\end{figure*}

Let us finally discuss the \emph{cov$\,\leftrightarrows$} system under the influence of a perpendicular external magnetic field.
Contrary to the case of a parallel external magnetic field, there are no particles that are aligned oppositely to the external field.
All particles can in principle rotate equally easily into the external magnetic field direction. 
However, the initial orientation of the magnetic moment of a particle determines the sense of rotation towards the field. 
Neighboring chains with opposing alignment of the magnetic moments show opposing sense of rotation.
As a consequence, in contrast to the \emph{cov$\,\rightrightarrows$} system, the rotations of complete chain segments towards the magnetic field are largely blocked.
Instead, the particles within the chains individually rotate towards the external field, as depicted in the snapshot of an unstretched sample in Fig.~\ref{fig.snapshot_covleftright-By10_equi}. 
Here, the external magnetic field of $B_y = 10 B_0$ has rotated the particles by a significant amount, but the chains are still relatively ordered and aligned along the stretching axis. 
Depending on their initial alignment, the magnetic moments together with their carrying particles rotate either clockwise or counterclockwise towards the field.
In this way, there are two competing magnetic polarities in the system, with roughly the same $y$-component but oppositely signed $x$-components. 
The resulting stress-strain behavior is plotted in Fig.~\ref{fig.covleftright_By_stresstrain} and reveals an influence of the external magnetic field very similar to the \emph{cov$\,\rightrightarrows$} system.
Increasing the magnetic field strength rotates the particles further and weakens their attraction along the stretching axis.
This gradually disables the detachment mechanism and, therefore, removes the superelastic plateau from the stress-strain curve.  
Again, we cannot observe significant stiffening of the system at high strains when increasing the external magnetic field strength, for the same reasons as in the \emph{cov$\,\rightrightarrows$} system.

The two competing magnetic polarities are reflected by the nematic order parameter $S_m$ plotted in Fig.~\ref{fig.covleftright_By_nematicOrder}. 
In the unstretched state, when neighboring particles in a chain are close to each other, their magnetic interaction intensifies an alignment of the magnetic moments parallel to the stretching axis. 
The magnetic field, however, urges the differently orientated magnetic moments and their carrying particles to rotate out of their common initial axis of alignment.
More precisely, for magnetic moments of opposite initial orientation, this leads to opposite senses of rotation, which destroys the overall nematic alignment. 
At low field strengths the particles rotate only slightly in the unstretched state, so that $S_m$ is initially high.
Stronger fields are able to rotate the particles further, see again Fig.~\ref{fig.snapshot_covleftright-By10_equi}, leading to a lower value of $S_m$ at zero strain. 
With increasing strain, the magnetic interactions between neighboring particles in a chain are weakened due to their increased separation. 
The particles become more susceptible to rotations by the magnetic field. 
Thus, a decline in $S_m$ can be observed.  
$S_m$ increases again when the $y$-direction becomes predominant for all magnetic moments so that they again align along a common axis. 
At even stronger fields of $B_y = 8 B_0$ and $B_y = 10 B_0$, the $y$-direction is prevalent at all strains, so that $S_m$ is monotoneously increasing. 
This is in agreement with the observation, that for these magnetic field strengths superelastic features in the stress-strain curve are absent. 

Finally, we show in Fig.~\ref{fig.covleftright_By_nematicOrder-nN} the nematic order parameter $S_r$ for the nearest-neighbor separation vectors. 
The minimum in each curve indicates the point where it becomes more likely for particles to find their nearest-neighbors in a direction perpendicular to the stretching axis than parallel.
For low field strengths, this structural bias along the perpendicular axis is not very distinctive.  
Increasing the field strength, however, shifts the minimum to lower strains and increases the value of $S_r$ at higher strains. 
This is intuitive, because for stronger magnetic fields there is simply less attraction within individual chains along the stretching axis and more attraction perpendicular to the stretching axis between reoriented particles belonging to different chains.

In summary, the main effect of the perpendicular external magnetic field in all systems is the gradual removal of the superelastic plateau from the stress-strain curves. 
This is mainly caused by the rotation of the magnetic moments into the direction of the magnetic field. 
When the magnetic attraction between neighboring particles along the stretching axis disappears, the detachment mechanism seizes to function. 
In the \emph{free} system, magnetic moment reorientations can be achieved exceptionally easily (see the different scales for $B_y$ in Figs.~\ref{fig.free_By}--\ref{fig.covleftright_By}), making this system highly susceptible to the perpendicular external magnetic field. 
Together with the detachment mechanism, also the flipping mechanism is gradually deactivated.   
In the \emph{cov$\,\rightrightarrows$} system rotations of the magnetic moments are harder to achieve and require significantly stronger magnetic fields. 
We can observe collective rotations of the particles such that global polar magnetic ordering is preserved with all magnetic moments aligned oblique to the external field. 
Furthermore, these systems avoid the energetically expensive rotations of individual particles by allowing whole segments of the chains to rotate towards the external magnetic field as one unit.
As a result, the chains buckle and undulate as a compromise between minimizing the magnetic and elastic energetic contributions.
Finally, the \emph{cov$\,\leftrightarrows$} system behaves quite similar concerning the influence of the external field on the stress-strain behavior.
However, here the particles do rotate individually towards the field, facilitated by the initially opposite magnetic alignment in different chains.
During the rotation process, the opposing magnetic alignments lead to two separate polarization directions of the magnetic moments.
Altogether, in both \emph{cov} systems, particle rotations induced by elastic inhomogeneities of the system are effectively superseded by particle rotations due to the external magnetic field.

\section{Conclusions}
\label{sec.conclusions}

We have numerically investigated the stress-strain behavior of uniaxial ferrogel systems. 
Our anisotropic numerical systems consist of chain-like aggregates of spherical colloidal magnetic particles that are embedded in an elastic matrix of a cross-linked polymer. 
The particles are rigid and of finite size, while the matrix is treated by continuum elasticity theory.  
In experimental situations, the chain-like aggregates can be generated by applying a strong homogeneous external magnetic field during synthesis. 
We have considered three different realizations of such uniaxial ferrogel systems.  
The \emph{free} system features magnetic moments that can freely reorient with respect to the frames of the carrying particles frames and the surrounding matrix. 
In contrast to that, in the \emph{cov$\,\rightrightarrows$} system the magnetic moments are fixed with respect to the axes of the carrying particles. 
Additionally, the particles are covalently embedded into the matrix: particle rotations require corotations of the directly surrounding elastic material, leading to matrix deformations and restoring torques.
Initially, all magnetic moments point into the same direction along the chain axes.  
The third system is the \emph{cov$\,\leftrightarrows$} system, where the magnetic moments are likewise firmly anchored.
However, initially the magnetic moments point into opposite directions along the chain axes. 

When we stretch these systems along the chain axes, a pronounced nonlinearity in the stress-strain behavior appears.
It has the form of a superelastic plateau, along which the samples can be strongly deformed while barely increasing the load. 
The deformation is reversible and the shape and intensity of the superelastic plateau can be reversibly tailored by external magnetic fields.
There are two stretching-induced mechanisms that enable superelasticty.
The main mechanism is a detachment mechanism and active in all systems. 
It relies on the strong magnetic dipolar attraction between neighboring particles within one chain as long as the magnetic moments align along the chain axis.
At certain threshold strains, parts of the chain can detach, leading to a local elongation of the system. 
This leaves the remainder of the chain intact until the next detachment event is triggered.
Besides, a flipping mechanism corresponding to reorientation events of magnetic moments is only active in the \emph{free} system, where the magnetic moments can easily rotate.  
A flip event occurs when elongation of the system causes positional rearrangements such that for a subset of magnetic moments a new orientation is suddenly rendered energetically more favorable.

The inhomogeneous distribution of the rigid inclusions in our samples results in regions of elevated stiffness. 
At high strains, this leads to local shears that rotate the embedded particles.
This is especially apparent in the \emph{cov$\,\rightrightarrows$} and \emph{cov$\,\leftrightarrows$} systems and influences their stress-strain behavior.

Our systems can be reversibly tuned by an external magnetic field as follows.
If the field is applied parallel to the chain axes, the detachment mechanism is not affected in the \emph{free} and \emph{cov$\,\rightrightarrows$} systems, so that the superelastic plateau remains intact. 
However, in the \emph{cov$\,\leftrightarrows$} system the particles carrying misaligned magnetic moments are forced to rotate.
The corresponding chains are strongly distorted, which perturbs the neighboring chains carrying aligned magnetic moments as well. 
This weakens the required magnetic attractions along the stretching axis that are vital for a pronounced detachment mechanism and removes the superelasticity from the stress-strain curve of the \emph{cov$\,\leftrightarrows$} system.
Moreover, in the \emph{free} system the flipping mechanism can be deactivated as well, as the aligning external magnetic field hinders reorientations of magnetic moments.  
Consequently, the related features are removed from the stress-strain behavior, leaving only a flat plateau caused by the detachment mechanism.
Finally, in the \emph{cov$\,\rightrightarrows$} system, the external field parallel to the chains has another interesting effect.  
We can observe a stiffening of the system when increasing the field strength at high strains beyond the superelastic plateau.
In this situation, all particles have been detached from their chains, leaving them particularly susceptible to rotations due to shears caused by the elastic inhomogeneity of the system.  
Since the external magnetic field introduces an energetic penalty for particle rotations, the intrinsic inhomogeneity-caused shear stresses cannot relax via particle rotations and the magnetic moments remain parallel to each other. 
The parallel magnetic moments repel each other in the direction perpendicular to the stretching axis and, thus, work against a volume-conserving stretching deformation.
In combination both effects increase the stiffness of the system.  

When instead the magnetic field is applied perpendicular to the stretching axis, the detachment mechanism is weakened in all three systems due to an induced rotation of the magnetic moments towards a configuration which is repulsive along the stretching axis. 
In this way, the superelastic plateau can be gradually removed from the stress-strain curve by increasing the field strength.
This works exceptionally well in the \emph{free} system, where  the magnetic moments are not anchored to the particle frames and the flipping mechanism is likewise weakened. 
In contrast to that, in the \emph{cov$\,\rightrightarrows$} and \emph{cov$\,\leftrightarrows$} systems, even a strong external magnetic field cannot rotate the magnetic moments completely.
While in the \emph{cov$\,\rightrightarrows$} system, the magnetic moments feature a global magnetic alignment oblique to the external magnetic field, the two opposite initial magnetic alignment directions in the \emph{cov$\,\leftrightarrows$} system lead to two separate polar alignment directions, each of them oblique to the external magnetic field. 

Our effects rely on the sufficiently strong magnetic interactions in our systems when compared to the elastic interactions.
To achieve this experimentally, the remnant magnetization of the particle material should be as high as possible. 
For example, \ce{NdFeB}, can easily exceed $2 \times 10^5 \, \textrm{A}/\textrm{m}$ \cite{Kramarenko2015_SmartMaterStruct}.
At the same time, the elastic matrix into which the particles are embedded should be soft.
Fabricating matrices with $E \lesssim 10^3 \, \textrm{Pa}$ is possible using silicone \cite{Hoang2009_SmartMaterStruct,Chertovich2010_MacromolMaterEng,Stoll2014_JApplPolymSci} or polydimethylsiloxane \cite{Huang2016_SoftMatter}. 
With these materials, our assumed value of $m = 10 \, m_0$ can be achieved and is, therefore, experimentally realistic.
Also the highest considered magnetic field strength of $B = 10 B_0$ corresponding to $100 \, \textrm{mT}$ is readily accessible.   
We stress that the behavior of our systems does not depend on the length scale of the problem. 
In an experiment, this freedom can for instance be exploited to adjust the particle size to the effect under investigation. 
For example, the \emph{free} system could be realized by relatively small particles where the N\'eel mechanism \cite{Neel1949_AnnGeophys} is active and the magnetic moments can rotate relatively to the particle frame. 
Increased particle size would be necessary to generate the \emph{cov$\,\rightrightarrows$} and \emph{cov$\,\leftrightarrows$} systems.

The \emph{free} and \emph{cov$\,\rightrightarrows$} systems can be generated by applying an external magnetic field during synthesis to form the embedded chains \cite{Zubarev2000_PhysRevE,Hynninen2005_PhysRevLett,Auernhammer2006_JChemPhys,Smallenburg2012_JPhysCondensMatter} from N\'eel-type particles \cite{Neel1949_AnnGeophys} and from monodomain particles of larger size, respectively, possibly by covalently anchoring appropriately sized particles into the matrix \cite{Frickel2011_JMaterChem,Ilg2013_SoftMatter,Roeder2014_Macromolecules,Roeder2015_PhysChemChemPhys}. 
For the small N\'eel-type particles, typically of sizes up to 10--15 nm, thermal fluctuations become important. These can suppress the hysteretic behavior as well as the negative slope associated with the dip in our stress-strain curves. 
Overall, these fluctuations will smoothen the bumps along the plateau, leading to a flatter appearance. \emph{Free} systems of larger particle size could be realized e.g. using so-called yolk-shell colloidal particles \cite{Liu2012_JMaterChem,Okada2013_Langmuir} that consist of a magnetic core rotatable within a shell.
To realize the \emph{cov$\,\leftrightarrows$} system, electro-magnetorheological fluids \cite{Fujita1999_PowderTechnol,Wen2003_PhysicaB,Wang2013_DaltonTrans} could be used as a precursor of the anisotropic ferrogel. 
In such a system, an external electrical field can be applied to induce the chain formation of the electrically polarizable magnetic particles, while still allowing for opposite alignments of the magnetic moments in separate chains. 
Subsequent cross-linking of the surrounding polymer with covalent embedding of the particles should lock the chain structures together with their oppositely directed magnetic alignments into the emerging matrix.
The result would be an anisotropic ferrogel with the desired \emph{cov$\,\leftrightarrows$} morphology. 

We have assumed permanent magnetic dipoles carried by spherical particles in this work. 
The particles are arranged in characteristic chain-like structures.
Possible quantitative refinements comprise extensions beyond the permanent point-dipole picture \cite{Biller2014_JApplPhys,Biller2015_JOptoelectronAdvMater,Allahyarov2015_PhysChemChemPhys} or to elongated, non-spherical particles \cite{Bender2011_JMagnMagnMater,Tierno2014_PhysChemChemPhys,Roeder2014_Macromolecules,Roeder2015_PhysChemChemPhys}.
However, the main mechanism leading to superelastic behavior in our systems is the detachment mechanism for which only strong attraction at short distances between the neighboring particles along the stretching axis is necessary. 
This kind of attraction can likewise be realized for soft magnetic particles magnetized by an external field. 
The same mechanism could also be realized for nonmagnetic attractive interaction forces, e.g., for particles sufficiently polarizable by an external electrical field.
Moreover, also the flipping mechanism could be initiated for soft magnetic particles, when the direction of a magnetizing external magnetic field is switched at the corresponding imposed strain. 
Furthermore, to observe the basic phenomenology, the chain-like aggregates do not necessarily need to span the whole system.
In the most basic opposite situation, embedded pair aggregates would be sufficient \cite{Biller2014_JApplPhys}. 
Also the chains do not need to be as perfectly straight as considered here but could for example be weakly wiggled \cite{Han2013_IntJSolidsStruct}. 
On the theoretical side, a connection to continuum descriptions on the macroscopic scale shall be established in the future \cite{Menzel2016_PhysRevE,Bohlius2004_PhysRevE,Jarkova2003_PhysRevE}.   

Exploiting the described reversibly tunable nonlinear stress-strain behavior of our systems should enables a manifold of applications. 
When a pre-stress is applied to the material, such that it is pre-strained to the superelastic regime, it becomes extremely deformable\cite{Menzel2009_EurPhysJE}. 
This is an interesting property for easily applicable gaskets, packagings, or valves \cite{Boese2012_JIntellMaterSystStruct}.
Moreover, in such a state, the ferrogel can be operated as a soft actuator \cite{Zhou2005_SmartMaterStruct,Zimmermann2006_JPhysCondensMatter,Kashima2012_IEEETransMagn,Galipeau2013_ProcRSocA,Allahyarov2014_SmartMaterStruct}, as external magnetic fields can trigger significant deformations. 
Passive dampers based on superelastic shape-memory alloys are already established \cite{Saadat2002_SmartMaterStruct,Ozbulut2011_JIntellMaterSystStruct} and utilize hysteretic losses under recoverable cyclic loading to dissipate the energy. 
Our results for the \emph{free} system might stimulate the construction of analogous soft passive dampers with the additional benefit of being reversibly tunable from outside. 
Finally, the typically elevated biocompatibility of polymeric materials allows for medical applications exploiting the above features, e.g., in the form of quickly fittable wound dressings, artificial muscles \cite{Ramanujan2006_SmartMaterStruct,Shahinpoor2007_book}, or tunable implants \cite{Cezar2014_AdvHealthcMater,Cezar2016_ProcNatlAcadSciUSA}. 

\section*{Acknowledgements}
The authors thank the Deutsche Forschungsgemeinschaft for support of this work through the priority program SPP 1681.

\bibliography{references} 

\begin{thebibliography}{97}%
\makeatletter
\providecommand \@ifxundefined [1]{%
 \@ifx{#1\undefined}
}%
\providecommand \@ifnum [1]{%
 \ifnum #1\expandafter \@firstoftwo
 \else \expandafter \@secondoftwo
 \fi
}%
\providecommand \@ifx [1]{%
 \ifx #1\expandafter \@firstoftwo
 \else \expandafter \@secondoftwo
 \fi
}%
\providecommand \natexlab [1]{#1}%
\providecommand \enquote  [1]{``#1''}%
\providecommand \bibnamefont  [1]{#1}%
\providecommand \bibfnamefont [1]{#1}%
\providecommand \citenamefont [1]{#1}%
\providecommand \href@noop [0]{\@secondoftwo}%
\providecommand \href [0]{\begingroup \@sanitize@url \@href}%
\providecommand \@href[1]{\@@startlink{#1}\@@href}%
\providecommand \@@href[1]{\endgroup#1\@@endlink}%
\providecommand \@sanitize@url [0]{\catcode `\\12\catcode `\$12\catcode
  `\&12\catcode `\#12\catcode `\^12\catcode `\_12\catcode `\%12\relax}%
\providecommand \@@startlink[1]{}%
\providecommand \@@endlink[0]{}%
\providecommand \url  [0]{\begingroup\@sanitize@url \@url }%
\providecommand \@url [1]{\endgroup\@href {#1}{\urlprefix }}%
\providecommand \urlprefix  [0]{URL }%
\providecommand \Eprint [0]{\href }%
\providecommand \doibase [0]{http://dx.doi.org/}%
\providecommand \selectlanguage [0]{\@gobble}%
\providecommand \bibinfo  [0]{\@secondoftwo}%
\providecommand \bibfield  [0]{\@secondoftwo}%
\providecommand \translation [1]{[#1]}%
\providecommand \BibitemOpen [0]{}%
\providecommand \bibitemStop [0]{}%
\providecommand \bibitemNoStop [0]{.\EOS\space}%
\providecommand \EOS [0]{\spacefactor3000\relax}%
\providecommand \BibitemShut  [1]{\csname bibitem#1\endcsname}%
\let\auto@bib@innerbib\@empty
\bibitem [{\citenamefont {Zr\'inyi}\ \emph {et~al.}(1995)\citenamefont
  {Zr\'inyi}, \citenamefont {Barsi},\ and\ \citenamefont
  {B\"uki}}]{Zrinyi1995_PolymGelsNetw}%
  \BibitemOpen
  \bibfield  {author} {\bibinfo {author} {\bibfnamefont {M.}~\bibnamefont
  {Zr\'inyi}}, \bibinfo {author} {\bibfnamefont {L.}~\bibnamefont {Barsi}}, \
  and\ \bibinfo {author} {\bibfnamefont {A.}~\bibnamefont {B\"uki}},\
  }\href@noop {} {\bibfield  {journal} {\bibinfo  {journal} {Polym. Gels
  Netw.}\ }\textbf {\bibinfo {volume} {5}},\ \bibinfo {pages} {415} (\bibinfo
  {year} {1995})}\BibitemShut {NoStop}%
\bibitem [{\citenamefont {Menzel}(2015)}]{Menzel2015_PhysRep}%
  \BibitemOpen
  \bibfield  {author} {\bibinfo {author} {\bibfnamefont {A.~M.}\ \bibnamefont
  {Menzel}},\ }\href@noop {} {\bibfield  {journal} {\bibinfo  {journal}
  {Physics Reports}\ }\textbf {\bibinfo {volume} {554}},\ \bibinfo {pages} {1}
  (\bibinfo {year} {2015})}\BibitemShut {NoStop}%
\bibitem [{\citenamefont {Odenbach}(2016)}]{Odenbach2016_ArchApplMech}%
  \BibitemOpen
  \bibfield  {author} {\bibinfo {author} {\bibfnamefont {S.}~\bibnamefont
  {Odenbach}},\ }\href@noop {} {\bibfield  {journal} {\bibinfo  {journal}
  {Arch. Appl. Mech.}\ }\textbf {\bibinfo {volume} {86}},\ \bibinfo {pages}
  {269} (\bibinfo {year} {2016})}\BibitemShut {NoStop}%
\bibitem [{\citenamefont {Jarkova}\ \emph {et~al.}(2003)\citenamefont
  {Jarkova}, \citenamefont {Pleiner}, \citenamefont {M\"uller},\ and\
  \citenamefont {Brand}}]{Jarkova2003_PhysRevE}%
  \BibitemOpen
  \bibfield  {author} {\bibinfo {author} {\bibfnamefont {E.}~\bibnamefont
  {Jarkova}}, \bibinfo {author} {\bibfnamefont {H.}~\bibnamefont {Pleiner}},
  \bibinfo {author} {\bibfnamefont {H.-W.}\ \bibnamefont {M\"uller}}, \ and\
  \bibinfo {author} {\bibfnamefont {H.~R.}\ \bibnamefont {Brand}},\ }\href@noop
  {} {\bibfield  {journal} {\bibinfo  {journal} {Phys. Rev. E}\ }\textbf
  {\bibinfo {volume} {68}},\ \bibinfo {pages} {041706} (\bibinfo {year}
  {2003})}\BibitemShut {NoStop}%
\bibitem [{\citenamefont {Filipcsei}\ \emph {et~al.}(2007)\citenamefont
  {Filipcsei}, \citenamefont {Csetneki}, \citenamefont {Szil\'agyi},\ and\
  \citenamefont {Zr\'inyi}}]{Filipcsei2007_AdvPolymSci}%
  \BibitemOpen
  \bibfield  {author} {\bibinfo {author} {\bibfnamefont {G.}~\bibnamefont
  {Filipcsei}}, \bibinfo {author} {\bibfnamefont {I.}~\bibnamefont {Csetneki}},
  \bibinfo {author} {\bibfnamefont {A.}~\bibnamefont {Szil\'agyi}}, \ and\
  \bibinfo {author} {\bibfnamefont {M.}~\bibnamefont {Zr\'inyi}},\ }\href@noop
  {} {\bibfield  {journal} {\bibinfo  {journal} {Adv. Polym. Sci.}\ }\textbf
  {\bibinfo {volume} {206}},\ \bibinfo {pages} {137} (\bibinfo {year}
  {2007})}\BibitemShut {NoStop}%
\bibitem [{\citenamefont {Mitsumata}\ and\ \citenamefont
  {Ohori}(2011)}]{Mitsumata2011_PolymChem}%
  \BibitemOpen
  \bibfield  {author} {\bibinfo {author} {\bibfnamefont {T.}~\bibnamefont
  {Mitsumata}}\ and\ \bibinfo {author} {\bibfnamefont {S.}~\bibnamefont
  {Ohori}},\ }\href@noop {} {\bibfield  {journal} {\bibinfo  {journal} {Polym.
  Chem.}\ }\textbf {\bibinfo {volume} {2}},\ \bibinfo {pages} {1063} (\bibinfo
  {year} {2011})}\BibitemShut {NoStop}%
\bibitem [{\citenamefont {Wood}\ and\ \citenamefont
  {Camp}(2011)}]{Wood2011_PhysRevE}%
  \BibitemOpen
  \bibfield  {author} {\bibinfo {author} {\bibfnamefont {D.~S.}\ \bibnamefont
  {Wood}}\ and\ \bibinfo {author} {\bibfnamefont {P.~J.}\ \bibnamefont
  {Camp}},\ }\href@noop {} {\bibfield  {journal} {\bibinfo  {journal} {Phys.
  Rev. E}\ }\textbf {\bibinfo {volume} {83}},\ \bibinfo {pages} {011402}
  (\bibinfo {year} {2011})}\BibitemShut {NoStop}%
\bibitem [{\citenamefont {Han}\ \emph {et~al.}(2013)\citenamefont {Han},
  \citenamefont {Hong},\ and\ \citenamefont
  {Faidley}}]{Han2013_IntJSolidsStruct}%
  \BibitemOpen
  \bibfield  {author} {\bibinfo {author} {\bibfnamefont {Y.}~\bibnamefont
  {Han}}, \bibinfo {author} {\bibfnamefont {W.}~\bibnamefont {Hong}}, \ and\
  \bibinfo {author} {\bibfnamefont {L.~E.}\ \bibnamefont {Faidley}},\
  }\href@noop {} {\bibfield  {journal} {\bibinfo  {journal} {Int. J. Solids
  Struct.}\ }\textbf {\bibinfo {volume} {50}},\ \bibinfo {pages} {2281}
  (\bibinfo {year} {2013})}\BibitemShut {NoStop}%
\bibitem [{\citenamefont {Mitsumata}\ \emph {et~al.}(2013)\citenamefont
  {Mitsumata}, \citenamefont {Ohori}, \citenamefont {Honda},\ and\
  \citenamefont {Kawai}}]{Mitsumata2013_SoftMatter}%
  \BibitemOpen
  \bibfield  {author} {\bibinfo {author} {\bibfnamefont {T.}~\bibnamefont
  {Mitsumata}}, \bibinfo {author} {\bibfnamefont {S.}~\bibnamefont {Ohori}},
  \bibinfo {author} {\bibfnamefont {A.}~\bibnamefont {Honda}}, \ and\ \bibinfo
  {author} {\bibfnamefont {M.}~\bibnamefont {Kawai}},\ }\href@noop {}
  {\bibfield  {journal} {\bibinfo  {journal} {Soft Matter}\ }\textbf {\bibinfo
  {volume} {9}},\ \bibinfo {pages} {904} (\bibinfo {year} {2013})}\BibitemShut
  {NoStop}%
\bibitem [{\citenamefont {Stoll}\ \emph {et~al.}(2014)\citenamefont {Stoll},
  \citenamefont {Mayer}, \citenamefont {Monkman},\ and\ \citenamefont
  {Shamonin}}]{Stoll2014_JApplPolymSci}%
  \BibitemOpen
  \bibfield  {author} {\bibinfo {author} {\bibfnamefont {A.}~\bibnamefont
  {Stoll}}, \bibinfo {author} {\bibfnamefont {M.}~\bibnamefont {Mayer}},
  \bibinfo {author} {\bibfnamefont {G.~J.}\ \bibnamefont {Monkman}}, \ and\
  \bibinfo {author} {\bibfnamefont {M.}~\bibnamefont {Shamonin}},\ }\href@noop
  {} {\bibfield  {journal} {\bibinfo  {journal} {J. Appl. Polym. Sci.}\
  }\textbf {\bibinfo {volume} {131}},\ \bibinfo {eid} {39793} (\bibinfo {year}
  {2014})}\BibitemShut {NoStop}%
\bibitem [{\citenamefont {Peroukidis}\ and\ \citenamefont
  {Klapp}(2015)}]{Peroukidis2015_PhysRevE}%
  \BibitemOpen
  \bibfield  {author} {\bibinfo {author} {\bibfnamefont {S.~D.}\ \bibnamefont
  {Peroukidis}}\ and\ \bibinfo {author} {\bibfnamefont {S.~H.~L.}\ \bibnamefont
  {Klapp}},\ }\href@noop {} {\bibfield  {journal} {\bibinfo  {journal} {Phys.
  Rev. E}\ }\textbf {\bibinfo {volume} {92}},\ \bibinfo {pages} {010501}
  (\bibinfo {year} {2015})}\BibitemShut {NoStop}%
\bibitem [{\citenamefont {Peroukidis}\ \emph {et~al.}(2015)\citenamefont
  {Peroukidis}, \citenamefont {Lichtner},\ and\ \citenamefont
  {Klapp}}]{Peroukidis2015_SoftMatter}%
  \BibitemOpen
  \bibfield  {author} {\bibinfo {author} {\bibfnamefont {S.~D.}\ \bibnamefont
  {Peroukidis}}, \bibinfo {author} {\bibfnamefont {K.}~\bibnamefont
  {Lichtner}}, \ and\ \bibinfo {author} {\bibfnamefont {S.~H.~L.}\ \bibnamefont
  {Klapp}},\ }\href@noop {} {\bibfield  {journal} {\bibinfo  {journal} {Soft
  Matter}\ }\textbf {\bibinfo {volume} {11}},\ \bibinfo {pages} {5999}
  (\bibinfo {year} {2015})}\BibitemShut {NoStop}%
\bibitem [{\citenamefont {Schubert}\ and\ \citenamefont
  {Harrison}(2016)}]{Schubert2016_SmartMaterStruct}%
  \BibitemOpen
  \bibfield  {author} {\bibinfo {author} {\bibfnamefont {G.}~\bibnamefont
  {Schubert}}\ and\ \bibinfo {author} {\bibfnamefont {P.}~\bibnamefont
  {Harrison}},\ }\href@noop {} {\bibfield  {journal} {\bibinfo  {journal}
  {Smart Mater. Struct.}\ }\textbf {\bibinfo {volume} {25}},\ \bibinfo {pages}
  {015015} (\bibinfo {year} {2016})}\BibitemShut {NoStop}%
\bibitem [{\citenamefont {Wang}\ \emph {et~al.}(2016)\citenamefont {Wang},
  \citenamefont {Xuan}, \citenamefont {Dong}, \citenamefont {Xu},\ and\
  \citenamefont {Gong}}]{Wang2016_SmartMaterStruct}%
  \BibitemOpen
  \bibfield  {author} {\bibinfo {author} {\bibfnamefont {Y.}~\bibnamefont
  {Wang}}, \bibinfo {author} {\bibfnamefont {S.}~\bibnamefont {Xuan}}, \bibinfo
  {author} {\bibfnamefont {B.}~\bibnamefont {Dong}}, \bibinfo {author}
  {\bibfnamefont {F.}~\bibnamefont {Xu}}, \ and\ \bibinfo {author}
  {\bibfnamefont {X.}~\bibnamefont {Gong}},\ }\href@noop {} {\bibfield
  {journal} {\bibinfo  {journal} {Smart Mater. Struct.}\ }\textbf {\bibinfo
  {volume} {25}},\ \bibinfo {pages} {025003} (\bibinfo {year}
  {2016})}\BibitemShut {NoStop}%
\bibitem [{\citenamefont {Sedlacik}\ \emph {et~al.}(2016)\citenamefont
  {Sedlacik}, \citenamefont {Mrlik}, \citenamefont {Babayan},\ and\
  \citenamefont {Pavlinek}}]{Sedlacik2016_ComposStruct}%
  \BibitemOpen
  \bibfield  {author} {\bibinfo {author} {\bibfnamefont {M.}~\bibnamefont
  {Sedlacik}}, \bibinfo {author} {\bibfnamefont {M.}~\bibnamefont {Mrlik}},
  \bibinfo {author} {\bibfnamefont {V.}~\bibnamefont {Babayan}}, \ and\
  \bibinfo {author} {\bibfnamefont {V.}~\bibnamefont {Pavlinek}},\ }\href@noop
  {} {\bibfield  {journal} {\bibinfo  {journal} {Compos. Struct.}\ }\textbf
  {\bibinfo {volume} {135}},\ \bibinfo {pages} {199} (\bibinfo {year}
  {2016})}\BibitemShut {NoStop}%
\bibitem [{\citenamefont {Sun}\ \emph {et~al.}(2008)\citenamefont {Sun},
  \citenamefont {Gong}, \citenamefont {Jiang}, \citenamefont {Li},
  \citenamefont {Xu},\ and\ \citenamefont {Li}}]{Sun2008_PolymTest}%
  \BibitemOpen
  \bibfield  {author} {\bibinfo {author} {\bibfnamefont {T.}~\bibnamefont
  {Sun}}, \bibinfo {author} {\bibfnamefont {X.}~\bibnamefont {Gong}}, \bibinfo
  {author} {\bibfnamefont {W.}~\bibnamefont {Jiang}}, \bibinfo {author}
  {\bibfnamefont {J.}~\bibnamefont {Li}}, \bibinfo {author} {\bibfnamefont
  {Z.}~\bibnamefont {Xu}}, \ and\ \bibinfo {author} {\bibfnamefont
  {W.}~\bibnamefont {Li}},\ }\href@noop {} {\bibfield  {journal} {\bibinfo
  {journal} {Polym. Test.}\ }\textbf {\bibinfo {volume} {27}},\ \bibinfo
  {pages} {520} (\bibinfo {year} {2008})}\BibitemShut {NoStop}%
\bibitem [{\citenamefont {Deng}\ \emph {et~al.}(2006)\citenamefont {Deng},
  \citenamefont {Gong},\ and\ \citenamefont
  {Wang}}]{Deng2006_SmartMaterStruct}%
  \BibitemOpen
  \bibfield  {author} {\bibinfo {author} {\bibfnamefont {H.-x.}\ \bibnamefont
  {Deng}}, \bibinfo {author} {\bibfnamefont {X.-l.}\ \bibnamefont {Gong}}, \
  and\ \bibinfo {author} {\bibfnamefont {L.-h.}\ \bibnamefont {Wang}},\
  }\href@noop {} {\bibfield  {journal} {\bibinfo  {journal} {Smart Mater.
  Struct.}\ }\textbf {\bibinfo {volume} {15}},\ \bibinfo {pages} {N111}
  (\bibinfo {year} {2006})}\BibitemShut {NoStop}%
\bibitem [{\citenamefont {Stepanov}\ \emph {et~al.}(2007)\citenamefont
  {Stepanov}, \citenamefont {Abramchuk}, \citenamefont {Grishin}, \citenamefont
  {Nikitin}, \citenamefont {Kramarenko},\ and\ \citenamefont
  {Khokhlov}}]{Stepanov2007_Polymer}%
  \BibitemOpen
  \bibfield  {author} {\bibinfo {author} {\bibfnamefont {G.~V.}\ \bibnamefont
  {Stepanov}}, \bibinfo {author} {\bibfnamefont {S.~S.}\ \bibnamefont
  {Abramchuk}}, \bibinfo {author} {\bibfnamefont {D.~A.}\ \bibnamefont
  {Grishin}}, \bibinfo {author} {\bibfnamefont {L.~V.}\ \bibnamefont
  {Nikitin}}, \bibinfo {author} {\bibfnamefont {E.~Y.}\ \bibnamefont
  {Kramarenko}}, \ and\ \bibinfo {author} {\bibfnamefont {A.~R.}\ \bibnamefont
  {Khokhlov}},\ }\href@noop {} {\bibfield  {journal} {\bibinfo  {journal}
  {Polymer}\ }\textbf {\bibinfo {volume} {48}},\ \bibinfo {pages} {488}
  (\bibinfo {year} {2007})}\BibitemShut {NoStop}%
\bibitem [{\citenamefont {Nguyen}\ and\ \citenamefont
  {Ramanujan}(2010)}]{Nguyen2010_MacromolChemPhys}%
  \BibitemOpen
  \bibfield  {author} {\bibinfo {author} {\bibfnamefont {V.~Q.}\ \bibnamefont
  {Nguyen}}\ and\ \bibinfo {author} {\bibfnamefont {R.~V.}\ \bibnamefont
  {Ramanujan}},\ }\href@noop {} {\bibfield  {journal} {\bibinfo  {journal}
  {Macromol. Chem. Phys.}\ }\textbf {\bibinfo {volume} {211}},\ \bibinfo
  {pages} {618} (\bibinfo {year} {2010})}\BibitemShut {NoStop}%
\bibitem [{\citenamefont {Snyder}\ \emph {et~al.}(2010)\citenamefont {Snyder},
  \citenamefont {Nguyen},\ and\ \citenamefont
  {Ramanujan}}]{Snyder2010_ActaMater}%
  \BibitemOpen
  \bibfield  {author} {\bibinfo {author} {\bibfnamefont {R.}~\bibnamefont
  {Snyder}}, \bibinfo {author} {\bibfnamefont {V.}~\bibnamefont {Nguyen}}, \
  and\ \bibinfo {author} {\bibfnamefont {R.}~\bibnamefont {Ramanujan}},\
  }\href@noop {} {\bibfield  {journal} {\bibinfo  {journal} {Acta Mater.}\
  }\textbf {\bibinfo {volume} {58}},\ \bibinfo {pages} {5620} (\bibinfo {year}
  {2010})}\BibitemShut {NoStop}%
\bibitem [{\citenamefont {Gong}\ \emph {et~al.}(2012)\citenamefont {Gong},
  \citenamefont {Liao},\ and\ \citenamefont {Xuan}}]{Gong2012_ApplPhysLett}%
  \BibitemOpen
  \bibfield  {author} {\bibinfo {author} {\bibfnamefont {X.}~\bibnamefont
  {Gong}}, \bibinfo {author} {\bibfnamefont {G.}~\bibnamefont {Liao}}, \ and\
  \bibinfo {author} {\bibfnamefont {S.}~\bibnamefont {Xuan}},\ }\href@noop {}
  {\bibfield  {journal} {\bibinfo  {journal} {Appl. Phys. Lett.}\ }\textbf
  {\bibinfo {volume} {100}},\ \bibinfo {eid} {211909} (\bibinfo {year}
  {2012})}\BibitemShut {NoStop}%
\bibitem [{\citenamefont {Zhou}\ and\ \citenamefont
  {Wang}(2005)}]{Zhou2005_SmartMaterStruct}%
  \BibitemOpen
  \bibfield  {author} {\bibinfo {author} {\bibfnamefont {G.}~\bibnamefont
  {Zhou}}\ and\ \bibinfo {author} {\bibfnamefont {Q.}~\bibnamefont {Wang}},\
  }\href@noop {} {\bibfield  {journal} {\bibinfo  {journal} {Smart Mater.
  Struct.}\ }\textbf {\bibinfo {volume} {14}},\ \bibinfo {pages} {504}
  (\bibinfo {year} {2005})}\BibitemShut {NoStop}%
\bibitem [{\citenamefont {Zimmermann}\ \emph {et~al.}(2006)\citenamefont
  {Zimmermann}, \citenamefont {Naletova}, \citenamefont {Zeidis}, \citenamefont
  {B\"ohm},\ and\ \citenamefont {Kolev}}]{Zimmermann2006_JPhysCondensMatter}%
  \BibitemOpen
  \bibfield  {author} {\bibinfo {author} {\bibfnamefont {K.}~\bibnamefont
  {Zimmermann}}, \bibinfo {author} {\bibfnamefont {V.~A.}\ \bibnamefont
  {Naletova}}, \bibinfo {author} {\bibfnamefont {I.}~\bibnamefont {Zeidis}},
  \bibinfo {author} {\bibfnamefont {V.}~\bibnamefont {B\"ohm}}, \ and\ \bibinfo
  {author} {\bibfnamefont {E.}~\bibnamefont {Kolev}},\ }\href@noop {}
  {\bibfield  {journal} {\bibinfo  {journal} {J. Phys.: Condens. Matter}\
  }\textbf {\bibinfo {volume} {18}},\ \bibinfo {pages} {S2973} (\bibinfo {year}
  {2006})}\BibitemShut {NoStop}%
\bibitem [{\citenamefont {B\"ose}\ \emph {et~al.}(2012)\citenamefont {B\"ose},
  \citenamefont {Rabindranath},\ and\ \citenamefont
  {Ehrlich}}]{Boese2012_JIntellMaterSystStruct}%
  \BibitemOpen
  \bibfield  {author} {\bibinfo {author} {\bibfnamefont {H.}~\bibnamefont
  {B\"ose}}, \bibinfo {author} {\bibfnamefont {R.}~\bibnamefont
  {Rabindranath}}, \ and\ \bibinfo {author} {\bibfnamefont {J.}~\bibnamefont
  {Ehrlich}},\ }\href@noop {} {\bibfield  {journal} {\bibinfo  {journal} {J.
  Intell. Mater. Syst. Struct.}\ }\textbf {\bibinfo {volume} {23}},\ \bibinfo
  {pages} {989} (\bibinfo {year} {2012})}\BibitemShut {NoStop}%
\bibitem [{\citenamefont {Kashima}\ \emph {et~al.}(2012)\citenamefont
  {Kashima}, \citenamefont {Miyasaka},\ and\ \citenamefont
  {Hirata}}]{Kashima2012_IEEETransMagn}%
  \BibitemOpen
  \bibfield  {author} {\bibinfo {author} {\bibfnamefont {S.}~\bibnamefont
  {Kashima}}, \bibinfo {author} {\bibfnamefont {F.}~\bibnamefont {Miyasaka}}, \
  and\ \bibinfo {author} {\bibfnamefont {K.}~\bibnamefont {Hirata}},\
  }\href@noop {} {\bibfield  {journal} {\bibinfo  {journal} {IEEE Trans.
  Magn.}\ }\textbf {\bibinfo {volume} {48}},\ \bibinfo {pages} {1649} (\bibinfo
  {year} {2012})}\BibitemShut {NoStop}%
\bibitem [{\citenamefont {Galipeau}\ and\ \citenamefont
  {Ponte~Casta{\~n}eda}(2013)}]{Galipeau2013_ProcRSocA}%
  \BibitemOpen
  \bibfield  {author} {\bibinfo {author} {\bibfnamefont {E.}~\bibnamefont
  {Galipeau}}\ and\ \bibinfo {author} {\bibfnamefont {P.}~\bibnamefont
  {Ponte~Casta{\~n}eda}},\ }\href@noop {} {\bibfield  {journal} {\bibinfo
  {journal} {Proc. R. Soc. A}\ }\textbf {\bibinfo {volume} {469}},\ \bibinfo
  {eid} {20130385} (\bibinfo {year} {2013})}\BibitemShut {NoStop}%
\bibitem [{\citenamefont {Allahyarov}\ \emph {et~al.}(2014)\citenamefont
  {Allahyarov}, \citenamefont {Menzel}, \citenamefont {Zhu},\ and\
  \citenamefont {L\"owen}}]{Allahyarov2014_SmartMaterStruct}%
  \BibitemOpen
  \bibfield  {author} {\bibinfo {author} {\bibfnamefont {E.}~\bibnamefont
  {Allahyarov}}, \bibinfo {author} {\bibfnamefont {A.~M.}\ \bibnamefont
  {Menzel}}, \bibinfo {author} {\bibfnamefont {L.}~\bibnamefont {Zhu}}, \ and\
  \bibinfo {author} {\bibfnamefont {H.}~\bibnamefont {L\"owen}},\ }\href@noop
  {} {\bibfield  {journal} {\bibinfo  {journal} {Smart Mater. Struct.}\
  }\textbf {\bibinfo {volume} {23}},\ \bibinfo {pages} {115004} (\bibinfo
  {year} {2014})}\BibitemShut {NoStop}%
\bibitem [{\citenamefont {Nikitin}\ \emph {et~al.}(2004)\citenamefont
  {Nikitin}, \citenamefont {Stepanov}, \citenamefont {Mironova},\ and\
  \citenamefont {Gorbunov}}]{Nikitin2004_JMagnMagnMater}%
  \BibitemOpen
  \bibfield  {author} {\bibinfo {author} {\bibfnamefont {L.~V.}\ \bibnamefont
  {Nikitin}}, \bibinfo {author} {\bibfnamefont {G.~V.}\ \bibnamefont
  {Stepanov}}, \bibinfo {author} {\bibfnamefont {L.~S.}\ \bibnamefont
  {Mironova}}, \ and\ \bibinfo {author} {\bibfnamefont {A.~I.}\ \bibnamefont
  {Gorbunov}},\ }\href@noop {} {\bibfield  {journal} {\bibinfo  {journal} {J.
  Magn. Magn. Mater.}\ }\textbf {\bibinfo {volume} {272-276}},\ \bibinfo
  {pages} {2072} (\bibinfo {year} {2004})}\BibitemShut {NoStop}%
\bibitem [{\citenamefont {Stepanov}\ \emph {et~al.}(2008)\citenamefont
  {Stepanov}, \citenamefont {Borin}, \citenamefont {Raikher}, \citenamefont
  {Melenev},\ and\ \citenamefont {Perov}}]{Stepanov2008_JPhysCondensMatter}%
  \BibitemOpen
  \bibfield  {author} {\bibinfo {author} {\bibfnamefont {H.}~\bibnamefont
  {Stepanov}}, \bibinfo {author} {\bibfnamefont {D.~Y.}\ \bibnamefont {Borin}},
  \bibinfo {author} {\bibfnamefont {Y.~L.}\ \bibnamefont {Raikher}}, \bibinfo
  {author} {\bibfnamefont {P.~V.}\ \bibnamefont {Melenev}}, \ and\ \bibinfo
  {author} {\bibfnamefont {N.~S.}\ \bibnamefont {Perov}},\ }\href@noop {}
  {\bibfield  {journal} {\bibinfo  {journal} {J. Phys.: Condens. Matter}\
  }\textbf {\bibinfo {volume} {20}},\ \bibinfo {pages} {204121} (\bibinfo
  {year} {2008})}\BibitemShut {NoStop}%
\bibitem [{\citenamefont {Melenev}\ \emph {et~al.}(2011)\citenamefont
  {Melenev}, \citenamefont {Raikher}, \citenamefont {Stepanov}, \citenamefont
  {Rusakov},\ and\ \citenamefont
  {Polygalova}}]{Melenev2011_JIntellMaterSystStruct}%
  \BibitemOpen
  \bibfield  {author} {\bibinfo {author} {\bibfnamefont {P.}~\bibnamefont
  {Melenev}}, \bibinfo {author} {\bibfnamefont {Y.}~\bibnamefont {Raikher}},
  \bibinfo {author} {\bibfnamefont {G.}~\bibnamefont {Stepanov}}, \bibinfo
  {author} {\bibfnamefont {V.}~\bibnamefont {Rusakov}}, \ and\ \bibinfo
  {author} {\bibfnamefont {L.}~\bibnamefont {Polygalova}},\ }\href@noop {}
  {\bibfield  {journal} {\bibinfo  {journal} {J. Intell. Mater. Syst. Struct.}\
  }\textbf {\bibinfo {volume} {22}},\ \bibinfo {pages} {531} (\bibinfo {year}
  {2011})}\BibitemShut {NoStop}%
\bibitem [{\citenamefont {Cremer}\ \emph {et~al.}(2015)\citenamefont {Cremer},
  \citenamefont {L\"owen},\ and\ \citenamefont
  {Menzel}}]{Cremer2015_ApplPhysLett}%
  \BibitemOpen
  \bibfield  {author} {\bibinfo {author} {\bibfnamefont {P.}~\bibnamefont
  {Cremer}}, \bibinfo {author} {\bibfnamefont {H.}~\bibnamefont {L\"owen}}, \
  and\ \bibinfo {author} {\bibfnamefont {A.~M.}\ \bibnamefont {Menzel}},\
  }\href@noop {} {\bibfield  {journal} {\bibinfo  {journal} {Appl. Phys.
  Lett.}\ }\textbf {\bibinfo {volume} {107}},\ \bibinfo {eid} {171903}
  (\bibinfo {year} {2015})}\BibitemShut {NoStop}%
\bibitem [{\citenamefont {Otsuka}\ and\ \citenamefont
  {Kakeshita}(2002)}]{Otsuka2002_MRSBull}%
  \BibitemOpen
  \bibfield  {author} {\bibinfo {author} {\bibfnamefont {K.}~\bibnamefont
  {Otsuka}}\ and\ \bibinfo {author} {\bibfnamefont {T.}~\bibnamefont
  {Kakeshita}},\ }\href@noop {} {\bibfield  {journal} {\bibinfo  {journal} {MRS
  Bull.}\ }\textbf {\bibinfo {volume} {27}},\ \bibinfo {pages} {91} (\bibinfo
  {year} {2002})}\BibitemShut {NoStop}%
\bibitem [{\citenamefont {Otsuka}\ and\ \citenamefont
  {Ren}(2005)}]{Otsuka2005_ProgMaterSci}%
  \BibitemOpen
  \bibfield  {author} {\bibinfo {author} {\bibfnamefont {K.}~\bibnamefont
  {Otsuka}}\ and\ \bibinfo {author} {\bibfnamefont {X.}~\bibnamefont {Ren}},\
  }\href@noop {} {\bibfield  {journal} {\bibinfo  {journal} {Prog. Mater.
  Sci.}\ }\textbf {\bibinfo {volume} {50}},\ \bibinfo {pages} {511} (\bibinfo
  {year} {2005})}\BibitemShut {NoStop}%
\bibitem [{\citenamefont {Liu}\ \emph {et~al.}(2013)\citenamefont {Liu},
  \citenamefont {Wang}, \citenamefont {Hao}, \citenamefont {Wang},
  \citenamefont {Nie}, \citenamefont {Wang}, \citenamefont {Ren}, \citenamefont
  {Lu}, \citenamefont {Wang}, \citenamefont {Wang}, \citenamefont {Hui},
  \citenamefont {Lu}, \citenamefont {Kim},\ and\ \citenamefont
  {Yang}}]{Liu2013_SciRep}%
  \BibitemOpen
  \bibfield  {author} {\bibinfo {author} {\bibfnamefont {J.-P.}\ \bibnamefont
  {Liu}}, \bibinfo {author} {\bibfnamefont {Y.-D.}\ \bibnamefont {Wang}},
  \bibinfo {author} {\bibfnamefont {Y.-L.}\ \bibnamefont {Hao}}, \bibinfo
  {author} {\bibfnamefont {Y.}~\bibnamefont {Wang}}, \bibinfo {author}
  {\bibfnamefont {Z.-H.}\ \bibnamefont {Nie}}, \bibinfo {author} {\bibfnamefont
  {D.}~\bibnamefont {Wang}}, \bibinfo {author} {\bibfnamefont {Y.}~\bibnamefont
  {Ren}}, \bibinfo {author} {\bibfnamefont {Z.-P.}\ \bibnamefont {Lu}},
  \bibinfo {author} {\bibfnamefont {J.}~\bibnamefont {Wang}}, \bibinfo {author}
  {\bibfnamefont {H.}~\bibnamefont {Wang}}, \bibinfo {author} {\bibfnamefont
  {X.}~\bibnamefont {Hui}}, \bibinfo {author} {\bibfnamefont {N.}~\bibnamefont
  {Lu}}, \bibinfo {author} {\bibfnamefont {M.~J.}\ \bibnamefont {Kim}}, \ and\
  \bibinfo {author} {\bibfnamefont {R.}~\bibnamefont {Yang}},\ }\href@noop {}
  {\bibfield  {journal} {\bibinfo  {journal} {Sci. Rep.}\ }\textbf {\bibinfo
  {volume} {3}},\ \bibinfo {pages} {2156} (\bibinfo {year} {2013})}\BibitemShut
  {NoStop}%
\bibitem [{\citenamefont {Collin}\ \emph {et~al.}(2003)\citenamefont {Collin},
  \citenamefont {Auernhammer}, \citenamefont {Gavat}, \citenamefont
  {Martinoty},\ and\ \citenamefont {Brand}}]{Collin2003_MacromolRapidCommun}%
  \BibitemOpen
  \bibfield  {author} {\bibinfo {author} {\bibfnamefont {D.}~\bibnamefont
  {Collin}}, \bibinfo {author} {\bibfnamefont {G.~K.}\ \bibnamefont
  {Auernhammer}}, \bibinfo {author} {\bibfnamefont {O.}~\bibnamefont {Gavat}},
  \bibinfo {author} {\bibfnamefont {P.}~\bibnamefont {Martinoty}}, \ and\
  \bibinfo {author} {\bibfnamefont {H.~R.}\ \bibnamefont {Brand}},\ }\href@noop
  {} {\bibfield  {journal} {\bibinfo  {journal} {Macromol. Rapid Commun.}\
  }\textbf {\bibinfo {volume} {24}},\ \bibinfo {pages} {737} (\bibinfo {year}
  {2003})}\BibitemShut {NoStop}%
\bibitem [{\citenamefont {Bohlius}\ \emph {et~al.}(2004)\citenamefont
  {Bohlius}, \citenamefont {Brand},\ and\ \citenamefont
  {Pleiner}}]{Bohlius2004_PhysRevE}%
  \BibitemOpen
  \bibfield  {author} {\bibinfo {author} {\bibfnamefont {S.}~\bibnamefont
  {Bohlius}}, \bibinfo {author} {\bibfnamefont {H.~R.}\ \bibnamefont {Brand}},
  \ and\ \bibinfo {author} {\bibfnamefont {H.}~\bibnamefont {Pleiner}},\
  }\href@noop {} {\bibfield  {journal} {\bibinfo  {journal} {Phys. Rev. E}\
  }\textbf {\bibinfo {volume} {70}},\ \bibinfo {pages} {061411} (\bibinfo
  {year} {2004})}\BibitemShut {NoStop}%
\bibitem [{\citenamefont {G\"unther}\ \emph {et~al.}(2012)\citenamefont
  {G\"unther}, \citenamefont {Borin}, \citenamefont {G\"unther},\ and\
  \citenamefont {Odenbach}}]{Guenther2012_SmartMaterStruct}%
  \BibitemOpen
  \bibfield  {author} {\bibinfo {author} {\bibfnamefont {D.}~\bibnamefont
  {G\"unther}}, \bibinfo {author} {\bibfnamefont {D.~Y.}\ \bibnamefont
  {Borin}}, \bibinfo {author} {\bibfnamefont {S.}~\bibnamefont {G\"unther}}, \
  and\ \bibinfo {author} {\bibfnamefont {S.}~\bibnamefont {Odenbach}},\
  }\href@noop {} {\bibfield  {journal} {\bibinfo  {journal} {Smart Mater.
  Struct.}\ }\textbf {\bibinfo {volume} {21}},\ \bibinfo {pages} {015005}
  (\bibinfo {year} {2012})}\BibitemShut {NoStop}%
\bibitem [{\citenamefont {Borin}\ \emph {et~al.}(2012)\citenamefont {Borin},
  \citenamefont {G\"unther}, \citenamefont {Hintze}, \citenamefont {Heinrich},\
  and\ \citenamefont {Odenbach}}]{Borin2012_JMagnMagnMater}%
  \BibitemOpen
  \bibfield  {author} {\bibinfo {author} {\bibfnamefont {D.}~\bibnamefont
  {Borin}}, \bibinfo {author} {\bibfnamefont {D.}~\bibnamefont {G\"unther}},
  \bibinfo {author} {\bibfnamefont {C.}~\bibnamefont {Hintze}}, \bibinfo
  {author} {\bibfnamefont {G.}~\bibnamefont {Heinrich}}, \ and\ \bibinfo
  {author} {\bibfnamefont {S.}~\bibnamefont {Odenbach}},\ }\href@noop {}
  {\bibfield  {journal} {\bibinfo  {journal} {J. Magn. Magn. Mater.}\ }\textbf
  {\bibinfo {volume} {324}},\ \bibinfo {pages} {3452} (\bibinfo {year}
  {2012})}\BibitemShut {NoStop}%
\bibitem [{\citenamefont {Tian}\ \emph {et~al.}(2013)\citenamefont {Tian},
  \citenamefont {Liu}, \citenamefont {He}, \citenamefont {Zhao},\ and\
  \citenamefont {Sun}}]{Tian2013_MaterResBull}%
  \BibitemOpen
  \bibfield  {author} {\bibinfo {author} {\bibfnamefont {Y.}~\bibnamefont
  {Tian}}, \bibinfo {author} {\bibfnamefont {Y.}~\bibnamefont {Liu}}, \bibinfo
  {author} {\bibfnamefont {M.}~\bibnamefont {He}}, \bibinfo {author}
  {\bibfnamefont {G.}~\bibnamefont {Zhao}}, \ and\ \bibinfo {author}
  {\bibfnamefont {Y.}~\bibnamefont {Sun}},\ }\href@noop {} {\bibfield
  {journal} {\bibinfo  {journal} {Mater. Res. Bull.}\ }\textbf {\bibinfo
  {volume} {48}},\ \bibinfo {pages} {2002} (\bibinfo {year}
  {2013})}\BibitemShut {NoStop}%
\bibitem [{\citenamefont {Zubarev}\ and\ \citenamefont
  {Iskakova}(2000)}]{Zubarev2000_PhysRevE}%
  \BibitemOpen
  \bibfield  {author} {\bibinfo {author} {\bibfnamefont {A.~Y.}\ \bibnamefont
  {Zubarev}}\ and\ \bibinfo {author} {\bibfnamefont {L.~Y.}\ \bibnamefont
  {Iskakova}},\ }\href@noop {} {\bibfield  {journal} {\bibinfo  {journal}
  {Phys. Rev. E}\ }\textbf {\bibinfo {volume} {61}},\ \bibinfo {pages} {5415}
  (\bibinfo {year} {2000})}\BibitemShut {NoStop}%
\bibitem [{\citenamefont {Hynninen}\ and\ \citenamefont
  {Dijkstra}(2005)}]{Hynninen2005_PhysRevLett}%
  \BibitemOpen
  \bibfield  {author} {\bibinfo {author} {\bibfnamefont {A.-P.}\ \bibnamefont
  {Hynninen}}\ and\ \bibinfo {author} {\bibfnamefont {M.}~\bibnamefont
  {Dijkstra}},\ }\href@noop {} {\bibfield  {journal} {\bibinfo  {journal}
  {Phys. Rev. Lett.}\ }\textbf {\bibinfo {volume} {94}},\ \bibinfo {pages}
  {138303} (\bibinfo {year} {2005})}\BibitemShut {NoStop}%
\bibitem [{\citenamefont {Auernhammer}\ \emph {et~al.}(2006)\citenamefont
  {Auernhammer}, \citenamefont {Collin},\ and\ \citenamefont
  {Martinoty}}]{Auernhammer2006_JChemPhys}%
  \BibitemOpen
  \bibfield  {author} {\bibinfo {author} {\bibfnamefont {G.~K.}\ \bibnamefont
  {Auernhammer}}, \bibinfo {author} {\bibfnamefont {D.}~\bibnamefont {Collin}},
  \ and\ \bibinfo {author} {\bibfnamefont {P.}~\bibnamefont {Martinoty}},\
  }\href@noop {} {\bibfield  {journal} {\bibinfo  {journal} {J. Chem. Phys}\
  }\textbf {\bibinfo {volume} {124}},\ \bibinfo {eid} {204907} (\bibinfo {year}
  {2006})}\BibitemShut {NoStop}%
\bibitem [{\citenamefont {Smallenburg}\ \emph {et~al.}(2012)\citenamefont
  {Smallenburg}, \citenamefont {Vutukuri}, \citenamefont {Imhof}, \citenamefont
  {van Blaaderen},\ and\ \citenamefont
  {Dijkstra}}]{Smallenburg2012_JPhysCondensMatter}%
  \BibitemOpen
  \bibfield  {author} {\bibinfo {author} {\bibfnamefont {F.}~\bibnamefont
  {Smallenburg}}, \bibinfo {author} {\bibfnamefont {H.~R.}\ \bibnamefont
  {Vutukuri}}, \bibinfo {author} {\bibfnamefont {A.}~\bibnamefont {Imhof}},
  \bibinfo {author} {\bibfnamefont {A.}~\bibnamefont {van Blaaderen}}, \ and\
  \bibinfo {author} {\bibfnamefont {M.}~\bibnamefont {Dijkstra}},\ }\href@noop
  {} {\bibfield  {journal} {\bibinfo  {journal} {J. Phys.: Condens. Matter}\
  }\textbf {\bibinfo {volume} {24}},\ \bibinfo {pages} {464113} (\bibinfo
  {year} {2012})}\BibitemShut {NoStop}%
\bibitem [{\citenamefont {El~Feninat}\ \emph {et~al.}(2002)\citenamefont
  {El~Feninat}, \citenamefont {Laroche}, \citenamefont {Fiset},\ and\
  \citenamefont {Mantovani}}]{ElFeninat2002_AdvEngMater}%
  \BibitemOpen
  \bibfield  {author} {\bibinfo {author} {\bibfnamefont {F.}~\bibnamefont
  {El~Feninat}}, \bibinfo {author} {\bibfnamefont {G.}~\bibnamefont {Laroche}},
  \bibinfo {author} {\bibfnamefont {M.}~\bibnamefont {Fiset}}, \ and\ \bibinfo
  {author} {\bibfnamefont {D.}~\bibnamefont {Mantovani}},\ }\href@noop {}
  {\bibfield  {journal} {\bibinfo  {journal} {Adv. Eng. Mater.}\ }\textbf
  {\bibinfo {volume} {4}},\ \bibinfo {pages} {91} (\bibinfo {year}
  {2002})}\BibitemShut {NoStop}%
\bibitem [{\citenamefont {Liu}\ \emph {et~al.}(2007)\citenamefont {Liu},
  \citenamefont {Qin},\ and\ \citenamefont {Mather}}]{Liu2007_JMaterChem}%
  \BibitemOpen
  \bibfield  {author} {\bibinfo {author} {\bibfnamefont {C.}~\bibnamefont
  {Liu}}, \bibinfo {author} {\bibfnamefont {H.}~\bibnamefont {Qin}}, \ and\
  \bibinfo {author} {\bibfnamefont {P.~T.}\ \bibnamefont {Mather}},\
  }\href@noop {} {\bibfield  {journal} {\bibinfo  {journal} {J. Mater. Chem.}\
  }\textbf {\bibinfo {volume} {17}},\ \bibinfo {pages} {1543} (\bibinfo {year}
  {2007})}\BibitemShut {NoStop}%
\bibitem [{\citenamefont {Sokolowski}\ \emph {et~al.}(2007)\citenamefont
  {Sokolowski}, \citenamefont {Metcalfe}, \citenamefont {Hayashi},
  \citenamefont {Yahia},\ and\ \citenamefont
  {Raymond}}]{Sokolowski2007_BiomedMater}%
  \BibitemOpen
  \bibfield  {author} {\bibinfo {author} {\bibfnamefont {W.}~\bibnamefont
  {Sokolowski}}, \bibinfo {author} {\bibfnamefont {A.}~\bibnamefont
  {Metcalfe}}, \bibinfo {author} {\bibfnamefont {S.}~\bibnamefont {Hayashi}},
  \bibinfo {author} {\bibfnamefont {L.}~\bibnamefont {Yahia}}, \ and\ \bibinfo
  {author} {\bibfnamefont {J.}~\bibnamefont {Raymond}},\ }\href@noop {}
  {\bibfield  {journal} {\bibinfo  {journal} {Biomed. Mater.}\ }\textbf
  {\bibinfo {volume} {2}},\ \bibinfo {pages} {S23} (\bibinfo {year}
  {2007})}\BibitemShut {NoStop}%
\bibitem [{\citenamefont {Leng}\ \emph {et~al.}(2009)\citenamefont {Leng},
  \citenamefont {Lu}, \citenamefont {Liu}, \citenamefont {Huang},\ and\
  \citenamefont {Du}}]{Leng2009_MRSBull}%
  \BibitemOpen
  \bibfield  {author} {\bibinfo {author} {\bibfnamefont {J.}~\bibnamefont
  {Leng}}, \bibinfo {author} {\bibfnamefont {H.}~\bibnamefont {Lu}}, \bibinfo
  {author} {\bibfnamefont {Y.}~\bibnamefont {Liu}}, \bibinfo {author}
  {\bibfnamefont {W.~M.}\ \bibnamefont {Huang}}, \ and\ \bibinfo {author}
  {\bibfnamefont {S.}~\bibnamefont {Du}},\ }\href@noop {} {\bibfield  {journal}
  {\bibinfo  {journal} {MRS Bull.}\ }\textbf {\bibinfo {volume} {34}},\
  \bibinfo {pages} {848} (\bibinfo {year} {2009})}\BibitemShut {NoStop}%
\bibitem [{\citenamefont {Behl}\ \emph {et~al.}(2010)\citenamefont {Behl},
  \citenamefont {Razzaq},\ and\ \citenamefont {Lendlein}}]{Behl2010_AdvMater}%
  \BibitemOpen
  \bibfield  {author} {\bibinfo {author} {\bibfnamefont {M.}~\bibnamefont
  {Behl}}, \bibinfo {author} {\bibfnamefont {M.~Y.}\ \bibnamefont {Razzaq}}, \
  and\ \bibinfo {author} {\bibfnamefont {A.}~\bibnamefont {Lendlein}},\
  }\href@noop {} {\bibfield  {journal} {\bibinfo  {journal} {Adv. Mater.}\
  }\textbf {\bibinfo {volume} {22}},\ \bibinfo {pages} {3388} (\bibinfo {year}
  {2010})}\BibitemShut {NoStop}%
\bibitem [{\citenamefont {Li}\ \emph {et~al.}(2013)\citenamefont {Li},
  \citenamefont {Huang}, \citenamefont {Zhang}, \citenamefont {Li},
  \citenamefont {Chen}, \citenamefont {Lu}, \citenamefont {Lu},\ and\
  \citenamefont {Xu}}]{Li2013_AdvFunctMater}%
  \BibitemOpen
  \bibfield  {author} {\bibinfo {author} {\bibfnamefont {Y.}~\bibnamefont
  {Li}}, \bibinfo {author} {\bibfnamefont {G.}~\bibnamefont {Huang}}, \bibinfo
  {author} {\bibfnamefont {X.}~\bibnamefont {Zhang}}, \bibinfo {author}
  {\bibfnamefont {B.}~\bibnamefont {Li}}, \bibinfo {author} {\bibfnamefont
  {Y.}~\bibnamefont {Chen}}, \bibinfo {author} {\bibfnamefont {T.}~\bibnamefont
  {Lu}}, \bibinfo {author} {\bibfnamefont {T.~J.}\ \bibnamefont {Lu}}, \ and\
  \bibinfo {author} {\bibfnamefont {F.}~\bibnamefont {Xu}},\ }\href@noop {}
  {\bibfield  {journal} {\bibinfo  {journal} {Adv. Funct. Mater.}\ }\textbf
  {\bibinfo {volume} {23}},\ \bibinfo {pages} {660} (\bibinfo {year}
  {2013})}\BibitemShut {NoStop}%
\bibitem [{\citenamefont {Cezar}\ \emph {et~al.}(2014)\citenamefont {Cezar},
  \citenamefont {Kennedy}, \citenamefont {Mehta}, \citenamefont {Weaver},
  \citenamefont {Gu}, \citenamefont {Vandenburgh},\ and\ \citenamefont
  {Mooney}}]{Cezar2014_AdvHealthcMater}%
  \BibitemOpen
  \bibfield  {author} {\bibinfo {author} {\bibfnamefont {C.~A.}\ \bibnamefont
  {Cezar}}, \bibinfo {author} {\bibfnamefont {S.~M.}\ \bibnamefont {Kennedy}},
  \bibinfo {author} {\bibfnamefont {M.}~\bibnamefont {Mehta}}, \bibinfo
  {author} {\bibfnamefont {J.~C.}\ \bibnamefont {Weaver}}, \bibinfo {author}
  {\bibfnamefont {L.}~\bibnamefont {Gu}}, \bibinfo {author} {\bibfnamefont
  {H.}~\bibnamefont {Vandenburgh}}, \ and\ \bibinfo {author} {\bibfnamefont
  {D.~J.}\ \bibnamefont {Mooney}},\ }\href@noop {} {\bibfield  {journal}
  {\bibinfo  {journal} {Adv. Healthc. Mater.}\ }\textbf {\bibinfo {volume}
  {3}},\ \bibinfo {pages} {1869} (\bibinfo {year} {2014})}\BibitemShut
  {NoStop}%
\bibitem [{\citenamefont {Cezar}\ \emph {et~al.}(2016)\citenamefont {Cezar},
  \citenamefont {Roche}, \citenamefont {Vandenburgh}, \citenamefont {Duda},
  \citenamefont {Walsh},\ and\ \citenamefont
  {Mooney}}]{Cezar2016_ProcNatlAcadSciUSA}%
  \BibitemOpen
  \bibfield  {author} {\bibinfo {author} {\bibfnamefont {C.~A.}\ \bibnamefont
  {Cezar}}, \bibinfo {author} {\bibfnamefont {E.~T.}\ \bibnamefont {Roche}},
  \bibinfo {author} {\bibfnamefont {H.~H.}\ \bibnamefont {Vandenburgh}},
  \bibinfo {author} {\bibfnamefont {G.~N.}\ \bibnamefont {Duda}}, \bibinfo
  {author} {\bibfnamefont {C.~J.}\ \bibnamefont {Walsh}}, \ and\ \bibinfo
  {author} {\bibfnamefont {D.~J.}\ \bibnamefont {Mooney}},\ }\href@noop {}
  {\bibfield  {journal} {\bibinfo  {journal} {Proc. Natl. Acad. Sci. USA}\
  }\textbf {\bibinfo {volume} {113}},\ \bibinfo {pages} {1534} (\bibinfo {year}
  {2016})}\BibitemShut {NoStop}%
\bibitem [{\citenamefont {Mody}\ \emph {et~al.}(2016)\citenamefont {Mody},
  \citenamefont {Hart}, \citenamefont {Romano}, \citenamefont {El-Magbri},
  \citenamefont {Esson}, \citenamefont {Ibeh}, \citenamefont {Knowlton},
  \citenamefont {Zhang}, \citenamefont {Wagner},\ and\ \citenamefont
  {Hartings}}]{Mody2016_JInorgBiochem}%
  \BibitemOpen
  \bibfield  {author} {\bibinfo {author} {\bibfnamefont {P.}~\bibnamefont
  {Mody}}, \bibinfo {author} {\bibfnamefont {C.}~\bibnamefont {Hart}}, \bibinfo
  {author} {\bibfnamefont {S.}~\bibnamefont {Romano}}, \bibinfo {author}
  {\bibfnamefont {M.}~\bibnamefont {El-Magbri}}, \bibinfo {author}
  {\bibfnamefont {M.~M.}\ \bibnamefont {Esson}}, \bibinfo {author}
  {\bibfnamefont {T.}~\bibnamefont {Ibeh}}, \bibinfo {author} {\bibfnamefont
  {E.~D.}\ \bibnamefont {Knowlton}}, \bibinfo {author} {\bibfnamefont
  {M.}~\bibnamefont {Zhang}}, \bibinfo {author} {\bibfnamefont {M.~J.}\
  \bibnamefont {Wagner}}, \ and\ \bibinfo {author} {\bibfnamefont {M.~R.}\
  \bibnamefont {Hartings}},\ }\href@noop {} {\bibfield  {journal} {\bibinfo
  {journal} {J. Inorg. Biochem.}\ }\textbf {\bibinfo {volume} {159}},\ \bibinfo
  {pages} {7} (\bibinfo {year} {2016})}\BibitemShut {NoStop}%
\bibitem [{\citenamefont {N{\'e}el}(1949)}]{Neel1949_AnnGeophys}%
  \BibitemOpen
  \bibfield  {author} {\bibinfo {author} {\bibfnamefont {L.}~\bibnamefont
  {N{\'e}el}},\ }\href@noop {} {\bibfield  {journal} {\bibinfo  {journal} {Ann.
  G{\'e}ophys}\ }\textbf {\bibinfo {volume} {5}},\ \bibinfo {pages} {99}
  (\bibinfo {year} {1949})}\BibitemShut {NoStop}%
\bibitem [{\citenamefont {Gundermann}\ and\ \citenamefont
  {Odenbach}(2014)}]{Gundermann2014_SmartMaterStruct}%
  \BibitemOpen
  \bibfield  {author} {\bibinfo {author} {\bibfnamefont {T.}~\bibnamefont
  {Gundermann}}\ and\ \bibinfo {author} {\bibfnamefont {S.}~\bibnamefont
  {Odenbach}},\ }\href@noop {} {\bibfield  {journal} {\bibinfo  {journal}
  {Smart Mater. Struct.}\ }\textbf {\bibinfo {volume} {23}},\ \bibinfo {pages}
  {105013} (\bibinfo {year} {2014})}\BibitemShut {NoStop}%
\bibitem [{\citenamefont {Liu}\ \emph {et~al.}(2012)\citenamefont {Liu},
  \citenamefont {Xu}, \citenamefont {Che}, \citenamefont {Chen}, \citenamefont
  {Liu},\ and\ \citenamefont {Xia}}]{Liu2012_JMaterChem}%
  \BibitemOpen
  \bibfield  {author} {\bibinfo {author} {\bibfnamefont {J.}~\bibnamefont
  {Liu}}, \bibinfo {author} {\bibfnamefont {J.}~\bibnamefont {Xu}}, \bibinfo
  {author} {\bibfnamefont {R.}~\bibnamefont {Che}}, \bibinfo {author}
  {\bibfnamefont {H.}~\bibnamefont {Chen}}, \bibinfo {author} {\bibfnamefont
  {Z.}~\bibnamefont {Liu}}, \ and\ \bibinfo {author} {\bibfnamefont
  {F.}~\bibnamefont {Xia}},\ }\href@noop {} {\bibfield  {journal} {\bibinfo
  {journal} {J. Mater. Chem.}\ }\textbf {\bibinfo {volume} {22}},\ \bibinfo
  {pages} {9277} (\bibinfo {year} {2012})}\BibitemShut {NoStop}%
\bibitem [{\citenamefont {Okada}\ \emph {et~al.}(2013)\citenamefont {Okada},
  \citenamefont {Nagao}, \citenamefont {Ueno}, \citenamefont {Ishii},\ and\
  \citenamefont {Konno}}]{Okada2013_Langmuir}%
  \BibitemOpen
  \bibfield  {author} {\bibinfo {author} {\bibfnamefont {A.}~\bibnamefont
  {Okada}}, \bibinfo {author} {\bibfnamefont {D.}~\bibnamefont {Nagao}},
  \bibinfo {author} {\bibfnamefont {T.}~\bibnamefont {Ueno}}, \bibinfo {author}
  {\bibfnamefont {H.}~\bibnamefont {Ishii}}, \ and\ \bibinfo {author}
  {\bibfnamefont {M.}~\bibnamefont {Konno}},\ }\href@noop {} {\bibfield
  {journal} {\bibinfo  {journal} {Langmuir}\ }\textbf {\bibinfo {volume}
  {29}},\ \bibinfo {pages} {9004} (\bibinfo {year} {2013})}\BibitemShut
  {NoStop}%
\bibitem [{\citenamefont {Fuhrer}\ \emph {et~al.}(2009)\citenamefont {Fuhrer},
  \citenamefont {Athanassiou}, \citenamefont {Luechinger},\ and\ \citenamefont
  {Stark}}]{Fuhrer2009_Small}%
  \BibitemOpen
  \bibfield  {author} {\bibinfo {author} {\bibfnamefont {R.}~\bibnamefont
  {Fuhrer}}, \bibinfo {author} {\bibfnamefont {E.~K.}\ \bibnamefont
  {Athanassiou}}, \bibinfo {author} {\bibfnamefont {N.~A.}\ \bibnamefont
  {Luechinger}}, \ and\ \bibinfo {author} {\bibfnamefont {W.~J.}\ \bibnamefont
  {Stark}},\ }\href@noop {} {\bibfield  {journal} {\bibinfo  {journal} {Small}\
  }\textbf {\bibinfo {volume} {5}},\ \bibinfo {pages} {383} (\bibinfo {year}
  {2009})}\BibitemShut {NoStop}%
\bibitem [{\citenamefont {Frickel}\ \emph {et~al.}(2011)\citenamefont
  {Frickel}, \citenamefont {Messing},\ and\ \citenamefont
  {Schmidt}}]{Frickel2011_JMaterChem}%
  \BibitemOpen
  \bibfield  {author} {\bibinfo {author} {\bibfnamefont {N.}~\bibnamefont
  {Frickel}}, \bibinfo {author} {\bibfnamefont {R.}~\bibnamefont {Messing}}, \
  and\ \bibinfo {author} {\bibfnamefont {A.~M.}\ \bibnamefont {Schmidt}},\
  }\href@noop {} {\bibfield  {journal} {\bibinfo  {journal} {J. Mater. Chem.}\
  }\textbf {\bibinfo {volume} {21}},\ \bibinfo {pages} {8466} (\bibinfo {year}
  {2011})}\BibitemShut {NoStop}%
\bibitem [{\citenamefont {Ilg}(2013)}]{Ilg2013_SoftMatter}%
  \BibitemOpen
  \bibfield  {author} {\bibinfo {author} {\bibfnamefont {P.}~\bibnamefont
  {Ilg}},\ }\href@noop {} {\bibfield  {journal} {\bibinfo  {journal} {Soft
  Matter}\ }\textbf {\bibinfo {volume} {9}},\ \bibinfo {pages} {3465} (\bibinfo
  {year} {2013})}\BibitemShut {NoStop}%
\bibitem [{\citenamefont {Roeder}\ \emph {et~al.}(2014)\citenamefont {Roeder},
  \citenamefont {Reckenth\"aler}, \citenamefont {Belkoura}, \citenamefont
  {Roitsch}, \citenamefont {Strey},\ and\ \citenamefont
  {Schmidt}}]{Roeder2014_Macromolecules}%
  \BibitemOpen
  \bibfield  {author} {\bibinfo {author} {\bibfnamefont {L.}~\bibnamefont
  {Roeder}}, \bibinfo {author} {\bibfnamefont {M.}~\bibnamefont
  {Reckenth\"aler}}, \bibinfo {author} {\bibfnamefont {L.}~\bibnamefont
  {Belkoura}}, \bibinfo {author} {\bibfnamefont {S.}~\bibnamefont {Roitsch}},
  \bibinfo {author} {\bibfnamefont {R.}~\bibnamefont {Strey}}, \ and\ \bibinfo
  {author} {\bibfnamefont {A.~M.}\ \bibnamefont {Schmidt}},\ }\href@noop {}
  {\bibfield  {journal} {\bibinfo  {journal} {Macromolecules}\ }\textbf
  {\bibinfo {volume} {47}},\ \bibinfo {pages} {7200} (\bibinfo {year}
  {2014})}\BibitemShut {NoStop}%
\bibitem [{\citenamefont {Roeder}\ \emph {et~al.}(2015)\citenamefont {Roeder},
  \citenamefont {Bender}, \citenamefont {Kundt}, \citenamefont {Tsch\"ope},\
  and\ \citenamefont {Schmidt}}]{Roeder2015_PhysChemChemPhys}%
  \BibitemOpen
  \bibfield  {author} {\bibinfo {author} {\bibfnamefont {L.}~\bibnamefont
  {Roeder}}, \bibinfo {author} {\bibfnamefont {P.}~\bibnamefont {Bender}},
  \bibinfo {author} {\bibfnamefont {M.}~\bibnamefont {Kundt}}, \bibinfo
  {author} {\bibfnamefont {A.}~\bibnamefont {Tsch\"ope}}, \ and\ \bibinfo
  {author} {\bibfnamefont {A.~M.}\ \bibnamefont {Schmidt}},\ }\href@noop {}
  {\bibfield  {journal} {\bibinfo  {journal} {Phys. Chem. Chem. Phys.}\
  }\textbf {\bibinfo {volume} {17}},\ \bibinfo {pages} {1290} (\bibinfo {year}
  {2015})}\BibitemShut {NoStop}%
\bibitem [{\citenamefont {Weeber}\ \emph {et~al.}(2015)\citenamefont {Weeber},
  \citenamefont {Kantorovich},\ and\ \citenamefont
  {Holm}}]{Weeber2015_JMagnMagnMater}%
  \BibitemOpen
  \bibfield  {author} {\bibinfo {author} {\bibfnamefont {R.}~\bibnamefont
  {Weeber}}, \bibinfo {author} {\bibfnamefont {S.}~\bibnamefont {Kantorovich}},
  \ and\ \bibinfo {author} {\bibfnamefont {C.}~\bibnamefont {Holm}},\
  }\href@noop {} {\bibfield  {journal} {\bibinfo  {journal} {J. Magn. Magn.
  Mater.}\ }\textbf {\bibinfo {volume} {383}},\ \bibinfo {pages} {262}
  (\bibinfo {year} {2015})}\BibitemShut {NoStop}%
\bibitem [{\citenamefont {Geuzaine}\ and\ \citenamefont
  {Remacle}(2009)}]{Geuzaine2009_IntJNumerMethEng}%
  \BibitemOpen
  \bibfield  {author} {\bibinfo {author} {\bibfnamefont {C.}~\bibnamefont
  {Geuzaine}}\ and\ \bibinfo {author} {\bibfnamefont {J.-F.}\ \bibnamefont
  {Remacle}},\ }\href@noop {} {\bibfield  {journal} {\bibinfo  {journal} {Int.
  J. Numer. Meth. Eng.}\ }\textbf {\bibinfo {volume} {79}},\ \bibinfo {pages}
  {1309} (\bibinfo {year} {2009})}\BibitemShut {NoStop}%
\bibitem [{\citenamefont {Delaunay}(1934)}]{Delaunay1934}%
  \BibitemOpen
  \bibfield  {author} {\bibinfo {author} {\bibfnamefont {B.~N.}\ \bibnamefont
  {Delaunay}},\ }\href@noop {} {\bibfield  {journal} {\bibinfo  {journal}
  {Bull. Acad. Sci. USSR VII: Class. Sci. Math. Nat.}\ ,\ \bibinfo {pages}
  {793}} (\bibinfo {year} {1934})}\BibitemShut {NoStop}%
\bibitem [{\citenamefont {Hartmann}\ and\ \citenamefont
  {Neff}(2003)}]{Hartmann2003_IntJSolidsStruct}%
  \BibitemOpen
  \bibfield  {author} {\bibinfo {author} {\bibfnamefont {S.}~\bibnamefont
  {Hartmann}}\ and\ \bibinfo {author} {\bibfnamefont {P.}~\bibnamefont
  {Neff}},\ }\href@noop {} {\bibfield  {journal} {\bibinfo  {journal} {Int. J.
  Solids Struct.}\ }\textbf {\bibinfo {volume} {40}},\ \bibinfo {pages} {2767}
  (\bibinfo {year} {2003})}\BibitemShut {NoStop}%
\bibitem [{\citenamefont {Landau}\ and\ \citenamefont
  {Lifshitz}(2012)}]{Landau2012_book}%
  \BibitemOpen
  \bibfield  {author} {\bibinfo {author} {\bibfnamefont {L.~D.}\ \bibnamefont
  {Landau}}\ and\ \bibinfo {author} {\bibfnamefont {E.~M.}\ \bibnamefont
  {Lifshitz}},\ }\href@noop {} {\emph {\bibinfo {title} {Theory of
  Elasticity}}}\ (\bibinfo  {publisher} {Elsevier Science},\ \bibinfo {year}
  {2012})\BibitemShut {NoStop}%
\bibitem [{\citenamefont {Irving}\ \emph {et~al.}(2006)\citenamefont {Irving},
  \citenamefont {Teran},\ and\ \citenamefont
  {Fedkiw}}]{Irving2006_GraphModels}%
  \BibitemOpen
  \bibfield  {author} {\bibinfo {author} {\bibfnamefont {G.}~\bibnamefont
  {Irving}}, \bibinfo {author} {\bibfnamefont {J.}~\bibnamefont {Teran}}, \
  and\ \bibinfo {author} {\bibfnamefont {R.}~\bibnamefont {Fedkiw}},\
  }\href@noop {} {\bibfield  {journal} {\bibinfo  {journal} {Graph. Models}\
  }\textbf {\bibinfo {volume} {68}},\ \bibinfo {pages} {66} (\bibinfo {year}
  {2006})}\BibitemShut {NoStop}%
\bibitem [{\citenamefont {Weeks}\ \emph {et~al.}(1971)\citenamefont {Weeks},
  \citenamefont {Chandler},\ and\ \citenamefont
  {Andersen}}]{Weeks1971_JChemPhys}%
  \BibitemOpen
  \bibfield  {author} {\bibinfo {author} {\bibfnamefont {J.~D.}\ \bibnamefont
  {Weeks}}, \bibinfo {author} {\bibfnamefont {D.}~\bibnamefont {Chandler}}, \
  and\ \bibinfo {author} {\bibfnamefont {H.~C.}\ \bibnamefont {Andersen}},\
  }\href@noop {} {\bibfield  {journal} {\bibinfo  {journal} {J. Chem. Phys.}\
  }\textbf {\bibinfo {volume} {54}},\ \bibinfo {pages} {5237} (\bibinfo {year}
  {1971})}\BibitemShut {NoStop}%
\bibitem [{\citenamefont {Bitzek}\ \emph {et~al.}(2006)\citenamefont {Bitzek},
  \citenamefont {Koskinen}, \citenamefont {G\"ahler}, \citenamefont {Moseler},\
  and\ \citenamefont {Gumbsch}}]{Bitzek2006_PhysRevLett}%
  \BibitemOpen
  \bibfield  {author} {\bibinfo {author} {\bibfnamefont {E.}~\bibnamefont
  {Bitzek}}, \bibinfo {author} {\bibfnamefont {P.}~\bibnamefont {Koskinen}},
  \bibinfo {author} {\bibfnamefont {F.}~\bibnamefont {G\"ahler}}, \bibinfo
  {author} {\bibfnamefont {M.}~\bibnamefont {Moseler}}, \ and\ \bibinfo
  {author} {\bibfnamefont {P.}~\bibnamefont {Gumbsch}},\ }\href@noop {}
  {\bibfield  {journal} {\bibinfo  {journal} {Phys. Rev. Lett.}\ }\textbf
  {\bibinfo {volume} {97}},\ \bibinfo {pages} {170201} (\bibinfo {year}
  {2006})}\BibitemShut {NoStop}%
\bibitem [{\citenamefont {Hager}\ and\ \citenamefont
  {Zhang}(2006)}]{Hager2006_PacJOptim}%
  \BibitemOpen
  \bibfield  {author} {\bibinfo {author} {\bibfnamefont {W.~W.}\ \bibnamefont
  {Hager}}\ and\ \bibinfo {author} {\bibfnamefont {H.}~\bibnamefont {Zhang}},\
  }\href@noop {} {\bibfield  {journal} {\bibinfo  {journal} {Pac. J. Optim.}\
  }\textbf {\bibinfo {volume} {2}},\ \bibinfo {pages} {35} (\bibinfo {year}
  {2006})}\BibitemShut {NoStop}%
\bibitem [{\citenamefont {Diguet}\ \emph {et~al.}(2010)\citenamefont {Diguet},
  \citenamefont {Beaugnon},\ and\ \citenamefont
  {Cavaill\'e}}]{Diguet2010_JMagnMagnMater}%
  \BibitemOpen
  \bibfield  {author} {\bibinfo {author} {\bibfnamefont {G.}~\bibnamefont
  {Diguet}}, \bibinfo {author} {\bibfnamefont {E.}~\bibnamefont {Beaugnon}}, \
  and\ \bibinfo {author} {\bibfnamefont {J.}~\bibnamefont {Cavaill\'e}},\
  }\href@noop {} {\bibfield  {journal} {\bibinfo  {journal} {J. Magn. Magn.
  Mater.}\ }\textbf {\bibinfo {volume} {322}},\ \bibinfo {pages} {3337}
  (\bibinfo {year} {2010})}\BibitemShut {NoStop}%
\bibitem [{\citenamefont {Zubarev}(2013)}]{Zubarev2013_PhysicaA}%
  \BibitemOpen
  \bibfield  {author} {\bibinfo {author} {\bibfnamefont {A.}~\bibnamefont
  {Zubarev}},\ }\href@noop {} {\bibfield  {journal} {\bibinfo  {journal}
  {Physica A}\ }\textbf {\bibinfo {volume} {392}},\ \bibinfo {pages} {4824}
  (\bibinfo {year} {2013})}\BibitemShut {NoStop}%
\bibitem [{\citenamefont {Pessot}\ \emph {et~al.}(2014)\citenamefont {Pessot},
  \citenamefont {Cremer}, \citenamefont {Borin}, \citenamefont {Odenbach},
  \citenamefont {L\"owen},\ and\ \citenamefont
  {Menzel}}]{Pessot2014_JChemPhys}%
  \BibitemOpen
  \bibfield  {author} {\bibinfo {author} {\bibfnamefont {G.}~\bibnamefont
  {Pessot}}, \bibinfo {author} {\bibfnamefont {P.}~\bibnamefont {Cremer}},
  \bibinfo {author} {\bibfnamefont {D.~Y.}\ \bibnamefont {Borin}}, \bibinfo
  {author} {\bibfnamefont {S.}~\bibnamefont {Odenbach}}, \bibinfo {author}
  {\bibfnamefont {H.}~\bibnamefont {L\"owen}}, \ and\ \bibinfo {author}
  {\bibfnamefont {A.~M.}\ \bibnamefont {Menzel}},\ }\href@noop {} {\bibfield
  {journal} {\bibinfo  {journal} {J. Chem. Phys.}\ }\textbf {\bibinfo {volume}
  {141}},\ \bibinfo {eid} {124904} (\bibinfo {year} {2014})}\BibitemShut
  {NoStop}%
\bibitem [{\citenamefont {Gundermann}\ \emph {et~al.}(2016)\citenamefont
  {Gundermann}, \citenamefont {Cremer}, \citenamefont {L\"owen}, \citenamefont
  {Menzel},\ and\ \citenamefont {Odenbach}}]{Gundermann2016_unpublished}%
  \BibitemOpen
  \bibfield  {author} {\bibinfo {author} {\bibfnamefont {T.}~\bibnamefont
  {Gundermann}}, \bibinfo {author} {\bibfnamefont {P.}~\bibnamefont {Cremer}},
  \bibinfo {author} {\bibfnamefont {H.}~\bibnamefont {L\"owen}}, \bibinfo
  {author} {\bibfnamefont {A.~M.}\ \bibnamefont {Menzel}}, \ and\ \bibinfo
  {author} {\bibfnamefont {S.}~\bibnamefont {Odenbach}},\ }\href@noop {}
  {\enquote {\bibinfo {title} {{Statistical analysis for the distribution of
  soft magnetic particles in magnetorheological elastomers}},}\ } (\bibinfo
  {year} {2016}),\ \bibinfo {note} {in preparation}\BibitemShut {NoStop}%
\bibitem [{\citenamefont {de~Gennes}\ and\ \citenamefont
  {Prost}(1995)}]{deGennes1995_book}%
  \BibitemOpen
  \bibfield  {author} {\bibinfo {author} {\bibfnamefont {P.~G.}\ \bibnamefont
  {de~Gennes}}\ and\ \bibinfo {author} {\bibfnamefont {J.}~\bibnamefont
  {Prost}},\ }\href@noop {} {\emph {\bibinfo {title} {The Physics of Liquid
  Crystals}}},\ International Series of Monographs on Physics\ (\bibinfo
  {publisher} {Clarendon Press},\ \bibinfo {year} {1995})\BibitemShut {NoStop}%
\bibitem [{\citenamefont {Chernenko}\ \emph {et~al.}(2003)\citenamefont
  {Chernenko}, \citenamefont {L'vov}, \citenamefont {Pons},\ and\ \citenamefont
  {Cesari}}]{Chernenko2003_JApplPhys}%
  \BibitemOpen
  \bibfield  {author} {\bibinfo {author} {\bibfnamefont {V.~A.}\ \bibnamefont
  {Chernenko}}, \bibinfo {author} {\bibfnamefont {V.}~\bibnamefont {L'vov}},
  \bibinfo {author} {\bibfnamefont {J.}~\bibnamefont {Pons}}, \ and\ \bibinfo
  {author} {\bibfnamefont {E.}~\bibnamefont {Cesari}},\ }\href@noop {}
  {\bibfield  {journal} {\bibinfo  {journal} {J. Appl. Phys.}\ }\textbf
  {\bibinfo {volume} {93}},\ \bibinfo {pages} {2394} (\bibinfo {year}
  {2003})}\BibitemShut {NoStop}%
\bibitem [{\citenamefont {Einstein}(1906)}]{Einstein1906_AnnPhysBerlin}%
  \BibitemOpen
  \bibfield  {author} {\bibinfo {author} {\bibfnamefont {A.}~\bibnamefont
  {Einstein}},\ }\href@noop {} {\bibfield  {journal} {\bibinfo  {journal} {Ann.
  Phys. (Berlin)}\ }\textbf {\bibinfo {volume} {324}},\ \bibinfo {pages} {289}
  (\bibinfo {year} {1906})}\BibitemShut {NoStop}%
\bibitem [{\citenamefont
  {Einstein}(1911)}]{Einstein1911_correction_AnnPhysBerlin}%
  \BibitemOpen
  \bibfield  {author} {\bibinfo {author} {\bibfnamefont {A.}~\bibnamefont
  {Einstein}},\ }\href@noop {} {\bibfield  {journal} {\bibinfo  {journal} {Ann.
  Phys. (Berlin)}\ }\textbf {\bibinfo {volume} {339}},\ \bibinfo {pages} {591}
  (\bibinfo {year} {1911})}\BibitemShut {NoStop}%
\bibitem [{\citenamefont {Lopez-Pamies}\ \emph {et~al.}(2013)\citenamefont
  {Lopez-Pamies}, \citenamefont {Goudarzi},\ and\ \citenamefont
  {Danas}}]{LopezPamies2013_JMechPhysSolids}%
  \BibitemOpen
  \bibfield  {author} {\bibinfo {author} {\bibfnamefont {O.}~\bibnamefont
  {Lopez-Pamies}}, \bibinfo {author} {\bibfnamefont {T.}~\bibnamefont
  {Goudarzi}}, \ and\ \bibinfo {author} {\bibfnamefont {K.}~\bibnamefont
  {Danas}},\ }\href@noop {} {\bibfield  {journal} {\bibinfo  {journal} {J.
  Mech. Phys. Solids}\ }\textbf {\bibinfo {volume} {61}},\ \bibinfo {pages}
  {19} (\bibinfo {year} {2013})}\BibitemShut {NoStop}%
\bibitem [{\citenamefont {Huang}\ \emph {et~al.}(2016)\citenamefont {Huang},
  \citenamefont {Pessot}, \citenamefont {Cremer}, \citenamefont {Weeber},
  \citenamefont {Holm}, \citenamefont {Nowak}, \citenamefont {Odenbach},
  \citenamefont {Menzel},\ and\ \citenamefont
  {Auernhammer}}]{Huang2016_SoftMatter}%
  \BibitemOpen
  \bibfield  {author} {\bibinfo {author} {\bibfnamefont {S.}~\bibnamefont
  {Huang}}, \bibinfo {author} {\bibfnamefont {G.}~\bibnamefont {Pessot}},
  \bibinfo {author} {\bibfnamefont {P.}~\bibnamefont {Cremer}}, \bibinfo
  {author} {\bibfnamefont {R.}~\bibnamefont {Weeber}}, \bibinfo {author}
  {\bibfnamefont {C.}~\bibnamefont {Holm}}, \bibinfo {author} {\bibfnamefont
  {J.}~\bibnamefont {Nowak}}, \bibinfo {author} {\bibfnamefont
  {S.}~\bibnamefont {Odenbach}}, \bibinfo {author} {\bibfnamefont {A.~M.}\
  \bibnamefont {Menzel}}, \ and\ \bibinfo {author} {\bibfnamefont {G.~K.}\
  \bibnamefont {Auernhammer}},\ }\href@noop {} {\bibfield  {journal} {\bibinfo
  {journal} {Soft Matter}\ }\textbf {\bibinfo {volume} {12}},\ \bibinfo {pages}
  {228} (\bibinfo {year} {2016})}\BibitemShut {NoStop}%
\bibitem [{\citenamefont {Kramarenko}\ \emph {et~al.}(2015)\citenamefont
  {Kramarenko}, \citenamefont {Chertovich}, \citenamefont {Stepanov},
  \citenamefont {Semisalova}, \citenamefont {Makarova}, \citenamefont {Perov},\
  and\ \citenamefont {Khokhlov}}]{Kramarenko2015_SmartMaterStruct}%
  \BibitemOpen
  \bibfield  {author} {\bibinfo {author} {\bibfnamefont {E.~Y.}\ \bibnamefont
  {Kramarenko}}, \bibinfo {author} {\bibfnamefont {A.~V.}\ \bibnamefont
  {Chertovich}}, \bibinfo {author} {\bibfnamefont {G.~V.}\ \bibnamefont
  {Stepanov}}, \bibinfo {author} {\bibfnamefont {A.~S.}\ \bibnamefont
  {Semisalova}}, \bibinfo {author} {\bibfnamefont {L.~A.}\ \bibnamefont
  {Makarova}}, \bibinfo {author} {\bibfnamefont {N.~S.}\ \bibnamefont {Perov}},
  \ and\ \bibinfo {author} {\bibfnamefont {A.~R.}\ \bibnamefont {Khokhlov}},\
  }\href@noop {} {\bibfield  {journal} {\bibinfo  {journal} {Smart Mater.
  Struct.}\ }\textbf {\bibinfo {volume} {24}},\ \bibinfo {pages} {035002}
  (\bibinfo {year} {2015})}\BibitemShut {NoStop}%
\bibitem [{\citenamefont {Hoang}\ \emph {et~al.}(2009)\citenamefont {Hoang},
  \citenamefont {Zhang},\ and\ \citenamefont
  {Du}}]{Hoang2009_SmartMaterStruct}%
  \BibitemOpen
  \bibfield  {author} {\bibinfo {author} {\bibfnamefont {N.}~\bibnamefont
  {Hoang}}, \bibinfo {author} {\bibfnamefont {N.}~\bibnamefont {Zhang}}, \ and\
  \bibinfo {author} {\bibfnamefont {N.}~\bibnamefont {Du}},\ }\href@noop {}
  {\bibfield  {journal} {\bibinfo  {journal} {Smart Mater. Struct.}\ }\textbf
  {\bibinfo {volume} {18}},\ \bibinfo {pages} {074009} (\bibinfo {year}
  {2009})}\BibitemShut {NoStop}%
\bibitem [{\citenamefont {Chertovich}\ \emph {et~al.}(2010)\citenamefont
  {Chertovich}, \citenamefont {Stepanov}, \citenamefont {Kramarenko},\ and\
  \citenamefont {Khokhlov}}]{Chertovich2010_MacromolMaterEng}%
  \BibitemOpen
  \bibfield  {author} {\bibinfo {author} {\bibfnamefont {A.~V.}\ \bibnamefont
  {Chertovich}}, \bibinfo {author} {\bibfnamefont {G.~V.}\ \bibnamefont
  {Stepanov}}, \bibinfo {author} {\bibfnamefont {E.~Y.}\ \bibnamefont
  {Kramarenko}}, \ and\ \bibinfo {author} {\bibfnamefont {A.~R.}\ \bibnamefont
  {Khokhlov}},\ }\href@noop {} {\bibfield  {journal} {\bibinfo  {journal}
  {Macromol. Mater. Eng.}\ }\textbf {\bibinfo {volume} {295}},\ \bibinfo
  {pages} {336} (\bibinfo {year} {2010})}\BibitemShut {NoStop}%
\bibitem [{\citenamefont {Fujita}\ \emph {et~al.}(1999)\citenamefont {Fujita},
  \citenamefont {Jeyadevan}, \citenamefont {Yamaguchi},\ and\ \citenamefont
  {Nishiyama}}]{Fujita1999_PowderTechnol}%
  \BibitemOpen
  \bibfield  {author} {\bibinfo {author} {\bibfnamefont {T.}~\bibnamefont
  {Fujita}}, \bibinfo {author} {\bibfnamefont {B.}~\bibnamefont {Jeyadevan}},
  \bibinfo {author} {\bibfnamefont {K.}~\bibnamefont {Yamaguchi}}, \ and\
  \bibinfo {author} {\bibfnamefont {H.}~\bibnamefont {Nishiyama}},\ }\href@noop
  {} {\bibfield  {journal} {\bibinfo  {journal} {Powder Technol.}\ }\textbf
  {\bibinfo {volume} {101}},\ \bibinfo {pages} {279} (\bibinfo {year}
  {1999})}\BibitemShut {NoStop}%
\bibitem [{\citenamefont {Wen}\ and\ \citenamefont
  {Sheng}(2003)}]{Wen2003_PhysicaB}%
  \BibitemOpen
  \bibfield  {author} {\bibinfo {author} {\bibfnamefont {W.}~\bibnamefont
  {Wen}}\ and\ \bibinfo {author} {\bibfnamefont {P.}~\bibnamefont {Sheng}},\
  }\href@noop {} {\bibfield  {journal} {\bibinfo  {journal} {Physica B}\
  }\textbf {\bibinfo {volume} {338}},\ \bibinfo {pages} {343} (\bibinfo {year}
  {2003})}\BibitemShut {NoStop}%
\bibitem [{\citenamefont {Wang}\ \emph {et~al.}(2013)\citenamefont {Wang},
  \citenamefont {Yin}, \citenamefont {Liu}, \citenamefont {Yu},\ and\
  \citenamefont {Chen}}]{Wang2013_DaltonTrans}%
  \BibitemOpen
  \bibfield  {author} {\bibinfo {author} {\bibfnamefont {B.}~\bibnamefont
  {Wang}}, \bibinfo {author} {\bibfnamefont {Y.}~\bibnamefont {Yin}}, \bibinfo
  {author} {\bibfnamefont {C.}~\bibnamefont {Liu}}, \bibinfo {author}
  {\bibfnamefont {S.}~\bibnamefont {Yu}}, \ and\ \bibinfo {author}
  {\bibfnamefont {K.}~\bibnamefont {Chen}},\ }\href@noop {} {\bibfield
  {journal} {\bibinfo  {journal} {Dalton Trans.}\ }\textbf {\bibinfo {volume}
  {42}},\ \bibinfo {pages} {10042} (\bibinfo {year} {2013})}\BibitemShut
  {NoStop}%
\bibitem [{\citenamefont {Biller}\ \emph {et~al.}(2014)\citenamefont {Biller},
  \citenamefont {Stolbov},\ and\ \citenamefont
  {Raikher}}]{Biller2014_JApplPhys}%
  \BibitemOpen
  \bibfield  {author} {\bibinfo {author} {\bibfnamefont {A.~M.}\ \bibnamefont
  {Biller}}, \bibinfo {author} {\bibfnamefont {O.~V.}\ \bibnamefont {Stolbov}},
  \ and\ \bibinfo {author} {\bibfnamefont {Y.~L.}\ \bibnamefont {Raikher}},\
  }\href@noop {} {\bibfield  {journal} {\bibinfo  {journal} {J. Appl. Phys.}\
  }\textbf {\bibinfo {volume} {116}},\ \bibinfo {pages} {114904} (\bibinfo
  {year} {2014})}\BibitemShut {NoStop}%
\bibitem [{\citenamefont {Biller}\ \emph {et~al.}(2015)\citenamefont {Biller},
  \citenamefont {Stolbov},\ and\ \citenamefont
  {Raikher}}]{Biller2015_JOptoelectronAdvMater}%
  \BibitemOpen
  \bibfield  {author} {\bibinfo {author} {\bibfnamefont {A.~M.}\ \bibnamefont
  {Biller}}, \bibinfo {author} {\bibfnamefont {O.~V.}\ \bibnamefont {Stolbov}},
  \ and\ \bibinfo {author} {\bibfnamefont {Y.~L.}\ \bibnamefont {Raikher}},\
  }\href@noop {} {\bibfield  {journal} {\bibinfo  {journal} {J. Optoelectron.
  Adv. Mater.}\ }\textbf {\bibinfo {volume} {17}},\ \bibinfo {pages} {1106}
  (\bibinfo {year} {2015})}\BibitemShut {NoStop}%
\bibitem [{\citenamefont {Allahyarov}\ \emph {et~al.}(2015)\citenamefont
  {Allahyarov}, \citenamefont {L\"owen},\ and\ \citenamefont
  {Zhu}}]{Allahyarov2015_PhysChemChemPhys}%
  \BibitemOpen
  \bibfield  {author} {\bibinfo {author} {\bibfnamefont {E.}~\bibnamefont
  {Allahyarov}}, \bibinfo {author} {\bibfnamefont {H.}~\bibnamefont {L\"owen}},
  \ and\ \bibinfo {author} {\bibfnamefont {L.}~\bibnamefont {Zhu}},\
  }\href@noop {} {\bibfield  {journal} {\bibinfo  {journal} {Phys. Chem. Chem.
  Phys.}\ }\textbf {\bibinfo {volume} {17}},\ \bibinfo {pages} {32479}
  (\bibinfo {year} {2015})}\BibitemShut {NoStop}%
\bibitem [{\citenamefont {Bender}\ \emph {et~al.}(2011)\citenamefont {Bender},
  \citenamefont {G\"unther}, \citenamefont {Tsch\"ope},\ and\ \citenamefont
  {Birringer}}]{Bender2011_JMagnMagnMater}%
  \BibitemOpen
  \bibfield  {author} {\bibinfo {author} {\bibfnamefont {P.}~\bibnamefont
  {Bender}}, \bibinfo {author} {\bibfnamefont {A.}~\bibnamefont {G\"unther}},
  \bibinfo {author} {\bibfnamefont {A.}~\bibnamefont {Tsch\"ope}}, \ and\
  \bibinfo {author} {\bibfnamefont {R.}~\bibnamefont {Birringer}},\ }\href@noop
  {} {\bibfield  {journal} {\bibinfo  {journal} {J. Magn. Magn. Mater.}\
  }\textbf {\bibinfo {volume} {323}},\ \bibinfo {pages} {2055} (\bibinfo {year}
  {2011})}\BibitemShut {NoStop}%
\bibitem [{\citenamefont {Tierno}(2014)}]{Tierno2014_PhysChemChemPhys}%
  \BibitemOpen
  \bibfield  {author} {\bibinfo {author} {\bibfnamefont {P.}~\bibnamefont
  {Tierno}},\ }\href@noop {} {\bibfield  {journal} {\bibinfo  {journal} {Phys.
  Chem. Chem. Phys.}\ }\textbf {\bibinfo {volume} {16}},\ \bibinfo {pages}
  {23515} (\bibinfo {year} {2014})}\BibitemShut {NoStop}%
\bibitem [{\citenamefont {Menzel}(2016)}]{Menzel2016_PhysRevE}%
  \BibitemOpen
  \bibfield  {author} {\bibinfo {author} {\bibfnamefont {A.~M.}\ \bibnamefont
  {Menzel}},\ }\href@noop {} {\bibfield  {journal} {\bibinfo  {journal} {Phys.
  Rev. E}\ }\textbf {\bibinfo {volume} {94}},\ \bibinfo {pages} {023003}
  (\bibinfo {year} {2016})}\BibitemShut {NoStop}%
\bibitem [{\citenamefont {Menzel}\ \emph {et~al.}(2009)\citenamefont {Menzel},
  \citenamefont {Pleiner},\ and\ \citenamefont {Brand}}]{Menzel2009_EurPhysJE}%
  \BibitemOpen
  \bibfield  {author} {\bibinfo {author} {\bibfnamefont {A.~M.}\ \bibnamefont
  {Menzel}}, \bibinfo {author} {\bibfnamefont {H.}~\bibnamefont {Pleiner}}, \
  and\ \bibinfo {author} {\bibfnamefont {H.~R.}\ \bibnamefont {Brand}},\
  }\href@noop {} {\bibfield  {journal} {\bibinfo  {journal} {Eur. Phys. J. E}\
  }\textbf {\bibinfo {volume} {30}},\ \bibinfo {pages} {371} (\bibinfo {year}
  {2009})}\BibitemShut {NoStop}%
\bibitem [{\citenamefont {Saadat}\ \emph {et~al.}(2002)\citenamefont {Saadat},
  \citenamefont {Salichs}, \citenamefont {Noori}, \citenamefont {Hou},
  \citenamefont {Davoodi}, \citenamefont {Bar-on}, \citenamefont {Suzuki},\
  and\ \citenamefont {Masuda}}]{Saadat2002_SmartMaterStruct}%
  \BibitemOpen
  \bibfield  {author} {\bibinfo {author} {\bibfnamefont {S.}~\bibnamefont
  {Saadat}}, \bibinfo {author} {\bibfnamefont {J.}~\bibnamefont {Salichs}},
  \bibinfo {author} {\bibfnamefont {M.}~\bibnamefont {Noori}}, \bibinfo
  {author} {\bibfnamefont {Z.}~\bibnamefont {Hou}}, \bibinfo {author}
  {\bibfnamefont {H.}~\bibnamefont {Davoodi}}, \bibinfo {author} {\bibfnamefont
  {I.}~\bibnamefont {Bar-on}}, \bibinfo {author} {\bibfnamefont
  {Y.}~\bibnamefont {Suzuki}}, \ and\ \bibinfo {author} {\bibfnamefont
  {A.}~\bibnamefont {Masuda}},\ }\href@noop {} {\bibfield  {journal} {\bibinfo
  {journal} {Smart Mater. Struct.}\ }\textbf {\bibinfo {volume} {11}},\
  \bibinfo {pages} {218} (\bibinfo {year} {2002})}\BibitemShut {NoStop}%
\bibitem [{\citenamefont {Ozbulut}\ \emph {et~al.}(2011)\citenamefont
  {Ozbulut}, \citenamefont {Hurlebaus},\ and\ \citenamefont
  {Desroches}}]{Ozbulut2011_JIntellMaterSystStruct}%
  \BibitemOpen
  \bibfield  {author} {\bibinfo {author} {\bibfnamefont {O.~E.}\ \bibnamefont
  {Ozbulut}}, \bibinfo {author} {\bibfnamefont {S.}~\bibnamefont {Hurlebaus}},
  \ and\ \bibinfo {author} {\bibfnamefont {R.}~\bibnamefont {Desroches}},\
  }\href@noop {} {\bibfield  {journal} {\bibinfo  {journal} {J. Intell. Mater.
  Syst. Struct.}\ }\textbf {\bibinfo {volume} {22}},\ \bibinfo {pages} {1531}
  (\bibinfo {year} {2011})}\BibitemShut {NoStop}%
\bibitem [{\citenamefont {Ramanujan}\ and\ \citenamefont
  {Lao}(2006)}]{Ramanujan2006_SmartMaterStruct}%
  \BibitemOpen
  \bibfield  {author} {\bibinfo {author} {\bibfnamefont {R.~V.}\ \bibnamefont
  {Ramanujan}}\ and\ \bibinfo {author} {\bibfnamefont {L.~L.}\ \bibnamefont
  {Lao}},\ }\href@noop {} {\bibfield  {journal} {\bibinfo  {journal} {Smart
  Mater. Struct.}\ }\textbf {\bibinfo {volume} {15}},\ \bibinfo {pages} {952}
  (\bibinfo {year} {2006})}\BibitemShut {NoStop}%
\bibitem [{\citenamefont {Shahinpoor}\ \emph {et~al.}(2007)\citenamefont
  {Shahinpoor}, \citenamefont {Kim},\ and\ \citenamefont
  {Mojarrad}}]{Shahinpoor2007_book}%
  \BibitemOpen
  \bibfield  {author} {\bibinfo {author} {\bibfnamefont {M.}~\bibnamefont
  {Shahinpoor}}, \bibinfo {author} {\bibfnamefont {K.~J.}\ \bibnamefont {Kim}},
  \ and\ \bibinfo {author} {\bibfnamefont {M.}~\bibnamefont {Mojarrad}},\
  }\href@noop {} {\emph {\bibinfo {title} {Artificial Muscles: Applications of
  Advanced Polymeric Nanocomposites}}}\ (\bibinfo  {publisher} {CRC Press},\
  \bibinfo {year} {2007})\BibitemShut {NoStop}%
\end{thebibliography}%

\end{document}